\pdfoutput=1

\documentclass{article}

\usepackage[utf8]{inputenc} % allow utf-8 input
\usepackage[T1]{fontenc}    % use 8-bit T1 fonts
\usepackage{hyperref}       % hyperlinks
\usepackage{url}            % simple URL typesetting
\usepackage{booktabs}       % professional-quality tables
\usepackage{amsfonts}       % blackboard math symbols
\usepackage{nicefrac}       % compact symbols for 1/2, etc.
\usepackage{microtype}      % microtypography
\usepackage{xcolor}         % colors

\usepackage[numbers]{natbib}

\usepackage{enumitem}
\usepackage{caption,bm}
\usepackage{subfigure}
\usepackage{graphicx}
\usepackage{xspace,color}
\usepackage{diagbox,multirow} 
\usepackage{subfigure}
\usepackage{algorithm}
\usepackage{algorithmic}
\usepackage{amsmath,amssymb,bbm}

\usepackage[english]{babel}
\usepackage{amsthm}

\newcommand{\std}[1]{\scriptsize{$\pm$#1}}

\newcommand{\fullname}{Reinforced Genetic Algorithm\ }
\newcommand{\mname}{\texttt{RGA}\xspace }

\newcommand{\bfz}{\mathbf{z}}
\newcommand{\bfZ}{\mathbf{Z}}
\newcommand{\bfa}{\mathbf{a}}

\newcommand{\bfh}{\mathbf{h}}

\newcommand{\bfD}{\mathbf{D}}
\newcommand{\bfH}{\mathbf{H}}

\newcommand{\bfw}{\mathbf{w}}

\newcommand{\bfv}{\mathbf{v}}

\newcommand{\EB}{\mathbb{E}}

\newcommand{\RB}{\mathbb{R}}

\newcommand{\calT}{\mathcal{T}}
\newcommand{\calR}{\mathcal{R}}

\newcommand{\calQ}{\mathcal{Q}}

\newcommand{\calV}{\mathcal{V}}

\newcommand{\calS}{\mathcal{S}}

\newcommand{\calA}{\mathcal{A}}

\newcommand{\calZ}{\mathcal{Z}}

\newcommand{\calY}{\mathcal{Y}}

\newcommand{\marked}{}

\title{Reinforced Genetic Algorithm for Structure-based Drug Design}

\author{Tianfan Fu$^{1*}$, Wenhao Gao$^{2*}$, 
Connor W. Coley$^{2, 3}$, Jimeng Sun$^{4, 5}$,  \\ 
$^{1}$Department of Computational Science and Engineering, \\
Georgia Institute of Technology, \\
$^{2}$Department of Chemical Engineering,\\ Massachusetts Institute of Technology,\\ 
$^{3}$Department of Electrical Engineering and Computer Science, \\Massachusetts Institute of Technology,\\ 
$^4$ Department of Computer Science, \\ 
University of Illinois at Urbana-Champaign,\\
$^5$ Carle Illinois College of Medicine, \\ 
University of Illinois at Urbana-Champaign,\\
$^{*}$Equal Contributions \\[2ex]%
\texttt{tfu42@gatech.edu},
\texttt{\{whgao,ccoley\}@mit.edu},
\texttt{jimeng@illinois.edu} 
}

\begin{document}

\maketitle

\begin{abstract}
% Structure-based drug design aims at generating novel molecules that could bind to the given target (related to the disease). %, which is a fundamental task and long-standing challenge in pharmaceutical industry. 
% Genetic algorithm (GA) often demonstrates state-of-the-art performance in structure-based drug design thanks to its flexible (molecular) assembling manner. 
% However, the existing GA method relies heavily on random exploration in chemical space through brute-force trial and error. 
% On the other hand, reinforcement learning (RL) methods have exhibited superior ability in finding prioritized searching branches when navigating the discrete space, but the existing RL-based drug design methods suffers from its auto-regressive assembling manner when growing molecules. 
% In this paper, we propose a \fullname (\mname) that inherits the advantage of RL and GA. Specifically, we use RL to prioritizes the promising GA operations and suppresses random-walk behaviour in GA. 
% We conduct thorough empirical studies on optimizing docking score for various targets to validate the effectiveness of the proposed methods, outperforming the best baseline method consistently on different targets in terms of docking score. 

Structure-based drug design (SBDD) aims to discover drug candidates by finding molecules (ligands) that bind tightly to a disease-related protein (targets), which is the primary approach to computer-aided drug discovery.
Recently, applying deep generative models for three-dimensional (3D) molecular design conditioned on protein pockets to solve SBDD has attracted much attention, but their formulation as probabilistic modeling often leads to unsatisfactory optimization performance. 
On the other hand, traditional combinatorial optimization methods such as genetic algorithms (GA) have demonstrated state-of-the-art performance in various molecular optimization tasks. However, they do not utilize protein target structure to inform design steps but rely on a random-walk-like exploration, which leads to unstable performance and no knowledge transfer between different tasks despite the similar binding physics.
To achieve a more stable and efficient SBDD, we propose \fullname (\mname) that uses neural models to prioritize the profitable design steps and suppress random-walk behavior. 
The neural models take the 3D structure of the targets and ligands as inputs and are pre-trained using native complex structures to utilize the knowledge of the shared binding physics from different targets and then fine-tuned during optimization. 
We conduct thorough empirical studies on optimizing binding affinity to various disease targets and show that \mname outperforms the baselines in terms of docking scores and is more robust to random initializations. The ablation study also indicates that the training on different targets helps improve the performance by leveraging the shared underlying physics of the binding processes. 
The code is available at \url{https://github.com/futianfan/reinforced-genetic-algorithm}. 
\end{abstract}

% Guiding deep molecular optimization with genetic exploration   https://arxiv.org/pdf/2007.04897.pdf  

% AutoGrow4: an open-source genetic algorithm for de novo drug design and lead optimization 
% https://jcheminf.biomedcentral.com/track/pdf/10.1186/s13321-020-00429-4.pdf 

% GCPN   https://arxiv.org/pdf/1806.02473.pdf 

% REINVENT  https://arxiv.org/pdf/1704.07555.pdf 

% A 3D Generative Model for Structure-Based Drug Design, Luo Shitong, https://openreview.net/pdf?id=yDwfVD_odRo  

% Structure-based de novo drug design using 3D deep generative models, Yibo Li, https://pubs.rsc.org/en/content/articlepdf/2021/sc/d1sc04444c  

\section{Introduction}
\label{sec:intro}

Rapid drug discovery that requires less time and cost is of significant interest in pharmaceutical science.
%, whose importance has been highlighted in the recent pandemic. 
Structure-based drug design (SBDD)~\cite{bohacek1996art} that leverages the three-dimensional (3D) structures of the disease-related proteins to design drug candidates is one primary approach to accelerate the drug discovery processes with physical simulation and data-driven modeling. 
According to the lock and key model~\cite{tripathi2017molecular}, the molecules that bind tighter to a disease target are more likely to expose bioactivity against the disease, which has been verified experimentally~\cite{alon2021structures}.
As AlphaFold2 has provided accurate predictions to most human proteins~\cite{jumper2021highly,varadi2022alphafold}, SBDD has a tremendous opportunity to discover new drugs for new targets that we cannot model before~\cite{ren2022alphafold}. 

% It utilizes the three-dimensional (3D) structures of the disease-related proteins and aims to find molecules (ligands) that bind tightly to the target proteins. 

% Designing new chemical compounds with desirable properties is an essential problem and long-standing challenge in chemical and pharmaceutical science~\cite{drews2000drug}. 
% Traditional drug design is a notoriously time-consuming and laborious process that takes tens of years and billions of dollars~\cite{dimasi2016innovation}. 
% For example, high-throughput screening (HTS) technologies, the dominating traditional approaches, search over the existing molecule databases exhaustively~\cite{wouters2020estimated}. 
% On the other hand, the number of the drug-like molecules is large as estimated to be $10^{60}$~\cite{polishchuk2013estimation} and it is prohibitive to enumerate all the possible molecules. 
% To address the issue, machine learning approaches have been developed to accelerate the process. 

SBDD could be formulated as an optimization problem where the objective function is the binding affinity estimated by simulations such as docking~\cite{tripathi2017molecular}. The most widely used design method is virtual screening, which exhaustively investigates every molecule in a library and ranks them. Lyu et al. successfully discovered new chemotypes for AmpC $\beta$-lactamase and the D$_4$ dopamine receptor by studying hundreds of millions of molecules with docking simulation~\cite{lyu2019ultra}. However, the number of the drug-like molecules is large as estimated to be $10^{60}$~\cite{bohacek1996art}, and it is computationally prohibitive to screen all of the possible molecules. Though machine learning approaches have been developed to accelerate screening~\cite{graff2021accelerating, gentile2022artificial}, it is still challenging to screen large enough chemical space within the foreseeable future.

Instead of naively screening a library, designing drug candidates with generative models has been highlighted as a promising strategy, exemplified by \cite{luo20213d,li2021structure}. This class of methods models the problem as the generation of ligands conditioned on the protein pockets. However, as generative models are trained to learn the distribution of known active compounds, they tend to produce molecules similar to training data~\cite{walters2020assessing}, which discourages finding novel molecules and leads to unsatisfactory optimization performance.

A more straightforward solution is a combinatorial optimization algorithm that searches the implicitly defined discrete chemical space. As shown in multiple standard molecule optimization benchmarks~\cite{brown2019guacamol, huang2021therapeutics,gao2022sample}, combinatorial optimization methods, especially genetic algorithms (GA)~\cite{jensen2019graph, spiegel2020autogrow4}, often perform better than deep generative models. 
The key to superior performance is GA's action definition. Specifically, in each generation (iteration), GA maintains a population of possible candidates (a.k.a. parents) and conducts the crossover between two candidates and mutation from a single candidate to generate new offspring. These two types of actions, crossover and mutation, enable global and local traversal over the chemical space, allowing a thorough exploration and superior optimization performance. 

However, most GAs select mutation and crossover operations randomly~\cite{jensen2019graph}, leading to significant variance between independent runs. Especially in SBDD, when the oracle functions are expensive molecular simulations, it is resource-consuming to ensure stability by running multiple times.
Further, most current combinatorial methods are designed for general-purpose molecular optimization and simply use a docking simulation as an oracle. It is challenging to leverage the structure of proteins in these methods, and we need to start from scratch whenever we change a protein target, even though the physics of ligand-protein interaction is shared. Ignoring the shared information across tasks leads to unnecessary exploration steps and, thus, demands for many more oracle calls, which require expensive and unnecessary simulations~\cite{tripp2021fresh}.

To overcome these issues in the GA method, we propose \fullname (\mname), which attempts to reformulate an evolutionary process as a Markov decision process and uses neural networks to make informed decisions and suppress the random-walk behavior. Specifically, we utilize an E(3)-equivariant neural network~\cite{satorras2021n} to choose parents and mutation types based on the 3D structure of the ligands and proteins. The networks are pre-trained with various native complex structures to utilize the knowledge of the shared binding physics between different targets and then fine-tuned with a reinforcement learning algorithm during optimizations. 
We test \mname's performance with various disease-related targets, including the main protease of SARS-CoV-2.

The main contributions of this paper can be summarized as follows:
\begin{itemize}[leftmargin=5mm]
  \item We propose an evolutionary Markov decision process (EMDP) that reformulates an evolutionary process as a Markov decision process, where the state is a population of molecules instead of a single molecule (Section~\ref{sec:mdp}). 
  \item We show the first successful attempt to use a neural model to guide the crossover and mutation operations in a genetic algorithm to suppress random-walk behavior and explore the chemical space intelligently (Section~\ref{sec:policy_net}). 
  \item We present a structure-based de novo drug design algorithm that outperforms baseline methods consistently through thorough empirical studies on optimizing binding affinity by leveraging the underlying binding physics (Section~\ref{sec:experiment}). 
\end{itemize}

\section{Related Works}

We will discuss the related works on methods of drug design and discuss the advantage of the proposed method over the existing works. 

% molecule generation from the following three aspects: task, assembling strategy and methodology. Also, we discuss the advantage of the proposed method over the existing works. 

\noindent\textbf{General Molecular Design}. Molecular generation methods offer a promising direction for the automated design of molecules with desired pharmaceutical properties such as synthesis accessibility and drug-likeliness. 
Based on how to generate or search molecules, these approaches can be categorized into two types, (1) deep generative models (DGMs) imitate the molecular data distribution, including variational autoencoder (VAE)~\cite{gomez2018automatic,jin2018junction}, generative adversarial network (GAN)~\cite{guimaraes2017objective,cao2018molgan}, normalizing flow model~\cite{shi2019graphaf,luo2021graphdf}, energy based model~\cite{liu2021graphebm};
% and diffusion model~\cite{xu2021geodiff}; 
and (2) combinatorial optimization methods directly search over the discrete chemical space, including genetic algorithm (GA)~\cite{jensen2019graph,nigam2019augmenting,gao2021amortized}, reinforcement learning approaches (RL)~\cite{Olivecrona,You2018-xh,zhou2019optimization,jin2020multi,ahn2020guiding}, Bayesian optimization (BO)~\cite{korovina2020chembo}, 
% Monte Carlo tree search (MCTS)~\cite{yang2020practical,jin2020multi}, 
Markov chain Monte Carlo (MCMC)~\cite{fu2021mimosa,xie2021mars,bengio2021gflownet} and gradient ascent~\cite{fu2021differentiable, shen2021deep}. 

General molecular design algorithms often use general black-box oracle functions, and some are only tested with trivial or self-designed oracles. For example, using penalized octanol-water partition coefficient (LogP) as the oracle function, it grows monotonically with the number of carbons, and thus there exists a trivial policy to optimize LogP. These oracles do not reflect the challenges of real drug discovery, and those algorithms have limited value for pharmaceutical discovery. Recent works are optimizing docking scores to simulate a more realistic discovery scenario~\cite{cieplinski2020we,steinmann2021using,tripp2021fresh,yang2021hit}, same as our work. However, they are still ignoring the information in the given protein structures that could potentially accelerate the design process. However, the extension to leveraging the structural knowledge is nontrivial.

\noindent\textbf{Structure-based Drug Design}. 
Structure-based drug design (SBDD) could utilize the structural information to guide the design of molecules, which are potentially more efficient in drug discovery tasks but poses additional challenges of how to leverage the structures. 
Since early 1990s, various SBDD algorithms have been proposed, mostly based on combinatorial optimization algorithms such as tree search~\cite{luo1996rasse,gillet1993sprout,pearlman1993concepts} and evolutionary algorithms~\cite{douguet2000genetic,durrant2013autogrow}.
Those methods typically optimize the ligands in the pockets according to a physical model characterizing the binding affinity.
For example, RASSE~\cite{luo1996rasse} used a force-field-like scoring function~\cite{wang1998score} to evaluate the partial solutions within a tree search. 
However, obtaining a fast and accurate model to quantify binding free energy itself is still an unsolved challenge.

Recently, generative modeling of 3D molecules conditioned on protein targets is attracting more attention~\cite{luo20213d, li2021structure}.
Similar to DGMs in general molecular design, those methods learn the atom's compositional and spatial distribution of native structure of protein-ligand complexes with neural models and design new ligands by complete the complex structure given targets.
Deep generative models are end-to-end and data-driven thus surpass the necessity of understanding the physics of interaction. 
However, as the training objective is to learn the distribution of known active compounds, the models tend to produce molecules close to the training set~\cite{walters2020assessing}, which is undesired in terms of patentability and leads to unsatisfactory optimization performance.

\section{Method}
\label{sec:method}

In this paper, we focus on structure-based drug design. The goal is to design drug molecules (a.k.a. ligands) that could bind tightly with the disease-related proteins (a.k.a. targets). Given the 3D structures of the target proteins, including binding site information, docking is a popular computational method for assessing the binding affinity, which can be roughly retrieved as the free energy changes during the binding processes. 
We present a variant of genetic algorithm that is guided by reinforcement learning and a docking oracle.
Next, we will first describe the general evolutionary process used in genetic algorithms (Section~\ref{sec:ga}); Then we will present how to model this evolutionary process as a Markov decision process (MDP) where RL framework can be constructed (Section~\ref{sec:mdp}); After that, we describe the detailed implementation of this MDP framework using multiple policy networks (Section~\ref{sec:policy_net}).

% \noindent\textbf{Docking simulation}. 
% The purpose of target-ligand docking is to find the optimal binding between a small molecule (ligand) and a target (target protein). 
% Docking can be conducted using well-commercialized software, such as AutoDock Vina~\cite{trott2010autodock}. The input is a 2D molecular graph, the 3D geometric shape of target and the corresponding binding site. The output is the 3D pose and relative position of the ligand (binds to target) that corresponds to the best binding affinity score. 
% In this paper, we use $X$ to denote the ligand (including its 3D pose that binds to the target), $\calT$ to denote 3D target structures. The mathematical notation table is available in Appendix. 
% In this paper, we leverage one 3D pose and relative coordinates of the ligand that corresponds to the best binding affinity score. 

\subsection{Evolutionary Process}
\label{sec:ga}

In this section, we introduce the primary setting of the evolutionary processes. With both optimization performance and synthetic accessibility taken into account~\cite{gao2020synthesizability,huang2021therapeutics}, we follow the action settings in Autogrow 4.0~\cite{spiegel2020autogrow4}. It demonstrated superior performance over other GA variants in the empirical validation of structure-based drug design~\cite{spiegel2020autogrow4}, and its mutation actions originated from chemical reactions so that the designed molecules are more likely to be synthesizable.
Specifically, an evolutionary processes starts by randomly sampling a \textit{population} of drug candidates from a library. In each \textit{generation} (iteration), it carries out (i) \textit{crossover} between parents selected from the last generation, and (ii) \textit{mutation} on a single child to obtain the offspring pool. An illustration of both crossover and mutation operations is available in Appendix. Note that we only adopted the action settings from Autogrow 4.0, without using other tricks such as elitism.

% an illustration, whose superiority over other GA variants is empirically validated in structure-based drug design~\cite{spiegel2020autogrow4}. 

% Specifically, GA starts from a \textit{population} of molecule candidates. In each \textit{generation} (iteration), it carries out (I) \textit{mutation} and (II) \textit{crossover} to obtain the offspring, as illustrated in Figure~\ref{fig:ga}. 

\noindent\textbf{Crossover}, also called recombination, combines the structure of two parents to generate new children. Following Autogrow 4.0~\cite{spiegel2020autogrow4}, we select two parents from the last generation and search for the largest common substructure shared between them. Then we generate two children by randomly switching their decorating moieties, i.e., the side chains attached to the common substructure. 

\noindent\textbf{Mutation} operates on a single parent molecule and modifies its structure slightly. Following Autogrow 4.0~\cite{spiegel2020autogrow4}, we adopt transformations based on chemical reactions. Unlike naively defined atom-editing actions, mutation steps based on chemical reactions could ensure all modification is reasonable in reality, leading to a larger probability of designing synthesizable molecules.
We included two types of chemical reactions: uni-molecular reactions, which only require one reactant, and bi-molecular reactions, which require two reactants. While uni-molecular reactions could be directly applied to the parent, we sample a purchasable compound to react with the parent when conducting a bi-molecular reaction. In both cases, the parent serves as one reactant, and we use the main product as the child molecule. 
We use the chemical reactions from \cite{spiegel2020autogrow4}, which was originally from \cite{durrant2013autogrow,hartenfeller2011collection}.

\noindent\textbf{Evolution}. At the $t$-th generation (iteration), given a population of molecules denoted as $\mathcal{S}^{(t)}$, we generate an offspring pool denoted as $\mathcal{Q}^{(t)}$ by applying crossover and mutation operations. Then we filter out the ones with undesirable physical and chemical properties (e.g., poor solubility, high toxicity) in the offspring pool and select the most promising $K$ to form the next generation pool ($\mathcal{S}^{(t+1)}$).

\begin{figure}[t]
\centering
\vspace{-15pt}
% \subfigure{\includegraphics[width=0.44\linewidth]{figure/ga_pipeline2.png}}
% \subfigure{\includegraphics[width=0.55\linewidth]{figure/pipeline.png}}
\includegraphics[width=0.86\linewidth]{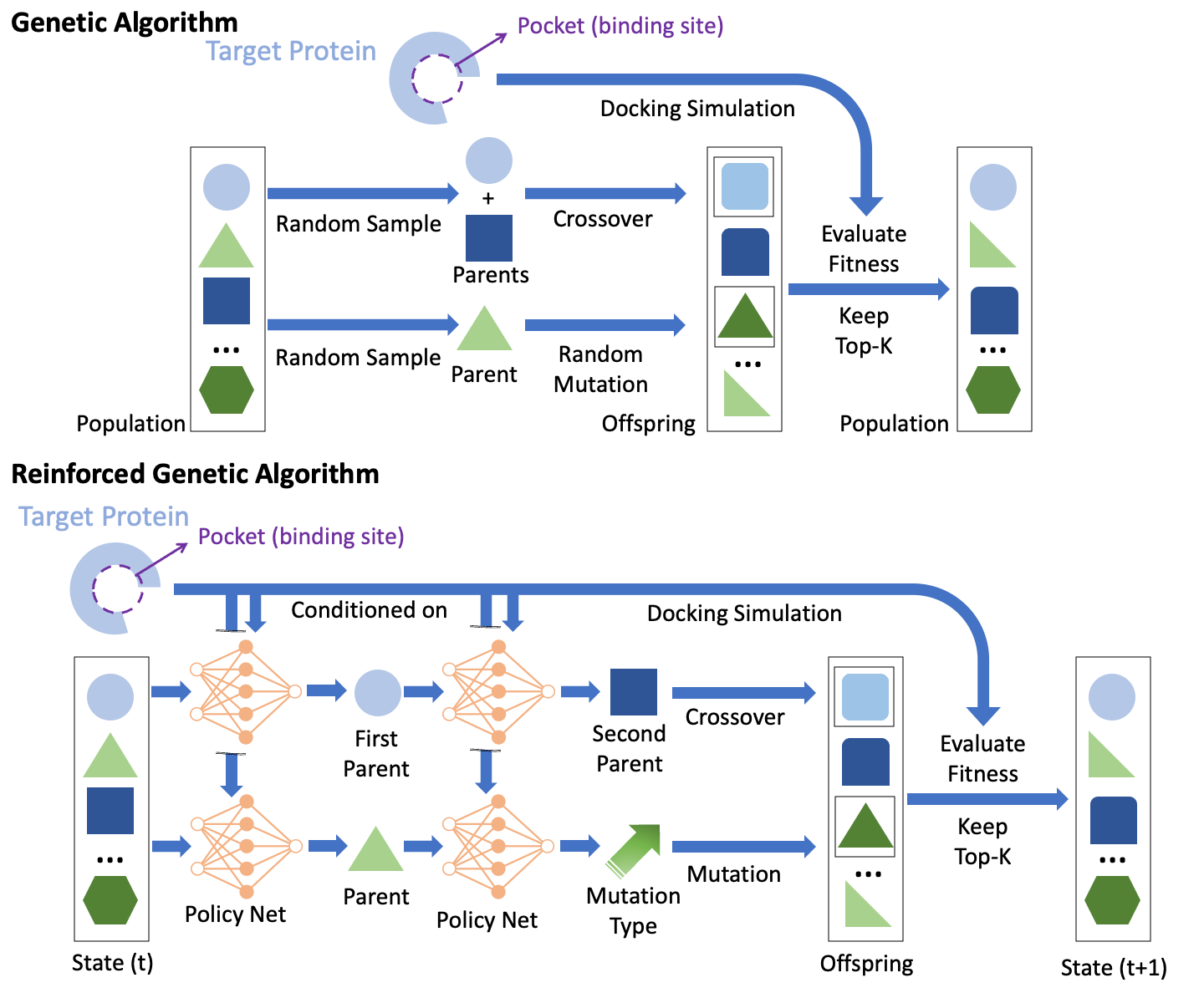}
\caption{
We illustrate one generation (iteration) of GA (top) and \mname pipeline (bottom). 
Specifically, we train policy networks that take the target and ligand as input to make informed choices on parents and mutation types in \mname.  
}
\vspace{-10pt}
\label{fig:pipeline}
\end{figure}

\subsection{Evolutionary Markov Decision Process}
\label{sec:mdp}

Next we propose the evolutionary Markov decision process (EMDP) that formulates an evolutionary process of genetic algorithm as a Markov decision process (MDP). The primary purpose is to utilize reinforcement learning algorithms to train networks to inform the decision steps to replace random selections. Taking a generation as a state, Markov property that requires $P(\mathcal{S}^{(t+1)} | \mathcal{S}^{(1)}, \cdots, \mathcal{S}^{(t)}) = P(\mathcal{S}^{(t+1)} | \mathcal{S}^{(t)}) $ is naturally satisfied by the evolutionary process described above, where $\mathcal{S}^{(t)}$ denotes the state at the $t$-th generation, which is the population of ligands. We use $X$ to denote a ligand. 
We elaborate essential components for Markov decision process as follows, and the EMDP pipeline is illustrated in Figure~\ref{fig:pipeline}. 

\noindent\textbf{State Space}. We define the population at the $t$-step generation, $\mathcal{S}^{(t)}$, in the evolutionary process as the state at the $t$-step in an EMDP. A state includes population of candidate molecules (i.e., ligand, denoted $X$) and their 3D poses docked to the target, fully observable to the RL agent. At the beginning of the EMDP, we randomly select a population of candidate molecules and use docking simulation to yield their 3D poses as the initial state. 

% In this section, we formulate the generation process of genetic algorithm as a Markov decision process (MDP), which requires that the state transition satisfies the Markov property: $P(\mathcal{S}^{(t+1)} | \mathcal{S}^{(1)}, \cdots, \mathcal{S}^{(t)}) = P(\mathcal{S}^{(t+1)} | \mathcal{S}^{(t)}) $, where $\mathcal{S}^{(t)}$ denotes the state at the $t$-th generation. 
% Markov property is naturally satisfied for GA described above. 
% % Figure~\ref{fig:framework} illustrates a generation process xxxxx
% Then we elaborate four essential components for Markov decision process as follows.  

% \noindent\textbf{State Space}. We define the state $\mathcal{S}^{(t)}$ at the $t$-step generation as the population of candidates and corresponding 3D poses with regard to the target, which is fully observable (by the RL agent). At the beginning of the generation, we randomly select a population of candidate molecules and use docking simulation to yield their 3D poses as the initial state. 

\noindent\textbf{Action Space}. The actions in an EMDP are to conduct the two evolutionary steps: crossover and mutation, in a population. For each evolutionary step, we need two actions to conduct it. Concretely, crossover ($X^{\text{parent 1}}_{\text{crossover}}, X^{\text{parent 2}}_{\text{crossover}} \xrightarrow[]{\text{crossover}} X^{\text{child 1}}_{\text{crossover}}, X^{\text{child 2}}_{\text{crossover}}$) can be divided to two steps:
% \begin{enumerate}
%  \item 
(1) select the first candidate ligand $X^{\text{parent 1}}_{\text{crossover}}$ from the current state (population $\mathcal{S}^{(t)}$);
%  \item 
(2) conditioned on the first selected candidate $X^{\text{parent 1}}_{\text{crossover}}$, select the second candidate ligand $X^{\text{parent 2}}_{\text{crossover}}$ from the remaining candidate ligand set $\calS^{(t)} - \{X^{\text{parent 1}}_{\text{crossover}} \}$ and apply crossover (Section~\ref{sec:ga}) to them. 
% \end{enumerate}

Mutation ($X^{\text{parent}}_{\text{mutation}} \xrightarrow[]{\text{mutated by}\ \xi} X^{\text{child}}_{\text{mutation}}$) can be divided to two steps:
% \begin{enumerate}
%  \item 
(1) select the candidate ligand $X^{\text{parent}}_{\text{mutated}}$ to be mutated from the current state (population $\mathcal{S}^{(t)}$); 
%  \item 
(2) conditioned on the selected candidate ligand $X^{\text{parent}}_{\text{mutated}}$, select the reaction $\xi$ from the reaction set $\mathcal{R}$ and apply it to $X^{\text{parent}}_{\text{mutated}}$. 
% \end{enumerate}

As applying the crossover and mutation steps are deterministic, the actions in an EMDP focus on selecting parents and mutation types. Upon finish the action, we could obtain offspring pool, $\mathcal{Q}^{(t)}$.

\noindent\textbf{State Transition Dynamics}. 
The state transition in an EMDP is identical to the evolution in an evolutionary process. Once we finish the actions and obtain the offspring pool, $\mathcal{Q}^{(t)}=\{X^{\text{child 1}}_{\text{}}, X^{\text{child 2}}_{\text{}}, \cdots \}$, we apply molecular quality filters to filter out the ones unlikely to be drug and then select the most promising $K$ to form the parent set for the next generation ($\mathcal{S}^{(t+1)}$). 

\noindent\textbf{Reward}. We define the reward as the binding affinity change (docking score). The actions leading to stronger binding score would be prioritized. As there is no ``episode'' concept in an EMDP, we treat every step equally.%, i.e., no intermediate and final reward.

\subsection{Target-Ligand Policy Network}
\label{sec:policy_net}

To utilize molecular structures' translational and rotational invariance, we adopt equivariance neural networks (ENNs)~\cite{satorras2021n} as the target-ligand policy neural networks to select the actions in both mutation and crossover steps. Each ligand has a 3D pose that binds to the target protein, and the complex serves as the input of ENN.

% We design target-ligand policy neural networks to select the actions in both mutation and crossover operations. Each ligand has 3D pose that binds to the 3D target structure. We use an equivariance neural network (ENN)~\cite{satorras2021n} to model the 3D geometric shape of target/ligand/target-ligand complex to learn the selection probability of the ligand. 
% The equivariance neural network achieves state-of-the-art performance in modeling 3D geometric molecule data~\cite{townshend2021atom3d} and is translation- and rotation- invariant with regard to the 3D graph. That is, the translation and rotation on the 3D graph (coordinates) would not change ENN's output. 

Specifically, we want to model a 3D graph $\calY$, which can be ligand, target, or target-ligand complex. The input feature can be described as $\calY = (\calA, \calZ)$, where $\calA$ represents atoms' categories (the vocabulary set $\calV =\{H, C, O, N, \cdots\}$) and $\calZ$ represents 3D coordinates of the atoms. 
Suppose $\bfD \in \RB^{|\calV|\times d}$ is the embedding matrix of all the categories of atoms in a vocabulary set $\calV$, is randomly initialized and learnable, $d$ is the hidden dimension in ENN. 
Each kind of atom corresponds to a row in $\bfD$. 
We suppose there are $N$ atoms, and each atom corresponds to a node in the 3D graph. 
Node embeddings at the $l$-th layer are denoted as $\bfH^{(l)} = \{\bfh^{(l)}_i\}_{i=1}^{N}$, where $l=0,1,\cdots,L$, $L$ is number of layers in ENN. 
The initial node embedding $\bfh_i^{(0)} = \bfD^\top {\bfa_i} \in \RB^{d}$ embeds the $i$-th node, where ${\bfa_i}$ is one-hot vector that encode the category of the $i$-th atom.  
Coordinate embeddings at the $l$-th layer are denoted $\bfZ^{(l)} = \{\bfz_i^{(l)}\}_{i=1}^{N}$. 
The initial coordinate embeddings $\bfZ^{(0)} = \{\bfz_i^{}\}_{i=1}^{N}$ are the real 3D coordinates of all the nodes.  
The following equation defines the feedforward rules of ENN, for $i,j=1,\cdots,N, \ i\neq j,\ l=0,1,\cdots,L-1$, we have 
\begin{equation}
\label{eqn:ENN}
\begin{aligned}
& \bfw^{(l+1)}_{ij} = \text{MLP}_{e} \Big( \bfh^{(l)}_i \oplus \bfh^{(l)}_j \oplus || \bfz_i^{(l)} - \bfz_j^{(l)} ||_2^2 \Big) \ \   \in \RB^{d}, 
% \\
& \bfv^{(l+1)}_i = \sum_{j=1,j\neq i}^{N} \bfw^{(l+1)}_{ij} \ \  \in \RB^{d}, \\
& \bfz_i^{(l+1)} = \bfz_i^{(l)} + \sum_{j=1,j\neq i}^{N} \Big(\bfz_i^{(l)} - \bfz_j^{(l)} \Big) \text{MLP}_{x} \Big(\bfw^{(l)}_{ij} \Big) \in \RB^{3}, 
% \\
& \bfh^{(l+1)}_i = \text{MLP}_h \Big(\bfh_i^{(l)} \oplus \bfv^{(l+1)}_i \Big) \in \RB^{d}, \\
& \bfh_{\calY} = \sum_{i=1}^{N} \bfh_{i}^{(L)} \ \in \RB^{d} \ \ \ \  \ \ \implies \underline{\bfh_{\calY} = \text{ENN}(\calY)}
\end{aligned}
\end{equation}
where $\oplus$ denotes the concatenation of vectors; $\text{MLP}_{e}(\cdot): \RB^{2d+1}\xrightarrow[]{}\RB^{d}; 
\text{MLP}_{x}(\cdot): \RB^{d}\xrightarrow[]{}\RB; 
\text{MLP}_{h}(\cdot): \RB^{2d}\xrightarrow[]{}\RB^{d}$ are all two-layer multiple layer perceptrons (MLPs) with Swish activation in the hidden layer \cite{ramachandran2017searching}. 
At the $l$-th layer, $\bfw^{(l)}_{ij}$ represents the message vector for the edge from node $i$ to node $j$; $\bfv^{(l)}_{i}$ represents the message vector for node $i$, $\bfz^{(l)}_i$ is the position embedding for node $i$; $\bfh_i^{(l)}$ is the node embedding for node $i$. 
$\bfH^{(L)} = [\bfh_1^{(L)}, \cdots, \bfh_N^{(L)}]$ are the node embeddings of the $L$-th (last) layer. 
We aggregate them using sum function as readout function to obtain a representation of the 3D graph, denoted $\bfh_{\calY}$. The whole process is written as $\bfh_{\calY} = \text{ENN}(\calY)$.

\noindent\textbf{Crossover Policy Network}. 
We design two policy networks for two corresponding actions in a crossover, as mentioned in Section~\ref{sec:mdp}. 
(1) the first action in crossover operation is to select the first parent ligand $X^{\text{parent 1}}_{\text{crossover}}$ from the population $\mathcal{S}^{(t)}$. 
Similar to the first action in mutation operation, we obtain a valid probability distribution over all the available ligands based on target-ligand complex as input feature and ENN as the neural network architecture, the selection probability of the ligand $X^{\text{parent 1}}_{\text{crossover}} \in \calS^{(t)} $ is  $p_{\text{crossover}}^{(1)}(X^{\text{parent 1}}_{\text{crossover}} | \calS^{(t)}) = \frac{\exp{\big( \text{MLP} (\bfh_{\calT \& X^{\text{parent 1}}_{\text{crossover}}} )\big)}}{\sum_{X'\in \calS^{(t)}} \exp{\big( \text{MLP}(\bfh_{\calT\& X'})\big)}}$, where $\calT$ and $X$ denotes target and ligand (including 3D pose), respectively, $\calT \& X$ denotes target-ligand complex. 
% It corresponds to the policy network in Figure~\ref{fig:pipeline}. 
(2) The second action is to select the second parent ligand conditioned on the first parent ligand selected in the first action. 
Specifically, for ligand in the remaining population set, we concatenate the ENN's embedding of the target, first parent ligand $X^{\text{parent 1}}_{\text{crossover}}$ and the second parent ligand \marked{$X^{\text{parent 2}}_{\text{mutation}}$}, and feed it into an MLP to estimate a scalar as an unnormalized probability. 
The unnormalized probabilities for all the ligands in the remaining population set are normalized via softmax function, i.e., 
$p_{\text{crossover}}^{(2)}(X^{\text{parent 2}}_{\text{crossover}} | X^{\text{parent 1}}_{\text{crossover}}, \mathcal{S}^{(t)} ) = \text{Softmax} \big\{\text{MLP}(\bfh_{\calT}\oplus \bfh_{X^{\text{parent 1}}_{\text{crossover}}} \oplus \bfh_{X^{\text{parent 2}}_{\text{crossover}}}), \cdots, \big\}_{X^{\text{parent 2}}_{\text{crossover}} \in \calS^{(t)}-\{X^{\text{parent 1}}_{\text{crossover}}\} }$. 
% It corresponds to the ``crossover policy network 2'' in Figure~\ref{fig:pipeline}. 
Given two parents ligands, crossover finds the largest substructure that the two parent compounds share and generates a child by combining their decorating moieties. Thus, the generation of child ligands are deterministic, and the probability of the generated ligands $X^{\text{child}}_{\text{crossover}}$ is 
\begin{equation}
\label{eqn:crossover_prob}
\begin{aligned}
& p_{\text{crossover}}(X_{\text{crossover}}^{\text{child 1}}, X_{\text{crossover}}^{\text{child 2}} | \mathcal{S}^{(t)}) 
= p_{\text{crossover}}(X_{\text{crossover}}^{\text{parent 1}}, X_{\text{crossover}}^{\text{parent 2}} | \mathcal{S}^{(t)}) \\ 
=\ & p_{\text{crossover}}^{(1)}(X^{\text{parent 1}}_{\text{crossover}} | \mathcal{S}^{(t)})
\cdot 
p_{\text{crossover}}^{(2)}(X^{\text{parent 2}}_{\text{crossover}} | X^{\text{parent 1}}_{\text{crossover}}, \mathcal{S}^{(t)} ). 
\end{aligned}
\end{equation}

\noindent\textbf{Mutation Policy Network}. We design two policy networks for two corresponding actions in mutation, as mentioned in Section~\ref{sec:mdp}. 
(1) the first action in mutation operation is to select a candidate ligand to be mutated from population $\mathcal{S}^{(t)}$. It models the 3D target-ligand complex to learn if there is improvement space in the current complex. 
Formally, we obtain a valid probability distribution over all the available ligands based on target-ligand complex as input feature and ENN as neural architecture, the selection probability of the ligand $X^{\text{parent}}_{\text{mutation}} \in \calS^{(t)} $ is  $p_{\text{mutation}}^{(1)}(X^{\text{parent}}_{\text{mutation}} | \calS^{(t)}) = \frac{\exp{\big( \text{MLP}( \bfh_{\calT \& X^{\text{parent}}_{\text{mutation}}} ) \big)}}{\sum_{X'\in \calS^{(t)}} \exp{\big(  \text{MLP}( \bfh_{\calT\& X'}) \big)}}$, where $\calT \& X$ denotes target-ligand complex, $\bfh_{\calT \& X} = \text{ENN}(\calT \& X)$ represents the ENN's embedding of target-ligand complex. 
(2) The second action is to select the SMARTS reaction from the reaction set conditioned on the selected ligand in the first action. Specifically, for each reaction, we generate the new ligand $X^{\text{child}}_{\text{mutation}}$, then obtain the embedding of target, first ligand $X^{\text{parent}}_{\text{mutation}}$ and the new ligand $X^{\text{child}}_{\text{mutation}}$ through ENN, concatenate these three embeddings and feed it into \marked{an} MLP to estimate a scalar as unnormalized probability. 
The unnormalized probabilities for all the reactions are normalized via softmax function, i.e.,  
$p_{\text{mutation}}^{(2)}(\xi | X^{\text{parent}}_{\text{mutation}}, \calS^{(t)}) = \text{Softmax} \big\{\text{MLP}(\bfh_{\calT} \oplus \bfh_{X^{\text{parent}}_{\text{mutation}}} \oplus \bfh_{X^{\text{child}}_{\text{mutation}}} ]), \cdots, \big\}_{\xi \in \mathcal{R}}^{}$, where $X^{\text{parent}}_{\text{mutation}} \xrightarrow[]{\text{mutated by}\ \xi} X^{\text{child}}_{\text{mutation}}$, $\calR$ is the reaction set. 
The probability of the generated ligand $X^{\text{child}}_{\text{mutation}}$ is 
\begin{equation}
\label{eqn:mutation_prob}
p_{\text{mutation}}(X^{\text{child}}_{\text{mutation}} | \mathcal{S}^{(t)}) = p_{\text{mutation}}^{(1)}(X^{\text{parent}}_{\text{mutation}} | \calS^{(t)}) \cdot 
p_{\text{mutation}}^{(2)}(\xi | X^{\text{parent}}_{\text{mutation}}, \calS^{(t)}). 
\end{equation}

\noindent\textbf{Policy Gradient}. 
We leverage policy gradient to train the target-ligand policy neural network. Specifically, we consider maximizing the expected reward as objective via REINFORCE~\cite{Olivecrona}, 
\begin{equation}
\label{eqn:policy}
\max\ \EB_{X \sim p(X | \mathcal{S}^{(t)})}\big[ \text{Reward}(X) \big], 
\end{equation}
where $p(X)$ is defined in Equation~\eqref{eqn:crossover_prob} and~\eqref{eqn:mutation_prob} for crossover and mutation, respectively. 
The whole pipeline is illustrated in Figure~\ref{fig:pipeline}. 
To provide a warm start and leverage the structural information, we pretrain the ENNs on 3D target-ligand binding affinity prediction task, where the inputs are the target-ligand complexes, and the outputs are their binding affinity.

% \noindent\textbf{Policy Gradient}. 
% We leverage policy gradient to train the target-ligand policy neural network. 
% Specifically, we consider maximizing the expected reward as objective via REINFORCE~\cite{Olivecrona}, 
% \begin{equation}
% \label{eqn:policy}
% \max\ \EB_{X \sim p(X | \mathcal{S}^{(t)})}\big[ \text{Reward}(X) \big], 
% \end{equation}
% where $p(X)$ is defined in Equation~\eqref{eqn:crossover_prob} and~\eqref{eqn:mutation_prob} for crossover and mutation mutation, respectively. 
% The reward is the binding affinity between the ligand and the target. Larger reward value are desirable. 
% Each episode corresponds to a generation in \mname and we decompose the whole reward into the sum of multiple intermediate reward in multiple generations. 
% The intermediate reward is the improvement of binding affinity over the last generation.  
% The whole pipeline is illustrated in Figure~\ref{fig:pipeline}. 
% % \noindent\textbf{Pretraining}. 
% To provide a warm start, we pretrain ENN on 3D target-ligand binding affinity prediction task, where the input are the target-ligand complex and the output is their binding affinity.  
% \noindent\textbf{Summarization}. 

\section{Experiment}
\label{sec:experiment}

In this section, we briefly describe the experimental setup and results. The Appendix includes more details, including software configuration, implementation details, dataset description \& processing, hyperparameter tuning, ablation study, and additional experimental results. 
The code is available at \url{https://github.com/futianfan/reinforced-genetic-algorithm}.

\subsection{Experimental Setup}
\label{sec:setup}

\noindent\textbf{Docking Simulation}. 
We adopt AutoDock Vina~\cite{trott2010autodock} to evaluate the binding affinity. The docking score estimated by AutoDock Vina is called Vina score and roughly characterizes the free energy changes of binding processes in kcal/mol. 
Thus lower Vina score means a stronger binding affinity between the ligand and target. 
%In the implementation, we maximize the negative value of the docking score. 
 We picked various disease-related proteins, including G-protein coupling receptors (GPCRs) and kinases from DUD-E~\cite{mysinger2012directory} and the SARS-CoV-2 main protease~\cite{zhang2021potent} as targets. Please see the Appendix for more information.

\noindent\textbf{Baselines}. 
The baseline methods cover traditional brute-force search methods (Screening), deep generative models (JTVAE and Gen3D), genetic algorithm (GA+D, graph-GA, Autogrow 4.0), reinforcement learning methods (MolDQN, RationaleRL, REINVENT, GEGL), and MCMC method (MARS).
Gen3D and Autogrow 4.0 are structure-based drug design methods, while others are general-purpose molecular design methods.
Although methods explicitly utilizing target structures are relatively few, we add general-purpose molecular design methods optimizing the same docking oracle scores as ours, which is a common use case, as baselines~\cite{jensen2019graph,huang2021therapeutics}. 
Concretely, 
\textbf{Screening} mimics high throughput screening via sampling from ZINC database randomly; 
\textbf{JTVAE} (Junction Tree Variational Auto-Encoder)~\cite{jin2018junction} uses a Bayesian optimization on the latent space to indirectly optimize molecules; 
\textbf{Gen3D}~\cite{luo20213d} is an auto-regressive generative model that grows 3D structures atom-wise inside the binding pocket; 
\textbf{GA+D}~\cite{nigam2019augmenting} represents molecule as SELFIES string~\cite{krenn2020self} and uses genetic algorithm enhanced by a discriminator neural network; 
\textbf{Graph-GA}~\cite{jensen2019graph} conduct genetic algorithm on molecular graph representation;   
\textbf{Autogrow 4.0}~\cite{spiegel2020autogrow4} is the state-of-the-art genetic algorithm in structure-based drug design;
\textbf{MolDQN} (Molecule Deep Q-Network)~\cite{zhou2019optimization}  leverages deep Q-value learning to grow molecules atom-wisely; 
\textbf{RationaleRL}~\cite{jin2020multi} uses rationale (e.g., functional groups or subgraphs) as the building block and a policy gradient method to guide the training of graph neural network-based generator; 
\textbf{REINVENT}~\cite{Olivecrona} represent molecules as SMILES string and uses policy gradient based reinforcement learning methods to guide the training of the RNN generator; 
 \textbf{GEGL} (genetic expert-guided learning)~\cite{ahn2020guiding} uses LSTM guided by reinforcement learning to imitate the GA exploration; 
\textbf{MARS} (Markov Molecule Sampling)~\cite{xie2021mars} leverages Markov chain Monte Carlo sampling (MCMC) with adaptive proposal and annealing scheme to search chemical space.
% \textbf{GCPN} (Graph Convolutional Policy Network)~\cite{You2018-xh}; 
To conduct a fair comparison, we limit the number of oracle calls to 1,000 times for each method. 
All the baselines can be run with one-line code using the software (\url{https://github.com/wenhao-gao/mol_opt}) in practical molecular optimization benchmark~\cite{gao2022sample}. 
% Implementation details of all the baseline methods are available in Appendix. 

\noindent\textbf{Dataset}: we randomly select molecules from ZINC~\cite{sterling2015zinc} database (around 250 thousands drug-like molecules) as 0-th generation of the genetic algorithms (\mname, Autogrow 4.0, GA+D). ZINC also serves as the training data for pretraining the model in JTVAE, REINVENT, RationaleRL, etc. We adopt {CrossDocked2020}~\cite{francoeur2020three} dataset that contains around 22 million ligand-protein complexes as the training data for pretraining the policy neural networks, as mentioned in Section~\ref{sec:policy_net}. More descriptions are available in Appendix. 

\noindent\textbf{Metrics}. 
The selection of evaluation metrics follows recent works in molecule optimization~\cite{jin2018junction,nigam2019augmenting,jin2020multi,xie2021mars} and \marked{structure}-based drug design~\cite{spiegel2020autogrow4,luo20213d,huang2021therapeutics}. 
For each method, we select top-100 molecules with the best docking scores for evaluation and consider the following metrics: 
{\textbf{TOP-1/10/100} (average docking score of top-1/10/100 molecules)}: docking score directly measures the binding affinity between the ligand and target and is the most informative metric in structure-based drug design; 
\textbf{Novelty (Nov)} {(\% of the generated molecules that are not in training set)}; 
\textbf{Diversity (Div)} {(average pairwise Tanimoto distance between the Morgan fingerprints)}; 
We also evaluate some simple {pharmaceutical properties}, including quantitative drug-likeness (\textbf{QED}) and synthetic accessibility (\textbf{SA}). QED score indicates drug-likeliness ranging from 0 to 1 (higher the better). SA score ranges from 1 to 10 (lower the better). All the evaluation functions are available at Therapeutics data commons (TDC, \url{https://tdcommons.ai/fct_overview})~\cite{huang2021therapeutics,Huang2022artificial}.

\subsection{Results}
\label{sec:results}

\begin{table*}[t!]
\centering
\vspace{-15pt}
\caption{The summarized performance of different methods. The mean and standard deviation across targets are reported. 
Arrows ($\uparrow$, $\downarrow$) indicate the direction of better performance. 
% Screening searches over the existing drug database, ZINC, so the novelty is 0.0\%. 
For each metric, the best method is underlined and the top-3 methods are bolded. 
\mname -pretrain and \mname -KT are two variants of \mname that without pretraining and without training on different target proteins, respectively.
% more discussions are in ``Knowledge trasfer between protein targets'' in Section~\ref{sec:results}. 
}
\label{table:result}
 \resizebox{0.98\textwidth}{!}{
\begin{tabular}{lcccccccc}
\toprule[0.6pt]
Method & TOP-100$\downarrow$ & TOP-10$\downarrow$ & TOP-1$\downarrow$ & Nov$\uparrow$ & Div$\uparrow$ & QED$\uparrow$ &  SA$\downarrow$  \\
\hline 
screening & -9.351\std{0.643} & -10.433\std{0.563} & -11.400\std{0.630} & 0.0\std{0.0}\% & 0.858\std{0.005} & 0.678\std{0.022} & 2.689\std{0.077} \\ 
MARS & -7.758\std{0.612} & -8.875\std{0.711} & -9.257\std{0.791} & 100.0\std{0.0}\% & {\bf\underline{0.877\std{0.001}}} &  {\bf 0.709\std{0.008}} & {\bf\underline{2.450\std{0.034}}} \\  
MolDQN & -6.287\std{0.396} & -7.043\std{0.487} & -7.501\std{0.402} & 100.0\std{0.0}\% & {\bf 0.877\std{0.009}} & 0.170\std{0.024} & 5.833\std{0.182}  \\ 
% \tabdashline 
GEGL & -9.064\std{0.920} & -9.91\std{0.990} & -10.45\std{1.040} & 100.0\std{0.0}\% & 0.853\std{0.003} & 0.643\std{0.014} & 2.99\std{0.054} \\ 
REINVENT & { -10.181\std{0.441}} & { -11.234\std{0.632}} & { -12.010\std{0.833}} & 100.0\std{0.0}\% & 0.857\std{0.011} &  0.445\std{0.058} & 2.596\std{0.116}  \\
RationaleRL & -9.233\std{0.920} & -10.834\std{0.856} & -11.642\std{1.102} & 100.0\std{0.0}\% & 0.717\std{0.025} & 0.315\std{0.023} & 2.919\std{0.126} \\ 
JTVAE & -9.291\std{0.702} & -10.242\std{0.839} & -10.963\std{1.133} & 98.0\std{0.027}\% &  0.867\std{0.001} & 0.593\std{0.035} & 3.222\std{0.136} \\ 
Gen3D & -8.686\std{0.450} & -9.285\std{0.584} & -9.832\std{0.324} & 100.0\std{0.0}\% & {\bf 0.870\std{0.006}} & 0.701\std{0.016} & 3.450\std{0.120} \\ 
GA+D & -7.487\std{0.757} & -8.305\std{0.803} & -8.760\std{0.796} & 99.2\std{0.011}\%  & 0.834\std{0.035} & 0.405\std{0.024} & 5.024\std{0.164} \\ 
Graph-GA & {\bf -10.848\std{0.860}} & {\bf -11.702\std{0.930}} & {\bf -12.302\std{1.010}} & 100.0\std{0.0}\% &  0.811\std{0.037} & 0.456\std{0.067} & 3.503\std{0.367} \\ 
Autogrow 4.0 & {\bf -11.371\std{0.398}} & {\bf -12.213\std{0.623}} & {\bf -12.474\std{0.839}} & 100.0\std{0.0}\% & 0.852\std{0.011} & {\bf 0.748\std{0.022}} & {\bf{2.497\std{0.049}}} \\ 
RGA (ours) & {\bf\underline{-11.867\std{0.170}}} & \underline{\bf{-12.564\std{0.287}}} & {\bf\underline{-12.869\std{0.473}}} & 100.0\std{0.0}\% & 0.857\std{0.020} & {\bf\underline{ 0.742\std{0.036}}} & {\bf 2.473\std{0.048}}  \\ \toprule 
RGA - pretrain & -11.443\std{0.219} & -12.424\std{0.386} & -12.435\std{0.654} & 100.0\std{0.0}\% & 0.854\std{0.035} & 0.750\std{0.034} & 2.494\std{0.043} \\ 
RGA - KT & -11.434\std{0.169} & -12.437\std{0.354} & -12.502\std{0.603} & 100.0\std{0.0}\% & 0.853\std{0.028} & 0.738\std{0.034} & 2.501\std{0.050} \\ 
\bottomrule[0.6pt]
\end{tabular}}
\end{table*}

\begin{figure}[t!]
\centering
\vspace{-15pt}
\subfigure[Example of 7l11, -10.8 kcal/mol]{
\includegraphics[width=0.31\linewidth]{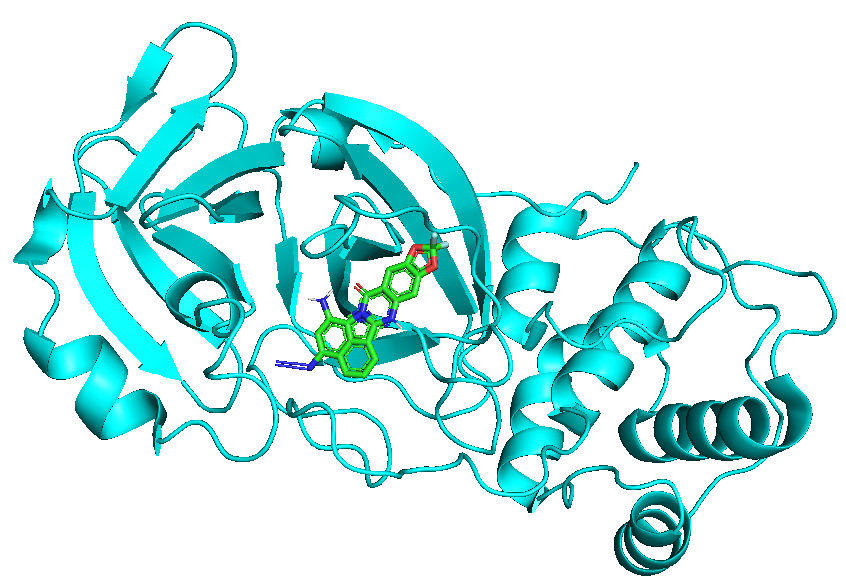}
\label{fig:7l11}}
\subfigure[Example of 3eml, -13.2 kcal/mol]{
\includegraphics[width=0.31\linewidth]{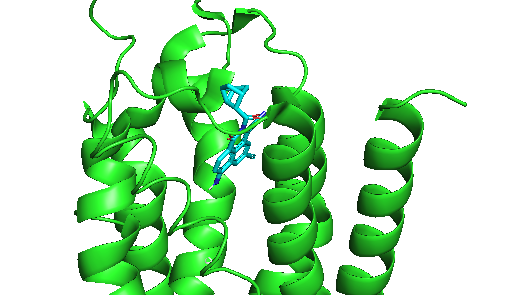}
\label{fig:3eml}}
\subfigure[TOP-100 vs \# oracle call]{
\includegraphics[width=0.42\linewidth]{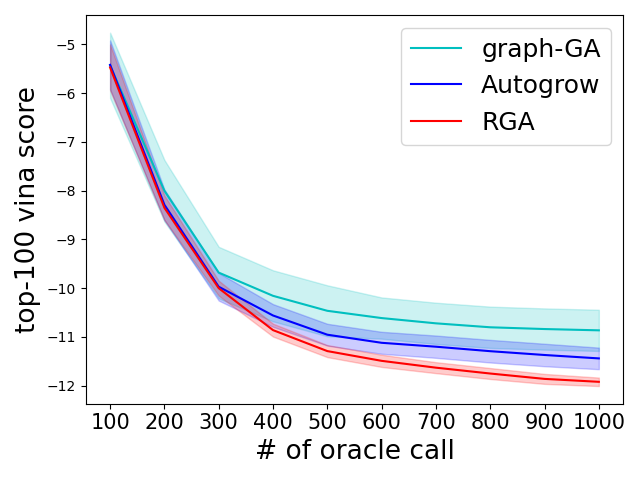}
\label{fig:randomwalk1}}
% \subfigure[TOP-10 vs \# oracle call]{
% \includegraphics[width=0.4\linewidth]{figure/all_top10.png}
% \label{fig:randomwalk2}}
\subfigure[Score bar for different runs]{
\includegraphics[width=0.42\linewidth]{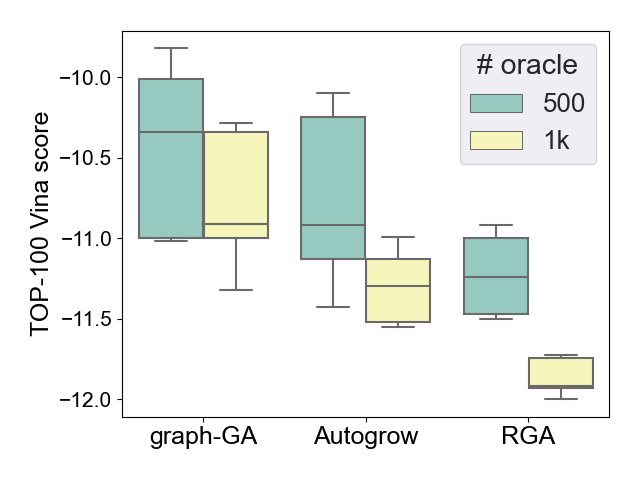}
\label{fig:randomwalk2}}
% \subfigure[1K oracle call]{
% \includegraphics[width=0.31\linewidth]{figure/errorbar_1k.png}
% \label{fig:randomwalk3}}
\caption{
(a) and (b): Example of ligand poses (generated by \mname) and binding sites of target structures. Example of 7l11: the PDB ID of target is 7l11, which is SARS-COV-2(2019-NCOV) main protease, the Vina score is -10.8 kcal/mol.  
Example of 3eml: the PDB ID is 3eml, which is a human A2A Adenosine receptor, the Vina score is -13.2 kcal/mol.  
(c) and (d): studies of suppressed random-walk behavior. (c) reports TOP-100 docking score as a function of oracle calls. The results are the means and standard deviations of 5 independent runs. (d) shows the bars of TOP-100 docking score for various independent runs. 
}
\label{fig:multiple_run}
\end{figure}

\noindent\textbf{Stronger Optimization Performance}. 
We summarized the main results of the structure-based drug design in Table~\ref{table:result}. 
We evaluate all the methods on all targets and report each metric's mean and standard deviations across all targets. 
Our result shows \mname achieves the best performance in TOP-100/10/1 scores among all methods we compared. 
Compared to Autogrow 4.0, \mname's better performance in docking score demonstrates that the policy networks contribute positively to the chemical space navigation and eventually help discover more potent binding molecules.
On the other hand, including longer-range navigation steps enabled by crossover leads to superior performance than other RL methods  (REINVENT, MolDQN, GEGL and RationaleRL) that only focus on local modifications. In addition, we also observed competitive structure quality measured by QED ($>0.7$) and SA\_Score ($<2.5$) in Autogrow 4.0 and \mname without involving them as optimization objectives, thanks to the mutation steps originating from chemical reactions. We visualize two designed ligands with optimal affinity for closer inspection in Figure~\ref{fig:7l11} and~\ref{fig:3eml}, and find both ligands bind tightly with the targets.

\noindent\textbf{Suppressed Random-Walk Behavior}.
Especially in SBDD, when the oracle functions are expensive molecular simulations, robustness to random seeds is essential for improving the worst-case performance of algorithms. One of the major issues in traditional GAs is that they have a significant variance between multiple independent runs as they randomly select parents for crossover and mutation types. To examine this behavior, we run five independent runs for \mname, Autogrow 4.0 and graph-GA (three best baselines, all are GA methods) on all targets and plot the standard deviations between runs in Figure~\ref{fig:randomwalk1} and \ref{fig:randomwalk2}. With policy networks guiding the action steps, we observed that the random-walk behavior in Autogrow 4.0 was suppressed in \mname, indicated by the smaller variance. Especially in the later learning phase (after 500 oracle calls), the policy networks are fine-tuned and guide the search more intelligently. This advantage leads to improved worst-case performance and a higher probability of successfully identifying bioactive drug candidates with constrained resources.

% \noindent\textbf{Suppressed Random-Walk Behavior}. 
% As mentioned, in the traditional GA, the crossover and mutation operation randomly selects the ligands and reactions, this kind of random-walk behavior usually leads to high variance and are undesirable in molecule optimization. 
% \mname is designed to suppress this random-walk behavior. Specifically, we conduct multiple independent runs to compare the \mname and its random-walk version, i.e., Autogrow 4.0. The average TOP-100 and TOP-10 vina score over 5 independent runs, and their standard deviations are reported in Figure~\ref{fig:randomwalk1} and~\ref{fig:randomwalk2}, respectively. We observe that \mname is able to significantly reduce the variance, especially in the later learning phase (after 500 oracle calls). 
% The observation validates \mname's ability to suppress random-walk behavior. 

\noindent\textbf{Knowledge Transfer Between Protein Targets}. 
To verify if \mname benefited from learning the shared physics of ligand-target interaction, we conducted an ablation study whose results are in the last two rows of Table~\ref{table:result}. Specifically, we compare \mname with two variants: (1) \mname-pretrain that does not pretrain the policy network with all native complex structures in the {CrossDocked2020}; (2) \mname-KT (knowledge transfer) that fine-tune the networks with data of individual target independently. We find that both strategies positively contribute to \mname on TOP-100/10/1 docking score. These results demonstrate the policy networks successfully learn the shared physics of ligand-target interactions and leverage the knowledge to improve their performance.

% \noindent\textbf{Knowledge Transfer Between Protein Targets}. 
% \mname is able to learn the knowledge of different protein targets using the ENN based target ligand policy network (Section~\ref{sec:policy_net}), which is pretrained on 3D target-ligand binding affinity prediction
%  task. To explore their empirical effect, we compare \mname with two variants: (1) \mname-pretrain does not pretrain the policy network. (2) \mname-KT (knowledge transfer) does not leverage the knowledge learned from other targets and optimize individual target independently. The results are also reported in the last two rows of Table~\ref{table:result}. We find that both strategies positively contribute to \mname on TOP-100/10/1 docking score. The reason is pretraining learns the 3D target-ligand structure information and provides a warm start for the policy network, knowledge transfer (KT) can learn the pattern of target-ligand binding physics from a large amount of data and promote the optimization performance. 
 
% \noindent\textbf{Case study}
% Also, we visualize examples of the binding sites and their top
% affinity ligands' poses for closer inspection in Figure~\ref{fig:7l11} and~\ref{fig:3eml}. 
% %%Figure~\ref{fig:demo}. 
% We observe that the ligand binds tightly with the target structure. 
% The example validates the ability of \mname to generate the high binding affinity molecules for designated target. 

% \begin{figure}[h!]
% \centering
% \subfigure{
% \includegraphics[width=0.86\linewidth]{figure/3eml_example.png}}
% % \subfigure{
% % \includegraphics[width=4.5cm]{figure/3eml_1.png}}
% \caption{Examples of ligand poses (generated by \mname) and binding sites of target structures. The PDB ID of the target is ``3eml''. We observe that the generated molecules have high binding affinity with the designated target. 
% }
% \label{fig:demo}
% \end{figure}

\section{Conclusion}
\label{sec:conclusion}

In this paper, we propose \fullname (\mname) to tackle the structure-based drug design problem. \mname reformulate the evolutionary process in genetic algorithms as a Markov decision process called evolutionary Markov decision process (EMDP) so that the searching processes could benefit from trained neural models. Specifically, we train policy networks to choose the parents to crossover and mutate instead of randomly sampling them. Further, we also leverage the common physics of the ligand-target interaction and adopt a knowledge-transfer strategy that uses data from other targets to train the networks. Through empirical study, we show that \mname has strong and robust optimization performance, consistently outperforming baseline methods in terms of docking score. 

Though we adopted mutations originating from chemical reactions and the structural quality metrics seem good, we need to emphasize that the designed molecules from \mname do not guarantee synthesizability~\cite{gao2020synthesizability}, as the crossover operations may break inheriting synthesizability. 
Directly working on synthetic pathways could solve the problem~\cite{gao2021amortized,bradshaw2020barking}, but the extension is not trivial. 
As for future direction, we expect to theoretically analyze the EMDP formulation and the performance of \mname. 
% We also expect to generalize \mname to other combinatorial optimization scenarios, such as symbolic laws discovery~\cite{schmidt2009distilling}, quantum circuits design~\cite{du2020quantum}), etc. 

% \begin{ack}
\section*{Acknowledgement}
T.F. and J.S. were supported by NSF award SCH-2205289, SCH-2014438, IIS-1838042, NIH award R01 1R01NS107291-01.
W.G. and C.C. were supported by the Office of Naval Research under grant number N00014-21-1-2195 and the Machine Learning for Pharmaceutical Discovery and Synthesis consortium. 
Any opinions, findings, and conclusions or recommendations expressed in this material are those of the author(s) and do not necessarily reflect the views of the Office of Naval Research.
% \end{ack}

% In this paper, to suppress the random-walk behaviour and scale up the exploration in genetic algorithm, we have proposed \fullname for structure based drug design. 
% It inherits the assembling strategy of genetic algorithm and use reinforcement learning to navigate the chemical space. 
% Specifically, we have designed target-ligand policy network to guide the crossover and mutation operation in genetic algorithm to suppress random-walk behaviour and explore the chemical space intelligently. 
% Thorough empirical studies that optimizes docking score on various targets have been conducted to corroborte the effectiveness of the proposed method, achieving consistent improvement in terms of docking score compared with the best baseline method. 

% This paper mainly has two limitations: (1) the lack of theoretical analysis on the superiority of GA's assembling strategy and the proposed method; (2) the effectiveness of the proposed methods on other scenarios (e.g., other GA (or combinatorial optimization) applications, symbolic laws discovery~\cite{schmidt2009distilling}, quantum circuits design~\cite{du2020quantum}) need to be validated. 

\bibliographystyle{unsrt}
\bibliography{main}

\begin{thebibliography}{10}

\bibitem{bohacek1996art}
Regine~S Bohacek, Colin McMartin, and Wayne~C Guida.
\newblock The art and practice of structure-based drug design: a molecular
  modeling perspective.
\newblock {\em Medicinal research reviews}, 16(1):3--50, 1996.

\bibitem{tripathi2017molecular}
Ashutosh Tripathi and Vytas~A Bankaitis.
\newblock Molecular docking: From lock and key to combination lock.
\newblock {\em Journal of molecular medicine and clinical applications}, 2(1),
  2017.

\bibitem{alon2021structures}
Assaf Alon, Jiankun Lyu, Joao~M Braz, Tia~A Tummino, Veronica Craik, Matthew~J
  O’Meara, Chase~M Webb, Dmytro~S Radchenko, Yurii~S Moroz, Xi-Ping Huang,
  et~al.
\newblock Structures of the $\sigma$2 receptor enable docking for bioactive
  ligand discovery.
\newblock {\em Nature}, 600(7890):759--764, 2021.

\bibitem{jumper2021highly}
John Jumper, Richard Evans, Alexander Pritzel, Tim Green, Michael Figurnov,
  Olaf Ronneberger, Kathryn Tunyasuvunakool, Russ Bates, Augustin
  {\v{Z}}{\'\i}dek, Anna Potapenko, et~al.
\newblock Highly accurate protein structure prediction with {AlphaFold}.
\newblock {\em Nature}, 596(7873):583--589, 2021.

\bibitem{varadi2022alphafold}
Mihaly Varadi, Stephen Anyango, Mandar Deshpande, Sreenath Nair, Cindy
  Natassia, Galabina Yordanova, David Yuan, Oana Stroe, Gemma Wood, Agata
  Laydon, et~al.
\newblock Alphafold protein structure database: Massively expanding the
  structural coverage of protein-sequence space with high-accuracy models.
\newblock {\em Nucleic acids research}, 50(D1):D439--D444, 2022.

\bibitem{ren2022alphafold}
Feng Ren, Xiao Ding, Min Zheng, Mikhail Korzinkin, Xin Cai, Wei Zhu, Alexey
  Mantsyzov, Alex Aliper, Vladimir Aladinskiy, Zhongying Cao, et~al.
\newblock Alphafold accelerates artificial intelligence powered drug discovery:
  Efficient discovery of a novel {C}yclin-dependent {K}inase 20 ({CDK20}) small
  molecule inhibitor.
\newblock {\em arXiv preprint arXiv:2201.09647}, 2022.

\bibitem{lyu2019ultra}
Jiankun Lyu, Sheng Wang, Trent~E Balius, Isha Singh, Anat Levit, Yurii~S Moroz,
  Matthew~J O’Meara, Tao Che, Enkhjargal Algaa, Kateryna Tolmachova, et~al.
\newblock Ultra-large library docking for discovering new chemotypes.
\newblock {\em Nature}, 566(7743):224--229, 2019.

\bibitem{graff2021accelerating}
David~E Graff, Eugene~I Shakhnovich, and Connor~W Coley.
\newblock Accelerating high-throughput virtual screening through molecular
  pool-based active learning.
\newblock {\em Chemical science}, 12(22):7866--7881, 2021.

\bibitem{gentile2022artificial}
Francesco Gentile, Jean~Charle Yaacoub, James Gleave, Michael Fernandez,
  Anh-Tien Ton, Fuqiang Ban, Abraham Stern, and Artem Cherkasov.
\newblock Artificial intelligence--enabled virtual screening of ultra-large
  chemical libraries with deep docking.
\newblock {\em Nature Protocols}, pages 1--26, 2022.

\bibitem{luo20213d}
Shitong Luo, Jiaqi Guan, Jianzhu Ma, and Jian Peng.
\newblock A {3D} generative model for structure-based drug design.
\newblock In {\em Thirty-Fifth Conference on Neural Information Processing
  Systems}, 2021.

\bibitem{li2021structure}
Yibo Li, Jianfeng Pei, and Luhua Lai.
\newblock Structure-based de novo drug design using {3D} deep generative
  models.
\newblock {\em Chemical science}, 12(41):13664--13675, 2021.

\bibitem{walters2020assessing}
W~Patrick Walters and Mark Murcko.
\newblock Assessing the impact of generative {AI} on medicinal chemistry.
\newblock {\em Nature biotechnology}, 38(2):143--145, 2020.

\bibitem{brown2019guacamol}
Nathan Brown, Marco Fiscato, Marwin~HS Segler, and Alain~C Vaucher.
\newblock {GuacaMol}: benchmarking models for de novo molecular design.
\newblock {\em Journal of chemical information and modeling}, 59(3):1096--1108,
  2019.

\bibitem{huang2021therapeutics}
Kexin Huang, Tianfan Fu, Wenhao Gao, Yue Zhao, Yusuf Roohani, Jure Leskovec,
  Connor~W Coley, Cao Xiao, Jimeng Sun, and Marinka Zitnik.
\newblock Therapeutics {D}ata {C}ommons: machine learning datasets and tasks
  for therapeutics.
\newblock {\em NeurIPS Track Datasets and Benchmarks}, 2021.

\bibitem{gao2022sample}
Wenhao Gao, Tianfan Fu, Jimeng Sun, and Connor~W Coley.
\newblock Sample efficiency matters: A benchmark for practical molecular
  optimization.
\newblock {\em NeurIPS Track Datasets and Benchmarks}, 2022.

\bibitem{jensen2019graph}
Jan~H Jensen.
\newblock A graph-based genetic algorithm and generative model/{M}onte {C}arlo
  {T}ree {S}earch for the exploration of chemical space.
\newblock {\em Chemical science}, 10(12):3567--3572, 2019.

\bibitem{spiegel2020autogrow4}
Jacob~O Spiegel and Jacob~D Durrant.
\newblock {AutoGrow4}: an open-source genetic algorithm for de novo drug design
  and lead optimization.
\newblock {\em Journal of cheminformatics}, 12(1):1--16, 2020.

\bibitem{tripp2021fresh}
Austin Tripp, Gregor~NC Simm, and Jos{\'e}~Miguel Hern{\'a}ndez-Lobato.
\newblock A fresh look at de novo molecular design benchmarks.
\newblock In {\em NeurIPS 2021 AI for Science Workshop}, 2021.

\bibitem{satorras2021n}
Victor~Garcia Satorras, Emiel Hoogeboom, and Max Welling.
\newblock E (n) equivariant graph neural networks.
\newblock {\em Proceedings of the 38th International Conference on Machine
  Learning, {ICML}}, 139:9323--9332, 2021.

\bibitem{gomez2018automatic}
Rafael G{\'o}mez-Bombarelli, Jennifer~N Wei, David Duvenaud, Jos{\'e}~Miguel
  Hern{\'a}ndez-Lobato, Benjam{\'\i}n S{\'a}nchez-Lengeling, Dennis Sheberla,
  Jorge Aguilera-Iparraguirre, Timothy~D Hirzel, Ryan~P Adams, and Al{\'a}n
  Aspuru-Guzik.
\newblock Automatic chemical design using a data-driven continuous
  representation of molecules.
\newblock {\em ACS central science}, 2018.

\bibitem{jin2018junction}
Wengong Jin, Regina Barzilay, and Tommi Jaakkola.
\newblock Junction tree variational autoencoder for molecular graph generation.
\newblock {\em ICML}, 2018.

\bibitem{guimaraes2017objective}
Gabriel~Lima Guimaraes, Benjamin Sanchez-Lengeling, Carlos Outeiral, Pedro
  Luis~Cunha Farias, and Al{\'a}n Aspuru-Guzik.
\newblock Objective-{R}einforced {G}enerative {A}dversarial {N}etworks
  ({ORGAN}) for sequence generation models.
\newblock {\em arXiv preprint arXiv:1705.10843}, 2017.

\bibitem{cao2018molgan}
Nicola~De Cao and Thomas Kipf.
\newblock {MolGAN}: An implicit generative model for small molecular graphs,
  2018.

\bibitem{shi2019graphaf}
Chence Shi, Minkai Xu, Zhaocheng Zhu, Weinan Zhang, Ming Zhang, and Jian Tang.
\newblock {GraphAF}: a flow-based autoregressive model for molecular graph
  generation.
\newblock In {\em ICLR}, 2020.

\bibitem{luo2021graphdf}
Youzhi Luo, Keqiang Yan, and Shuiwang Ji.
\newblock {GraphDF}: A discrete flow model for molecular graph generation.
\newblock {\em Proceedings of the 38th International Conference on Machine
  Learning, {ICML}}, 139:7192--7203, 2021.

\bibitem{liu2021graphebm}
Meng Liu, Keqiang Yan, Bora Oztekin, and Shuiwang Ji.
\newblock {GraphEBM}: Molecular graph generation with energy-based models.
\newblock {\em arXiv preprint arXiv:2102.00546}, 2021.

\bibitem{nigam2019augmenting}
AkshatKumar Nigam, Pascal Friederich, Mario Krenn, and Al{\'a}n Aspuru-Guzik.
\newblock Augmenting genetic algorithms with deep neural networks for exploring
  the chemical space.
\newblock In {\em ICLR}, 2020.

\bibitem{gao2021amortized}
Wenhao Gao, Roc{\'\i}o Mercado, and Connor~W Coley.
\newblock Amortized tree generation for bottom-up synthesis planning and
  synthesizable molecular design.
\newblock {\em International Conference on Learning Representations}, 2022.

\bibitem{Olivecrona}
M.~Olivecrona, T.~Blaschke, O.~Engkvist, and H.~Chen.
\newblock Molecular de-novo design through deep reinforcement learning.
\newblock {\em Journal of Cheminformatics}, 2017.

\bibitem{You2018-xh}
Jiaxuan You, Bowen Liu, Rex Ying, Vijay Pande, and Jure Leskovec.
\newblock Graph convolutional policy network for goal-directed molecular graph
  generation.
\newblock In {\em NIPS}, 2018.

\bibitem{zhou2019optimization}
Zhenpeng Zhou, Steven Kearnes, Li~Li, Richard~N Zare, and Patrick Riley.
\newblock Optimization of molecules via deep reinforcement learning.
\newblock {\em Scientific reports}, 9(1):1--10, 2019.

\bibitem{jin2020multi}
Wengong Jin, Regina Barzilay, and Tommi Jaakkola.
\newblock Multi-objective molecule generation using interpretable
  substructures.
\newblock In {\em International Conference on Machine Learning}, pages
  4849--4859. PMLR, 2020.

\bibitem{ahn2020guiding}
Sungsoo Ahn, Junsu Kim, Hankook Lee, and Jinwoo Shin.
\newblock Guiding deep molecular optimization with genetic exploration.
\newblock {\em Advances in neural information processing systems},
  33:12008--12021, 2020.

\bibitem{korovina2020chembo}
Ksenia Korovina, Sailun Xu, Kirthevasan Kandasamy, Willie Neiswanger, Barnabas
  Poczos, Jeff Schneider, and Eric Xing.
\newblock {ChemBO}: Bayesian optimization of small organic molecules with
  synthesizable recommendations.
\newblock In {\em International Conference on Artificial Intelligence and
  Statistics}, pages 3393--3403. PMLR, 2020.

\bibitem{fu2021mimosa}
Tianfan Fu, Cao Xiao, Xinhao Li, Lucas~M Glass, and Jimeng Sun.
\newblock {MIMOSA}: Multi-constraint molecule sampling for molecule
  optimization.
\newblock {\em AAAI}, 2020.

\bibitem{xie2021mars}
Yutong Xie, Chence Shi, Hao Zhou, Yuwei Yang, Weinan Zhang, Yong Yu, and Lei
  Li.
\newblock {MARS}: Markov molecular sampling for multi-objective drug discovery.
\newblock In {\em ICLR}, 2021.

\bibitem{bengio2021gflownet}
Yoshua Bengio, Tristan Deleu, Edward~J. Hu, Salem Lahlou, Mo~Tiwari, and
  Emmanuel Bengio.
\newblock {GFlowNet} foundations.
\newblock {\em CoRR}, abs/2111.09266, 2021.

\bibitem{fu2021differentiable}
Tianfan Fu, Wenhao Gao, Cao Xiao, Jacob Yasonik, Connor~W Coley, and Jimeng
  Sun.
\newblock Differentiable scaffolding tree for molecular optimization.
\newblock {\em International Conference on Learning Representations}, 2022.

\bibitem{shen2021deep}
Cynthia Shen, Mario Krenn, Sagi Eppel, and Alan Aspuru-Guzik.
\newblock Deep molecular dreaming: Inverse machine learning for de-novo
  molecular design and interpretability with surjective representations.
\newblock {\em Machine Learning: Science and Technology}, 2021.

\bibitem{cieplinski2020we}
Tobiasz Cieplinski, Tomasz Danel, Sabina Podlewska, and Stanislaw Jastrzebski.
\newblock We should at least be able to design molecules that dock well.
\newblock {\em arXiv:2006.16955}, 2020.

\bibitem{steinmann2021using}
Casper Steinmann and Jan~H Jensen.
\newblock Using a genetic algorithm to find molecules with good docking scores.
\newblock {\em PeerJ Physical Chemistry}, 3:e18, 2021.

\bibitem{yang2021hit}
Soojung Yang, Doyeong Hwang, Seul Lee, Seongok Ryu, and Sung~Ju Hwang.
\newblock Hit and lead discovery with explorative {RL} and fragment-based
  molecule generation.
\newblock {\em Advances in Neural Information Processing Systems}, 34, 2021.

\bibitem{luo1996rasse}
Zhaowen Luo, Renxiao Wang, and Luhua Lai.
\newblock {RASSE}: a new method for structure-based drug design.
\newblock {\em Journal of chemical information and computer sciences},
  36(6):1187--1194, 1996.

\bibitem{gillet1993sprout}
Valerie Gillet, A~Peter Johnson, Pauline Mata, Sandor Sike, and Philip
  Williams.
\newblock {SPROUT}: a program for structure generation.
\newblock {\em Journal of computer-aided molecular design}, 7(2):127--153,
  1993.

\bibitem{pearlman1993concepts}
David~A Pearlman and Mark~A Murcko.
\newblock {CONCEPTS}: New dynamic algorithm for de novo drug suggestion.
\newblock {\em Journal of computational chemistry}, 14(10):1184--1193, 1993.

\bibitem{douguet2000genetic}
Dominique Douguet, Etienne Thoreau, and G{\'e}rard Grassy.
\newblock A genetic algorithm for the automated generation of small organic
  molecules: drug design using an evolutionary algorithm.
\newblock {\em Journal of computer-aided molecular design}, 14(5):449--466,
  2000.

\bibitem{durrant2013autogrow}
Jacob~D Durrant, Steffen Lindert, and J~Andrew McCammon.
\newblock {AutoGrow} 3.0: an improved algorithm for chemically tractable,
  semi-automated protein inhibitor design.
\newblock {\em Journal of Molecular Graphics and Modelling}, 44:104--112, 2013.

\bibitem{wang1998score}
Renxiao Wang, Liang Liu, Luhua Lai, and Youqi Tang.
\newblock {SCORE}: A new empirical method for estimating the binding affinity
  of a protein-ligand complex.
\newblock {\em Molecular modeling annual}, 4(12):379--394, 1998.

\bibitem{gao2020synthesizability}
Wenhao Gao and Connor~W Coley.
\newblock The synthesizability of molecules proposed by generative models.
\newblock {\em Journal of chemical information and modeling},
  60(12):5714--5723, 2020.

\bibitem{hartenfeller2011collection}
Markus Hartenfeller, Martin Eberle, Peter Meier, Cristina Nieto-Oberhuber,
  Karl-Heinz Altmann, Gisbert Schneider, Edgar Jacoby, and Steffen Renner.
\newblock A collection of robust organic synthesis reactions for in silico
  molecule design.
\newblock {\em Journal of chemical information and modeling},
  51(12):3093--3098, 2011.

\bibitem{ramachandran2017searching}
Prajit Ramachandran, Barret Zoph, and Quoc~V Le.
\newblock Searching for activation functions.
\newblock {\em arXiv preprint arXiv:1710.05941}, 2017.

\bibitem{trott2010autodock}
Oleg Trott and Arthur~J Olson.
\newblock Autodock {Vina}: improving the speed and accuracy of docking with a
  new scoring function, efficient optimization, and multithreading.
\newblock {\em Journal of computational chemistry}, 31(2):455--461, 2010.

\bibitem{mysinger2012directory}
Michael~M Mysinger, Michael Carchia, John~J Irwin, and Brian~K Shoichet.
\newblock Directory of useful decoys, enhanced ({DUD-E}): better ligands and
  decoys for better benchmarking.
\newblock {\em Journal of medicinal chemistry}, 55(14):6582--6594, 2012.

\bibitem{zhang2021potent}
Chun-Hui Zhang, Elizabeth~A Stone, Maya Deshmukh, Joseph~A Ippolito, Mohammad~M
  Ghahremanpour, Julian Tirado-Rives, Krasimir~A Spasov, Shuo Zhang, Yuka
  Takeo, Shalley~N Kudalkar, et~al.
\newblock Potent noncovalent inhibitors of the main protease of sars-cov-2 from
  molecular sculpting of the drug perampanel guided by free energy perturbation
  calculations.
\newblock {\em ACS central science}, 7(3):467--475, 2021.

\bibitem{krenn2020self}
Mario Krenn, Florian H{\"a}se, AkshatKumar Nigam, Pascal Friederich, and Alan
  Aspuru-Guzik.
\newblock Self-referencing embedded strings ({SELFIES}): A 100\% robust
  molecular string representation.
\newblock {\em Machine Learning: Science and Technology}, 1(4):045024, 2020.

\bibitem{sterling2015zinc}
Teague Sterling and John~J Irwin.
\newblock {ZINC} 15--ligand discovery for everyone.
\newblock {\em Journal of chemical information and modeling},
  55(11):2324--2337, 2015.

\bibitem{francoeur2020three}
Paul~G Francoeur, Tomohide Masuda, Jocelyn Sunseri, Andrew Jia, Richard~B
  Iovanisci, Ian Snyder, and David~R Koes.
\newblock Three-dimensional convolutional neural networks and a cross-docked
  data set for structure-based drug design.
\newblock {\em Journal of Chemical Information and Modeling}, 60(9):4200--4215,
  2020.

\bibitem{Huang2022artificial}
Kexin Huang, Tianfan Fu, Wenhao Gao, Yue Zhao, Yusuf Roohani, Jure Leskovec,
  Connor~W Coley, Cao Xiao, Jimeng Sun, and Marinka Zitnik.
\newblock Artificial intelligence foundation for therapeutic science.
\newblock {\em Nature Chemical Biology}, 2022.

\bibitem{bradshaw2020barking}
John Bradshaw, Brooks Paige, Matt~J Kusner, Marwin Segler, and Jos{\'e}~Miguel
  Hern{\'a}ndez-Lobato.
\newblock Barking up the right tree: an approach to search over molecule
  synthesis dags.
\newblock {\em Advances in Neural Information Processing Systems},
  33:6852--6866, 2020.

\bibitem{landrum2006rdkit}
Greg Landrum et~al.
\newblock {RDKit}: Open-source cheminformatics, 2006.

\bibitem{ertl2009estimation}
Peter Ertl and Ansgar Schuffenhauer.
\newblock Estimation of synthetic accessibility score of drug-like molecules
  based on molecular complexity and fragment contributions.
\newblock {\em Journal of cheminformatics}, 1(1):8, 2009.

\bibitem{peters2010policy}
Jan Peters and J~Andrew Bagnell.
\newblock Policy gradient methods.
\newblock {\em Scholarpedia}, 5(11):3698, 2010.

\bibitem{zang2020moflow}
Chengxi Zang and Fei Wang.
\newblock {MoFlow}: an invertible flow model for generating molecular graphs.
\newblock In {\em ACM SIGKDD}, pages 617--626, 2020.

\bibitem{xu2021geodiff}
Minkai Xu, Lantao Yu, Yang Song, Chence Shi, Stefano Ermon, and Jian Tang.
\newblock {GeoDiff}: A geometric diffusion model for molecular conformation
  generation.
\newblock In {\em International Conference on Learning Representations}, 2021.

\bibitem{moss2020boss}
Henry Moss, David Leslie, Daniel Beck, Javier Gonzalez, and Paul Rayson.
\newblock {BOSS}: Bayesian optimization over string spaces.
\newblock {\em Advances in neural information processing systems},
  33:15476--15486, 2020.

\bibitem{yang2017chemts}
Xiufeng Yang, Jinzhe Zhang, Kazuki Yoshizoe, Kei Terayama, and Koji Tsuda.
\newblock {ChemTS}: an efficient python library for de novo molecular
  generation.
\newblock {\em Science and technology of advanced materials}, 18(1):972--976,
  2017.

\bibitem{yang2020practical}
Xiufeng Yang, Tanuj~Kr Aasawat, and Kazuki Yoshizoe.
\newblock Practical massively parallel {M}onte-{C}arlo {T}ree {S}earch applied
  to molecular design.
\newblock {\em arXiv preprint arXiv:2006.10504}, 2020.

\bibitem{conti2018improving}
Edoardo Conti, Vashisht Madhavan, Felipe Petroski~Such, Joel Lehman, Kenneth
  Stanley, and Jeff Clune.
\newblock Improving exploration in evolution strategies for deep reinforcement
  learning via a population of novelty-seeking agents.
\newblock {\em Advances in neural information processing systems}, 31, 2018.

\bibitem{liu2021decoupling}
Evan~Z Liu, Aditi Raghunathan, Percy Liang, and Chelsea Finn.
\newblock Decoupling exploration and exploitation for meta-reinforcement
  learning without sacrifices.
\newblock In {\em International Conference on Machine Learning}, pages
  6925--6935. PMLR, 2021.

\bibitem{silver2017mastering}
David Silver, Julian Schrittwieser, Karen Simonyan, Ioannis Antonoglou, Aja
  Huang, Arthur Guez, Thomas Hubert, Lucas Baker, Matthew Lai, Adrian Bolton,
  et~al.
\newblock Mastering the game of go without human knowledge.
\newblock {\em nature}, 550(7676):354--359, 2017.

\bibitem{kwon2021evolutionary}
Youngchun Kwon, Seokho Kang, Youn-Suk Choi, and Inkoo Kim.
\newblock Evolutionary design of molecules based on deep learning and a genetic
  algorithm.
\newblock {\em Scientific reports}, 11(1):1--11, 2021.

\end{thebibliography}

\appendix

\tableofcontents

\section{Mathematical Notation}

For ease of exposition, we list the mathematical notations in Table~\ref{table:notation}. 
All the mathematical notations are divided into three parts: (1) notation for genetic algorithm (Section~\ref{sec:ga}); (2) notation for equivariance neural networks (ENN)~\cite{satorras2021n} (Section~\ref{sec:policy_net}); (3) notations for policy network (Section~\ref{sec:policy_net}). 

\begin{table}[h!]
\small 
\centering
\caption{Mathematical Notations. All the mathematical notations are divided into three parts: (1) notation for genetic algorithm (Section~\ref{sec:ga}); (2) notation for equivariance neural networks (ENN)~\cite{satorras2021n} (Section~\ref{sec:policy_net}); (3) notations for policy network (Section~\ref{sec:policy_net}). 
}
\resizebox{1.0\columnwidth}{!}{
\begin{tabular}{c|l}
\toprule[1pt]
Notations & Descriptions \\ 
\hline 
$X$ & ligand (drug molecule, including 3D pose) \\
$\calT$ & target (target protein related to the disease) \\ 
$\calS^{(t)}$ & the state (population of molecule) at the $t$-th generation. \\ 
$\calQ^{(t)}$ & offspring pool at the $t$-th generation. \\ 
$K$ & the number of molecule in the state, i.e., size of population. \\ 
$X^{\text{parent 1/2}}_{\text{crossover}}$ & the first/second parent molecule in the crossover. \\ 
$X^{\text{child 1/2}}_{\text{crossover}}$ & the first/second child molecule in the crossover. \\ 
% $X^{\text{parent 1}}_{\text{crossover}}, X^{\text{parent 2}}_{\text{crossover}} \xrightarrow[]{\text{crossover}} X^{\text{child 1}}_{\text{crossover}}, X^{\text{child 2}}_{\text{crossover}}$ & \\ 
$X^{\text{parent}}_{\text{mutation}}$ & parent molecule in the mutation \\ 
$X^{\text{child}}_{\text{mutation}}$ & child molecule in the mutation \\ 
% $X^{\text{parent}}_{\text{mutation}} \xrightarrow[]{\text{mutated by}\ \xi} X^{\text{child}}_{\text{mutation}}$ &  \\ 
$\xi \in \calR$ & the selection reaction in the mutation \\ 
$\calR$ & the reaction set (library) for mutation \\ 
\hline 
ENN & equivariance neural networks~\cite{satorras2021n} \\ 
$\calV =\{H, C, O, N, \cdots\}$ & vocabulary set of atoms \\ 
$\calY = (\calA, \calZ)$ & 3D structure \\ 
$\calA$ & categories of all the atoms \\ 
${\bfa_i}$ & one-hot vector that encode category of $i$-th atom \\ 
$\calZ$ & 3D coordinates of the atoms \\ 
$\bfD \in \RB^{|\calV|\times d}$ & the embedding matrix of all the categories of atoms \\ 
 $d$ & the hidden dimension in ENN. \\ 
 $N$ & number of atoms in the input of ENN. \\ 
 $L$ & number of layers in ENN \\ 
 $l=0,1,\cdots,L$ & index of layer in ENN \\
MLP &  multiple layer perceptrons \\ 
$\text{MLP}_{e}(\cdot),  
\text{MLP}_{x}(\cdot), 
\text{MLP}_{h}(\cdot)$ & two-layer MLP in ENN with Swish activation~\cite{ramachandran2017searching} in hidden layer \\ 
$\oplus$ &  the concatenation of vectors \\ 
$\bfZ^{(0)} = \{\bfz_i^{}\}_{i=1}^{N}$ & initial coordinate embeddings, real 3D coordinates of all the nodes.  \\
 $\bfH^{(l)} = \{\bfh^{(l)}_i\}_{i=1}^{N}$ & Node embeddings at the $l$-th layer \\
$\bfh_i^{(0)} = \bfD^\top {\bfa_i} \in \RB^{d}$ & The initial node embedding that embeds the $i$-th node  \\
$\bfZ^{(l)} = \{\bfz_i^{(l)}\}_{i=1}^{N}$ & Coordinate embeddings at the $l$-th layer \\ 
$\bfw^{(l)}_{ij}$ & message vector for the edge from node $i$ to node $j$ at $l$-th layer \\
$\bfv^{(l)}_{i}$ & message vector for node $i$ at $l$-th layer \\ 
$\bfz^{(l)}_i$ & the position embedding for node $i$ at $l$-th layer \\ 
$\bfh_i^{(l)}$ & the node embedding for node $i$ at $l$-th layer \\ 
$\bfh_{\calY} = \text{ENN}(\calY)$ & ENN representation of the 3D graph $\calY$ (Equation~\ref{eqn:ENN}) \\
\hline 
$p_{\text{crossover}}^{(1)}(X^{\text{parent 1}}_{\text{crossover}} | \calS^{(t)})$ & probability to select the first parent molecule in crossover \\ 
$p_{\text{crossover}}^{(2)}(X^{\text{parent 2}}_{\text{crossover}} | X^{\text{parent 1}}_{\text{crossover}}, \mathcal{S}^{(t)} )$ & probability to select the second parent molecule in crossover \\ 
$p_{\text{crossover}}(X_{\text{crossover}}^{\text{child 1}}, X_{\text{crossover}}^{\text{child 2}} | \mathcal{S}^{(t)}) $ & probability of two generated child molecules in crossover (Eq~\ref{eqn:crossover_prob}) \\ 
$ p_{\text{mutation}}^{(1)}(X^{\text{parent}}_{\text{mutation}} | \calS^{(t)})$ & probability to select the parent molecule in mutation \\ 
$p_{\text{mutation}}^{(2)}(\xi | X^{\text{parent}}_{\text{mutation}}, \calS^{(t)}) $ & probability to select the reaction in mutation \\ 
$p_{\text{mutation}}(X^{\text{child}}_{\text{mutation}} | \mathcal{S}^{(t)})$ & probability of generated child molecule in mutation (Eq~\ref{eqn:mutation_prob})\\ 
\bottomrule[1pt]
\end{tabular}
}
\label{table:notation}
\end{table}

\section{Illustration of Genetic Algorithm}

\begin{figure}[h!]
\centering
\subfigure[crossover]{
\includegraphics[width=4.5cm]{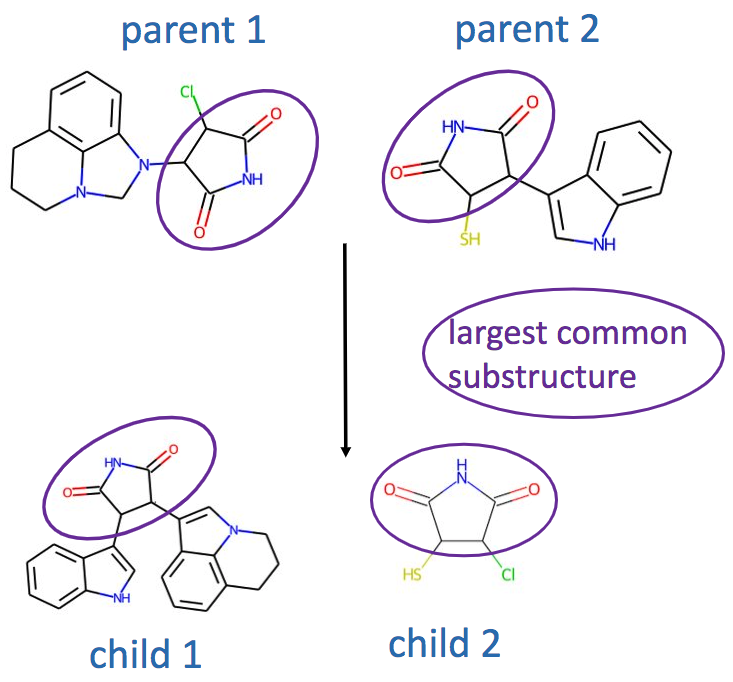}
\label{fig:crossover}}
\subfigure[mutation]{
\includegraphics[width=4.99cm]{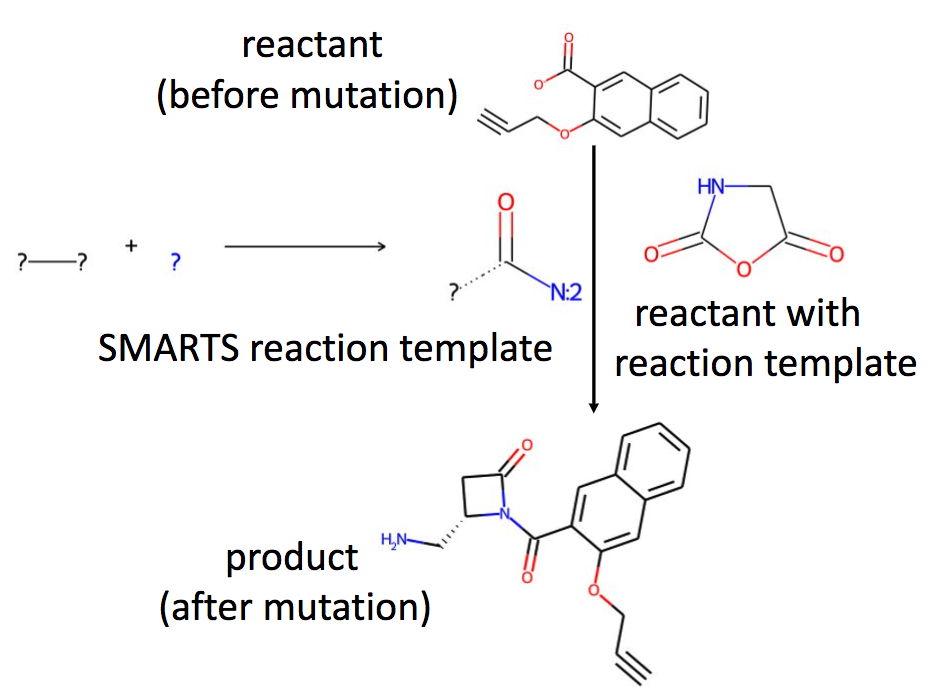}
\label{fig:mutation}}
\caption{Illustration of GA operations: (a) \textbf{crossover} finds the largest substructure that the two parent compounds share and generates a child by combining their decorating moieties. 
(b) \textbf{mutation}: given a reactant (i.e., parent), {mutation} operator uses SMARTS-reaction template (with another reactant) to performs an \textit{in silico} chemical reaction to generate child compound (i.e., product). 
}
\label{fig:ga}
\end{figure}

Figure~\ref{fig:ga} provides two examples to illustrate crossover and mutation operations in genetic algorithm described in Section~\ref{sec:ga}.

\noindent\textbf{Crossover}, also called recombination, combines the structure of two parents to generate new children. Following Autogrow 4.0~\cite{spiegel2020autogrow4}, as shown in Figure~\ref{fig:crossover}, we select two parents from the last generation and search for the largest common substructure shared between them. Then we generate two children by randomly switching their decorating moieties, i.e., the side chains attached to the common substructure. 

\noindent\textbf{Mutation} operator performs an \textit{in silico} chemical reaction to generate an altered child compound (i.e., product in the chemical reaction) derived from a parent (reactant in the chemical reaction), as shown in Figure~\ref{fig:mutation}. The chemical reaction here contains two reactants, one is parent molecule, another is from reaction set. 
The reaction set $\mathcal{R}$ is generated via merging two public reaction libraries: (1) the AutoClickChemRxn set (36 reactions)~\cite{durrant2013autogrow} and (2) RobustRxn set (58 reactions~\cite{hartenfeller2011collection}). 
Each reaction $\xi$ in reaction set $\mathcal{R}$ contains a SMARTS string based reaction template and a reactant. 
It uses SMARTS reaction template, together with RDKit~\cite{landrum2006rdkit}, to perform chemical mutations efficiently. 
The process is written as $X^{\text{parent}}_{\text{mutation}} \xrightarrow[]{\text{mutated by}\ \xi} X^{\text{child}}_{\text{mutation}}$, where $\xi$ is selected reaction (with a reaction template and another reactant).
The ligand to be mutated and the reaction used for mutation are both randomly selected from previous generation and reaction set $\mathcal{R}$, respectively. 
Compared with the mutation operator in conventional GA that randomly flipping an arbitrary bit, the reaction-based mutation enhance synthesizability of the generated molecules~\cite{spiegel2020autogrow4}. 
{Mutation} operator performs an \textit{in silico} chemical reaction to generate an altered child compound (i.e., product in the chemical reaction) derived from a parent (reactant in the chemical reaction). The chemical reaction here contains two reactants, one is parent molecule, another is from reaction set.

% \section{Algorithm}

% The essential steps of the whole pipeline are summarized in Algorithm~\ref{alg:main}. 

% \begin{algorithm}[h!]
% \caption{\fullname} 
% \label{alg:main}
% \begin{algorithmic}[1]
% \STATE \textbf{Input}: multiple target proteins $\{\calT\}$, initial state (population of molecule) $\calS^{(0)}$. 
% \STATE \textbf{Output}: 
% \FOR{$X^{(t)} \in \text{MLP}$}
% \STATE Initialize 
% \ENDFOR
% \end{algorithmic}
% \end{algorithm}

\section{Baseline Setup}
\label{sec:baseline}

In this section, we describe the experimental setting for baseline methods. 
Most of the settings follow the original papers. 
\begin{itemize}
\item \textbf{GA+D} (genetic algorithm enhanced by discriminator neural network)~\cite{nigam2019augmenting} utilizes SELFIES string as the representation of molecules, thus guaranteeing the 100\% chemical validity of the generated molecules. Following their original paper, the discriminator neural network is a two-layer fully connected neural network with ReLU activation and sigmoid output layer. The hidden size is 100. the size of the output layer is 1. The input feature of discriminator neural network is a vector of chemical and geometrical properties characterizing the molecules. 
The population size is set to 300. Maximal generation number is set to 1000. The patience is set to 5. When the property does not improve when the patience exhausts, the process early stops. We used Adam optimizer with 1e-3 as the initial learning rate. 
beta ($\beta$) is the weight of discriminator neural network's score in fitness evaluation, which is used to select most promising molecules in each generation. We set $\beta=1$. 
\item \textbf{Graph-GA}~\cite{jensen2019graph} uses molecular graph to represent molecules and uses crossover and mutation operations to edit the molecular graph. 
After tuning, the size of population is set to 120. The size of offspring is set to 70. The mutation rate is set to 0.067. Graph-GA do not have learnable parameters and is easy to implement. We use the implementation in GuacaMol~\cite{brown2019guacamol}. 
\item \textbf{MolDQN} (Molecule Deep Q-Networks)~\cite{zhou2019optimization} uses molecular graph to represent molecules, formulate the molecule optimization process as a Markov Decision Process (MDP) and utilize Deep Q-value learning to optimize it. It grows molecular graph atom-wise, that is, in each episode, it adds one atom to the partially generated molecular graph. 
The reward function is the negative value of the Vina score. 
Following the original paper, maximal step in each episode is 40. Each step calls oracle once. 
The discount factor is 0.9. 
Deep Q-network is a multilayer perceptron (MLP) whose hidden dimensions are 1024, 512, 128, 32, respectively. The model size is 6.4 M.
The input of the Q-network is the concatenation of the molecule feature (2048-bit Morgan fingerprint, with a radius of 3) and the number of left steps. 
Adam is used as an optimizer with 1e-4 as the initial learning rate.
Only rings with a size of 5 and 6 are allowed. 
It leverages $\epsilon$-greedy together with randomized value functions (bootstrapped-DQN) as an exploration policy, $\epsilon$ is annealed from 1 to 0.01 in a piecewise linear way. 
\item \textbf{RationaleRL}~\cite{jin2020multi}. 
The architecture of the generator is a message-passing network (MPN) followed by MLPs applied in breadth-first order. The generator is pre-trained on general molecules combined with an encoder and then fine-tuned to maximize the reward function using policy gradient. 
The encoder and decoder MPNs both have hidden dimensions of 400. The dimension of the latent variable is 50. Adam optimizer is used on both pre-training and fine-tuning with initial learning rates of 1e-3, 5e-4, respectively. The annealing rate is 0.9. We pre-trained the model with 20 epochs. 

% These rationales are identified from molecules as substructures that are likely responsible for each property of interest. We then learn to expand rationales into a full molecule using graph generative models. Our final generative model composes molecules as mixtures of multiple rationale completions, and this mixture is fine-tuned to preserve the properties of interest. 

\item \textbf{MARS}~\cite{xie2021mars} leverage Markov chain Monte Carlo sampling (MCMC) on molecules with an annealing scheme and an adaptive proposal. The proposal is parameterized by a graph neural network, which is trained on MCMC samples. 
We follow most of the settings in the original paper. 
The message passing network has six layers, where the node embedding size is set to 64. 
Adam is used as an optimizer with 3e-4 initial learning rate. To generate a basic unit, top-1000 frequent fragments are drawn from ZINC database~\cite{sterling2015zinc} by enumerating single bonds to break.  
During the annealing process, the temperature $T = 0.95^{\lfloor t/5\rfloor}$ would gradually decrease to 0.
\item \textbf{Autogrow 4.0}~\cite{spiegel2020autogrow4} is the base model for \mname and have been briefly described in Section~\ref{sec:ga}. The setup of Autogrow is the same as \mname for fair comparison, the only difference is that \mname use policy network to guide the selection of ligands and reaction for crossover and mutation while Autogrow randomly selects them.  
The reaction set $\mathcal{R}$ is generated via merging two public reaction libraries: (1) the AutoClickChemRxn set (36 reactions)~\cite{durrant2013autogrow} and (2) RobustRxn set (58 reactions~\cite{hartenfeller2011collection}). 
In each generation, It generates 200 offspring (100 from crossover and 100 from mutation) and keep 50 most promising (with lowest Vina scores) ones for the next generation. 
\item \textbf{Screening} exhaustively searches the ZINC database~\cite{sterling2015zinc} within oracle budget. It is the traditional high-throughput screening approach.  
\item \textbf{REINVENT}~\cite{Olivecrona} is a reinforcement learning approach, represent molecule as SMILES string and uses recurrent neural network to model SMILES string. It pretrains a prior model using molecules on ZINC and finetune the model using the reward function. It uses REINFORCE to maximize the expected reward function. The learning rate is set to 0.0005; the batch size is set to 64; The hyperparameter $\sigma$ weighs the pretrained prior model and the reward function, and is set to 60. The model size is 16.3M. 
\item \textbf{JTVAE}~\cite{jin2018junction} build a junction tree to represent molecule via using substructure (either ring or atom) to represent molecule. It uses both molecular graph-leven and junction tree-level encoder and decoder. The VAE model is pretrained on ZINC databases. Then Bayesian Optimization is used to optimize the docking score on the continuous latent space. We use ``botorch'', the python's Bayesian optimization package, to implement the Bayesian optimization process. It has 703 substructures in vocabulary, extracted from ZINC. The hidden size is 450. 
The latent size of VAE is set to 56. The model size is 21.8 M. 
\item \textbf{Gen3D}~\cite{luo20213d} uses 3D deep generative models and grow the molecule via adding atoms auto-regressively. 
It train a universal model for all the targets. 
The number of message passing layers in context encoder is 6, and the hidden dimension is 256. We train the model using the Adam optimizer at learning rate 0.0001. 
The model size is 17.4 M. 
\item \textbf{GEGL} (Genetic Expert Guided Learning)~\cite{ahn2020guiding} uses LSTM (guided by RL agent) to imitate GA process, however, it is unable to inherit the GA's flexible assembling manner due to the auto-regressive essence of LSTM. It use Adam as optimizer with initial learning rate 1e-3. The batch size during sampling is 512, the batch size during optimization is 256. In GA, mutation rate is 0.01. The similarity threshold is 0.4, which constrain the similarity between the original molecule and the edited molecule. The maximal SMILES length is set to 120. 
\end{itemize}

\section{Additional Experimental Setup}

\subsection{Docking Simulation}

Molecular docking is a computational method which predicts the preferred orientation of one molecule to a second when a ligand and a target are bound to each other to form a stable complex. Knowledge of the preferred orientation in turn may be used to predict the strength of association or binding affinity between two molecules using, for example, scoring functions.
We adopt AutoDock Vina~\cite{trott2010autodock} to evaluate the binding affinity. The docking score estimated by AutoDock Vina is called Vina score and roughly characterizes the free energy changes of binding processes in kcal/mol. 
Vina score is usually smaller than 0 and lower Vina score means a stronger binding affinity between the ligand and target. 
We leverage the negative value of the docking score as reward function (Equation~\ref{eqn:policy}).

\subsection{Dataset}

In this paper, we use ZINC~\cite{sterling2015zinc} database and {CrossDocked2020}~\cite{francoeur2020three} dataset. 
ZINC is a free database of 250 thousands commercially-available drug-like chemical compounds for virtual screening~\cite{sterling2015zinc}. 
We randomly select molecules from ZINC~\cite{sterling2015zinc} database (around 250 thousands drug-like molecules) as 0-th generation of the genetic algorithms (\mname, Autogrow 4.0, graph-GA GA+D). 
Other baseline methods also use ZINC to either pretrain the models, e.g., JTVAE, REINVENT, RationaleRL or provide searching database, e.g., screening. 
We adopt {CrossDocked2020}~\cite{francoeur2020three} dataset that contains around 22 million ligand-protein complexes as the training data for pretraining the policy neural networks, as mentioned in Section~\ref{sec:policy_net}. 

Regarding the target proteins, we picked various disease-related proteins, including G-protein coupling receptors (GPCRs) and kinases from DUD-E~\cite{mysinger2012directory} and the SARS-CoV-2 main protease~\cite{zhang2021potent} as targets. 
for all the selected target protein, the binding pocket size for all the targets are set to (15.0, 15.0, 15.0). 
The units of coordinate are Angstrom $\mathring{A}$ ($10^{-10}$ m). 
Detailed descriptions of these targets are available at \url{https://www.rcsb.org/}.

\subsection{Evaluation metrics}
\label{sec:metrics}
We leverage the following {evaluation metrics} to measure the optimization performance:  
\begin{itemize}
\item \textbf{Novelty} is the fraction of the generated molecules that do not appear in the training set. 
\item  \textbf{Diversity} of generated molecules is defined as the average pairwise Tanimoto distance between the Morgan fingerprints~\cite{You2018-xh,jin2020multi,xie2021mars}. 
\begin{equation}
\text{diversity} = 1 - \frac{1}{|\mathcal{M}|(|\mathcal{M}|-1)}\sum_{m_1,m_2 \in \mathcal{M}} \text{sim}(m_1,m_2),
\end{equation}
where $\mathcal{M} $ is the set of generated molecules that we want to evaluate. $\text{sim}(m_1,m_2)$ is the Tanimoto similarity between molecule $m_1$ and $m_2$, where {(Tanimoto) Similarity} measures the similarity between the input molecule and generated molecules. 
It is defined as
$\text{sim}(X,Y) = \frac{\text{FP}_{X}^\top \text{FP}_{Y}}{\Vert \text{FP}_{X}\Vert_2 \Vert \text{FP}_{Y}\Vert_2}$, 
$\text{FP}_{X}$ is the binary Morgan fingerprint vector for the molecule $X$. In this paper, it is a 2048-bit binary vector. 
% \item \textbf{Molecular Properties} are the main target of optimization and would be evaluated. Detailed description of these properties are in Section~\ref{sec:property}. 
\item \textbf{QED} represents a quantitative estimate of drug-likeness. QED score ranges from 0 to 1. It can be evaluated by the RDKit package (\url{https://www.rdkit.org/}). 
\item \textbf{SA} (Synthetic Accessibility) score measures how hard it is to synthesize a given molecule, based on a combination of the molecule’s fragments contributions~\cite{ertl2009estimation}. 
It is evaluated via RDKit~\cite{landrum2006rdkit}. 
The raw SA score ranges from 1 to 10. A higher SA score means the molecule is hard to be synthesized and is not desirable. 
\item \textbf{Run Time}. Unlike optimizing some simple oracles such as QED and LogP scores, the docking simulation need to search 3D molecular conformation docked in the target, which is computationally expensive. Thus run time is an important metric to measure the efficiency of the methods. 
\end{itemize}

\section{Implementation Details}
\label{sec:implementation_details}

\subsection{Software/Hardware Configuration}
\label{sec:hardware_config}
We implemented \mname using Pytorch 1.10.2, Python 3.7, RDKit v2020.09.1.0 on an Intel Xeon E5-2690 machine with 256G RAM and NVIDIA Pascal Titan X GPUs.

\subsection{Hyperparameter Setup}
\label{sec:hyperparam}

The neural architectures of policy networks are E(3)-equivariant neural network (ENN)~\cite{satorras2021n}. 
The vocabulary set $\calV = \{C, N, O, S, H, \text{other}\}$. In ENN, the number of layers is set to 3, i.e., $L=3$; the hidden dimension is set to 100, i.e., $d=100$. 
In Equation~\ref{eqn:ENN}, $\text{MLP}_{e}(\cdot),  
\text{MLP}_{x}(\cdot), \text{MLP}_{h}(\cdot)$ are two-layer MLP in ENN with Swish activation~\cite{ramachandran2017searching} in hidden layer. 
Summation function is used as aggregation function to aggregate the last-layer's node embedding into graph-level embedding. 
All the atoms that within the binding site are used as the input of ENN. 
REINFORCE is used to implement policy gradient~\cite{peters2010policy,Olivecrona}. Adam is utilized as optimizer with learning rate $0.001$ for both crossover and mutation policy networks. 
The reaction set $\mathcal{R}$ is generated via merging two public reaction libraries: (1) the AutoClickChemRxn set (36 reactions)~\cite{durrant2013autogrow} and (2) RobustRxn set (58 reactions~\cite{hartenfeller2011collection}). 
We use RDKit~\cite{landrum2006rdkit} to perform \textit{in silico} chemical reaction based on SMARTS reaction template, \marked{then we have $|\calR| = 36+58=94$}.
In each generation, we generate up to 200 offspring (100 from crossover and 100 from mutation) and keep 50 most promising (with lowest Vina scores) ones for the next generation, \marked{i.e., $K=50$.}

\subsection{Code Repository}

The code repository is uploaded in supplementary material for reproducibility.

\section{Additional Experiment}

\subsection{Efficiency Study}

\begin{figure}[h!]
\centering
\subfigure[Slower Group ($>$10 hours)]{
\includegraphics[width=7.5cm]{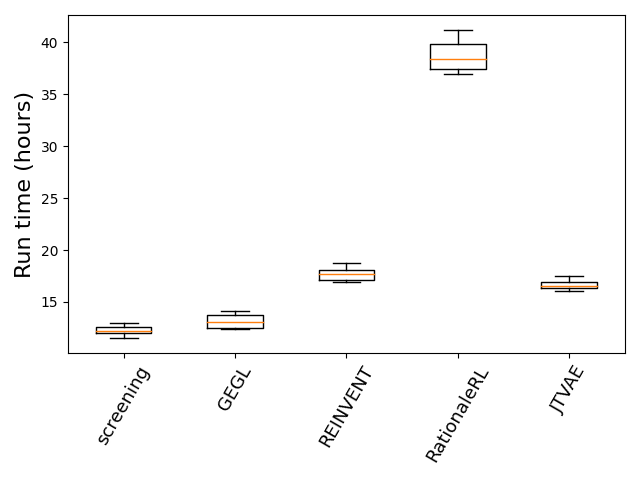}
\label{fig:runtime1}}
\subfigure[Fast Group ($<$10 hours)]{
\includegraphics[width=7.99cm]{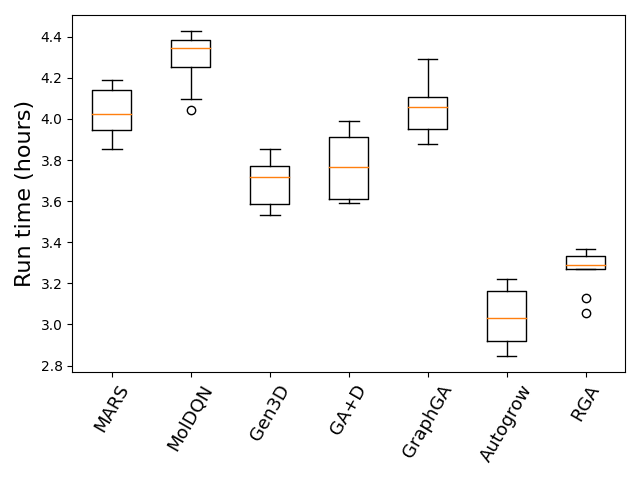}
\label{fig:runtime2}}
\caption{Efficiency evaluation measured by run time for all the methods. The unit of run time is hours. Due to the big variance in run time, for ease of visualization, we divide all the methods into two groups. One is slow group, containing 5 methods: screening, GEGL, REINVENT, RationaleRL and JTVAE. 
Another is fast group, containing 7 methods, MARS, MolDQN, Gen3D, GA+D, GraphGA, Autogrow, and RGA. 
}
\label{fig:runtime}
\end{figure}

As mentioned before, unlike optimizing some simple oracles such as QED and LogP scores, the docking simulation need to search 3D molecular conformation docked in the target, which is time-consuming. Thus run time is an important measurement to evaluate the efficiency of the methods. 
We report the bar of run time over different targets for all the compared methods in Figure~\ref{fig:runtime}. The run times varies greatly over different methods. Thus, for ease of visualization, we divided all the methods into two groups. One is slow group, containing 5 methods: screening, GEGL, REINVENT, RationaleRL and JTVAE, where all the methods take more than 10 hours. 
Another is fast group ($<$10 hours), containing 7 methods, MARS, MolDQN, Gen3D, GA+D, GraphGA, Autogrow, and RGA. We find that both Autogrow and RGA are efficient compared with other methods. 
This attributes to the unique design of genetic algorithm (both crossover and mutation operations) and the usage of filter after GA operators, as described in Section~\ref{sec:ga}. 
RGA is only slightly slower than Autogrow because it requires additional computation to pretrain/train the policy neural networks. 

\subsection{Pretraining Equivariant neural network}
\label{sec:pretraining}

\marked{
Pretraining equivariant neural network is a crucial step to \mname. 
We adopt {CrossDocked2020}~\cite{francoeur2020three} dataset that contains around 22 million ligand-protein complexes as the training data for pretraining the policy neural networks, as mentioned in Section~\ref{sec:policy_net}. 
We split the whole dataset into training/validation dataset with ratio of 9:1. 
Each data point is a target-ligand complex and the binding affinity (scalar). 
We report the validation loss of ENN on target-ligand binding affinity prediction task. The validation loss function is root mean square error (RMSE). We find the learning process converges rapidly when passing 150K data points (within an epoch) in terms of validation RMSE loss. 
}

\begin{figure}[h!]
\centering
\includegraphics[width=0.47\linewidth]{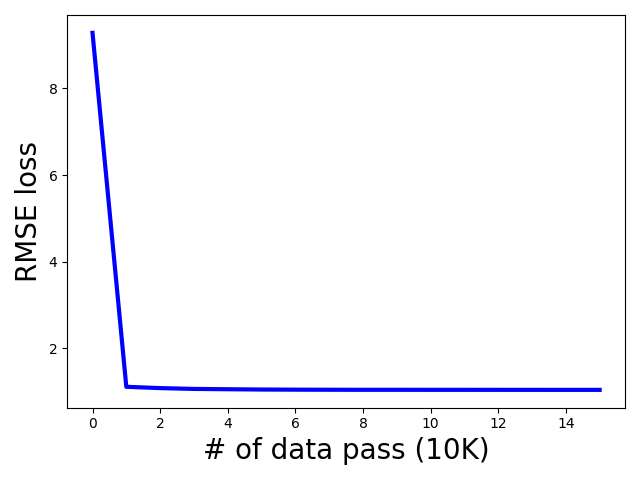}
\caption{\marked{Learning curve of pretraining Equivariant neural network based on target-ligand binding affinity prediction. We plot the root-mean-square error (RMSE) loss as a function of number of passed data. We use early stop strategy to terminate the learning process earlier when the validation loss would not decrease to save computational resource and avoid overfitting. We found the learning process would converges when passing 150K data samples (within an epoch), RMSE loss decreases from more than 8 to less than 1. 
}}
\label{fig:binding_affinity_regression}
\end{figure}

\subsection{Additional Ablation Study}

To further understand our model and GA process, we conduct an ablation study to investigate the impact of each component/strategy to optimization performance. Specifically, we consider the following four variants of RGA. 
\mname -pretrain is a variant of \mname that does not pretrain the policy neural network. 
\mname -KT (Knowledge Transfer) is a variant of \mname that does not training policy neural network on different target proteins, i.e., optimizing ligand for one target at a time. 
\mname -MU (mutation) is a variant of \mname that does not involve mutation operation in GA. That is, all the ligands are generated via crossover operator. 
Correspondingly, \mname -CO (crossover) is a variant that does not use crossover operation in GA, which means no mutation operator. 
The results are reported in Table~\ref{table:ablation}. 
We find that removing either component/strategy will cause a drop in optimization performance (i.e., increase in TOP-100/10/1 scores). Both crossover and mutation are critical to the optimization performance. Also, both pretraining the policy networks and knowledge transfer between different target have positive contribution to the performance. The ablation study furtherly validates the effectiveness of the proposed RGA method. 

\begin{table*}[t!]
\centering
\caption{Ablation studies.  
Arrows ($\uparrow$, $\downarrow$) indicate the direction of better performance. 
For each metric, the best method is underlined. 
\mname -pretrain is a variant of \mname that does not pretrain the policy neural network. 
\mname -KT (Knowledge Transfer) is a variant of \mname that does not training policy neural network on different target proteins, i.e., optimizing ligand for one target at a time. 
\mname -MU (mutation) is a variant of \mname that does not involve mutation operation in GA. That is, all the ligands are generated via crossover operator. 
Correspondingly, \mname -CO (crossover) is a variant that does not use crossover operation in GA, which means no mutation operator. 
Via comparing the results with RGA (full) in the first line, we observe that removing either component would cause a drop in optimization performance (i.e., increase in TOP-100/10/1 scores). 
}
\label{table:ablation}
 \resizebox{0.98\textwidth}{!}{
\begin{tabular}{lcccccccc}
\toprule[0.6pt]
Method & TOP-100$\downarrow$ & TOP-10$\downarrow$ & TOP-1$\downarrow$ & Nov$\uparrow$ & Div$\uparrow$ & QED$\uparrow$ &  SA$\downarrow$ \\
\hline 
RGA (full) & {\underline{-11.867\std{0.170}}} & \underline{{-12.564\std{0.287}}} & {\underline{-12.869\std{0.473}}} & 100.0\std{0.0}\% & \underline{0.857\std{0.020}} & {{ 0.742\std{0.036}}} & { 2.473\std{0.048}}  \\  
RGA - pretrain & -11.443\std{0.219} & -12.424\std{0.386} & -12.435\std{0.654} & 100.0\std{0.0}\% & 0.854\std{0.035} & \underline{0.750\std{0.034}} & 2.494\std{0.043} \\ 
RGA - KT & -11.434\std{0.169} & -12.437\std{0.354} & -12.502\std{0.603} & 100.0\std{0.0}\% & 0.853\std{0.028} & 0.738\std{0.034} & 2.501\std{0.050} \\ 
RGA - MU & -10.919\std{0.166} & -11.135\std{0.362} & -11.747\std{0.455} & 100.0\std{0.0}\% & 0.812\std{0.032} & 0.702\std{0.050} & 2.970\std{0.048} \\
RGA - CO & -9.866\std{0.169} & -10.320\std{0.296} & -10.793\std{0.501} & 100.0\std{0.0}\% & 0.737\std{0.048} & 0.748\std{0.067} & \underline{2.467\std{0.034}} \\
\bottomrule[0.6pt]
\end{tabular}}
\end{table*}

% \subsection{Examples}

% In this section, we provide more examples of ligand poses (generated by \mname) and binding sites for different target proteins in Figure~\ref{fig:examples}. 

% \begin{figure}[h!]
% \centering
% \subfigure[]{
% \includegraphics[width=4.5cm]{figure/crossover.png}
% \label{fig:example1}}
% \subfigure[]{
% \includegraphics[width=4.99cm]{figure/mutation.png}
% \label{fig:example2}}
% \caption{Case studies
% }
% \label{fig:examples}
% \end{figure}

\subsection{Scalability}
\label{sec:scalability}

\marked{
As mentioned before, the population size in \mname is $K$. 
Then we analyze the computational complexity within each generation. 
There are two operations, including crossover and mutation operations, as described in Section~\ref{sec:policy_net}. }

\marked{
For crossover, the first step is to select the first parent molecule, we need to evaluate the probability over all the molecules in current population, whose complexity is $O(K)$. 
The second step is to select the second parent molecule based on the first parent molecule, we need to evaluate the probability over all the remaining molecules in current population, as shown in Equation~\eqref{eqn:crossover_prob}, whose complexity is $O(K-1)$. The complexity of crossover operation is $O(K)$. 
}

\marked{
On the other hand, for mutation operation, the first step is to select the parent molecule (only one parent) via evaluating the probability over all the molecules in the current population, i.e.,  $p_{\text{mutation}}^{(1)}(X^{\text{parent}}_{\text{mutation}} | \calS^{(t)})$, whose complexity is $O(K)$. 
The second step is to select a mutated molecules (generated by chemical reaction), as shown in Equation~\eqref{eqn:mutation_prob}. The complexity is $O(|\calR|)$, where $\calR$ is the reaction set (see Table~\ref{table:notation}). The complexity of mutation operation is $O(|\calR|+K)$. 
}

\marked{
To summarize, the complexity of \mname is $O(K+|\calR|)$. That is, \mname scales linearly in population size $K$ and the size of the chemical reaction set $|\calR|$. As mentioned in Section~\ref{sec:hyperparam}, $|\calR|=94$, $K=50$. Thus, \mname owns desired scalability and is not computational expensive. 
}

\subsection{Significance Studies}
\label{sec:statistical_test}

\marked{In this section, we present the significance studies. 
Specifically, we conduct the hypothesis testing to show the statistical significance of our method over the other GA methods. 
We compare the significance of the improvement over other GA methods (including AutoGrow4, Graph-GA, GA+D, and GEGL) via running 5 independent trials with different random seeds and then evaluating the \textbf{p-value}. 
We consider the top-100 and top-10 score as the major metrics, which are the most important metrics for optimization performance. 
We compare \mname versus AutoGrow 4, Graph-GA and GA+D. The results (p-value) are shown in Table~\ref{table:statistical_test}. 
We find almost all the p-values are less than 0.05 (except one value), which indicates that the improvements of RGA over other GA methods are statistically significant. 
}

\begin{table*}[t!]
\centering
\caption{\marked{Results of hypothesis testing. 
We conduct hypothesis testing to show the statistical significance of our method over other GA-based methods. Specifically, we compare the significance of the improvement over GA methods (AutoGrow4, Graph-GA, GA+D, GEGL) via running 5 independent trials with different random seeds and then evaluating p-value. We consider top-100/10/1 scores, the most important metrics for optimization performance. 
}}
\label{table:statistical_test}
 \resizebox{0.98\textwidth}{!}{
\begin{tabular}{lccc}
\toprule[0.6pt]
 & p-value on TOP-100 & p-value on TOP-10 & p-value on TOP-1 \\
\hline 
\mname v.s. AutoGrow 4 & 0.002 & 0.005 & 0.07 \\ 
\mname v.s. Graph-GA & 0.003 & 0.010 & 0.046 \\ 
\mname v.s. GA+D & 1.0e-7 & 5.0e-5 & 3e-4 \\ 
\mname v.s. GEGL & 2.5e-4 & 3.7e-3 & 5.0e-3 \\ 
\bottomrule[0.6pt]
\end{tabular}}
\end{table*}

\begin{figure}[t!]
\centering
\subfigure[-12.8 kcal/mol]{
\includegraphics[width=0.31\linewidth]{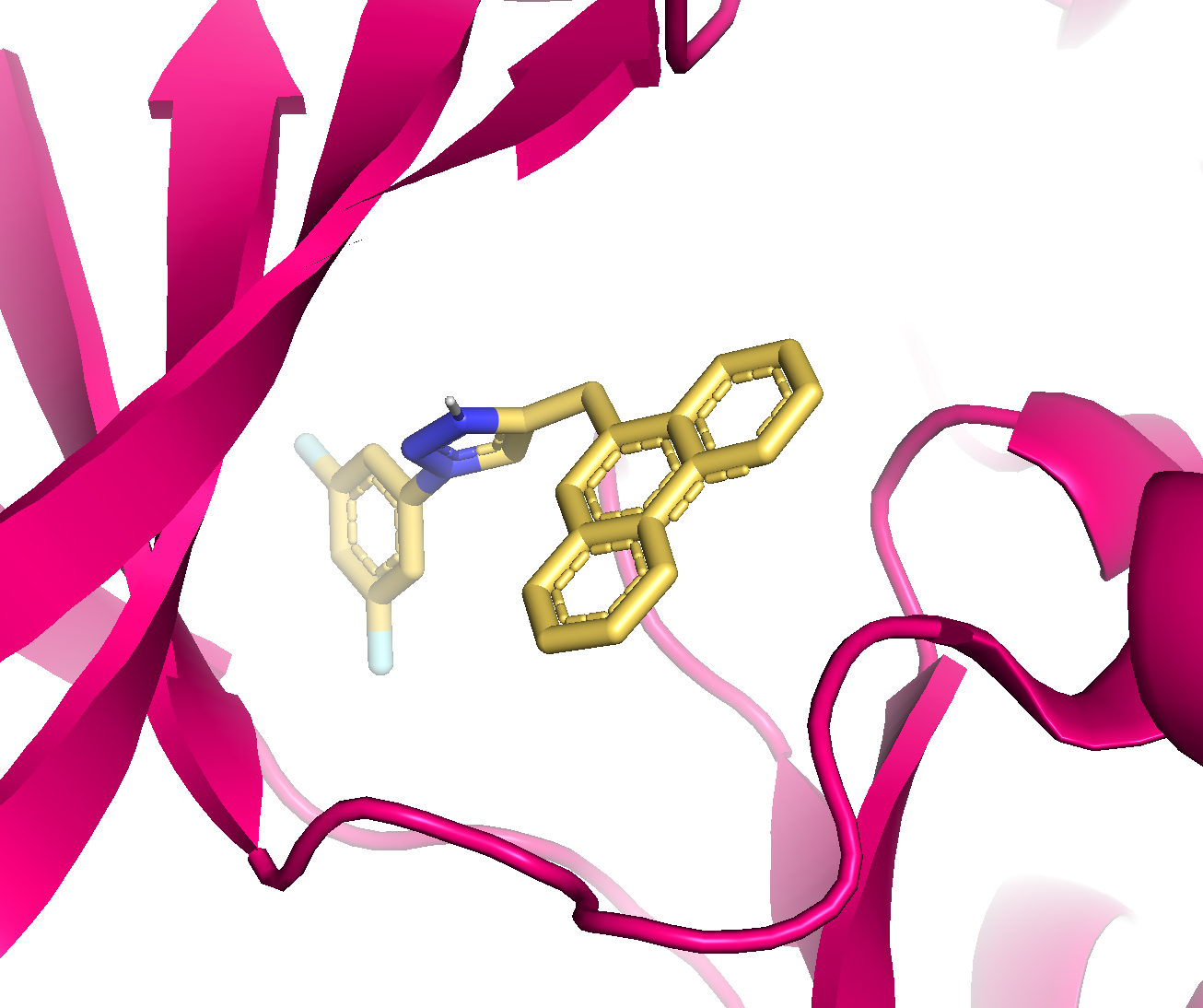}}
\subfigure[-12.7 kcal/mol]{
\includegraphics[width=0.31\linewidth]{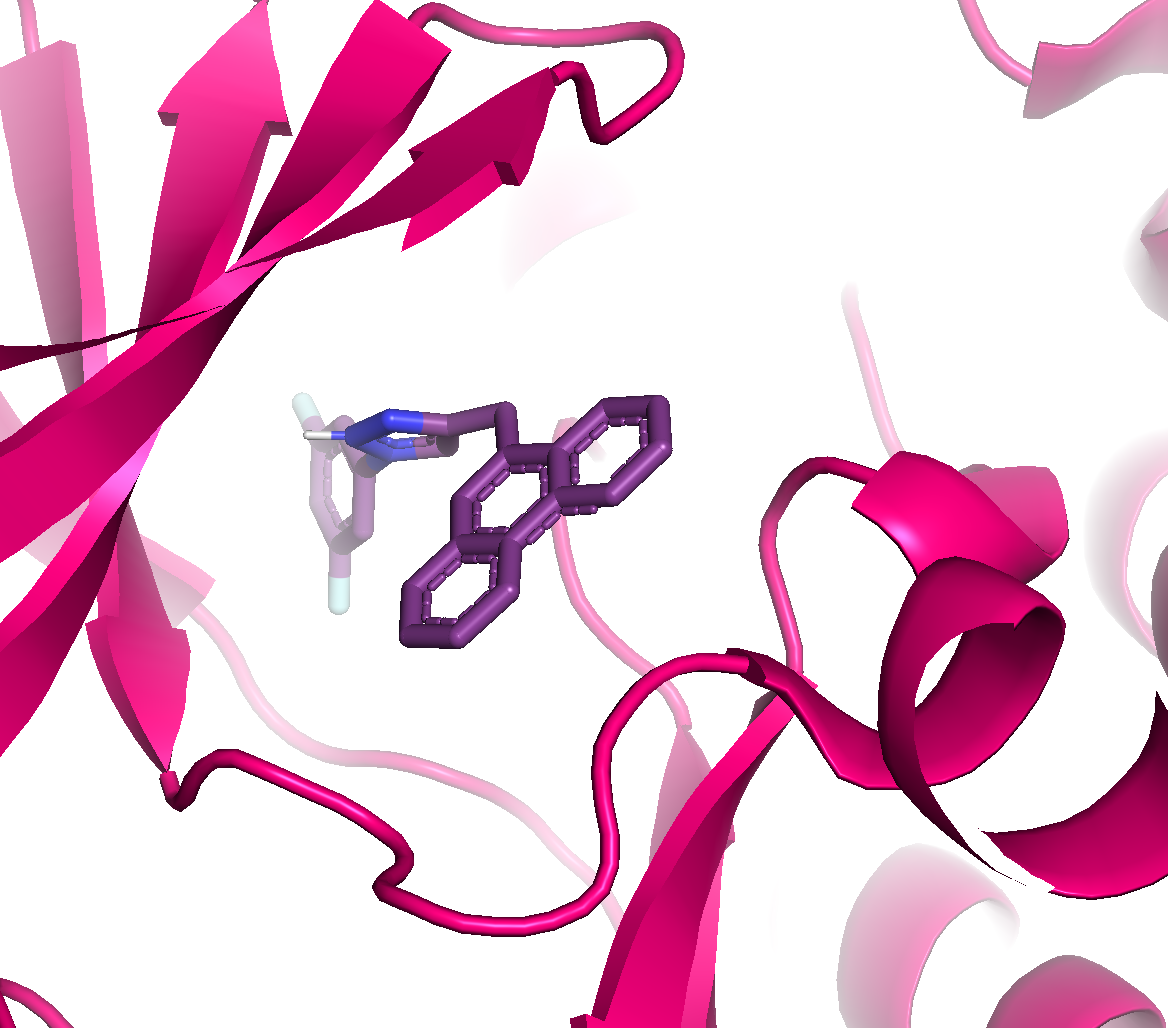}}
\subfigure[-12.7 kcal/mol]{
\includegraphics[width=0.31\linewidth]{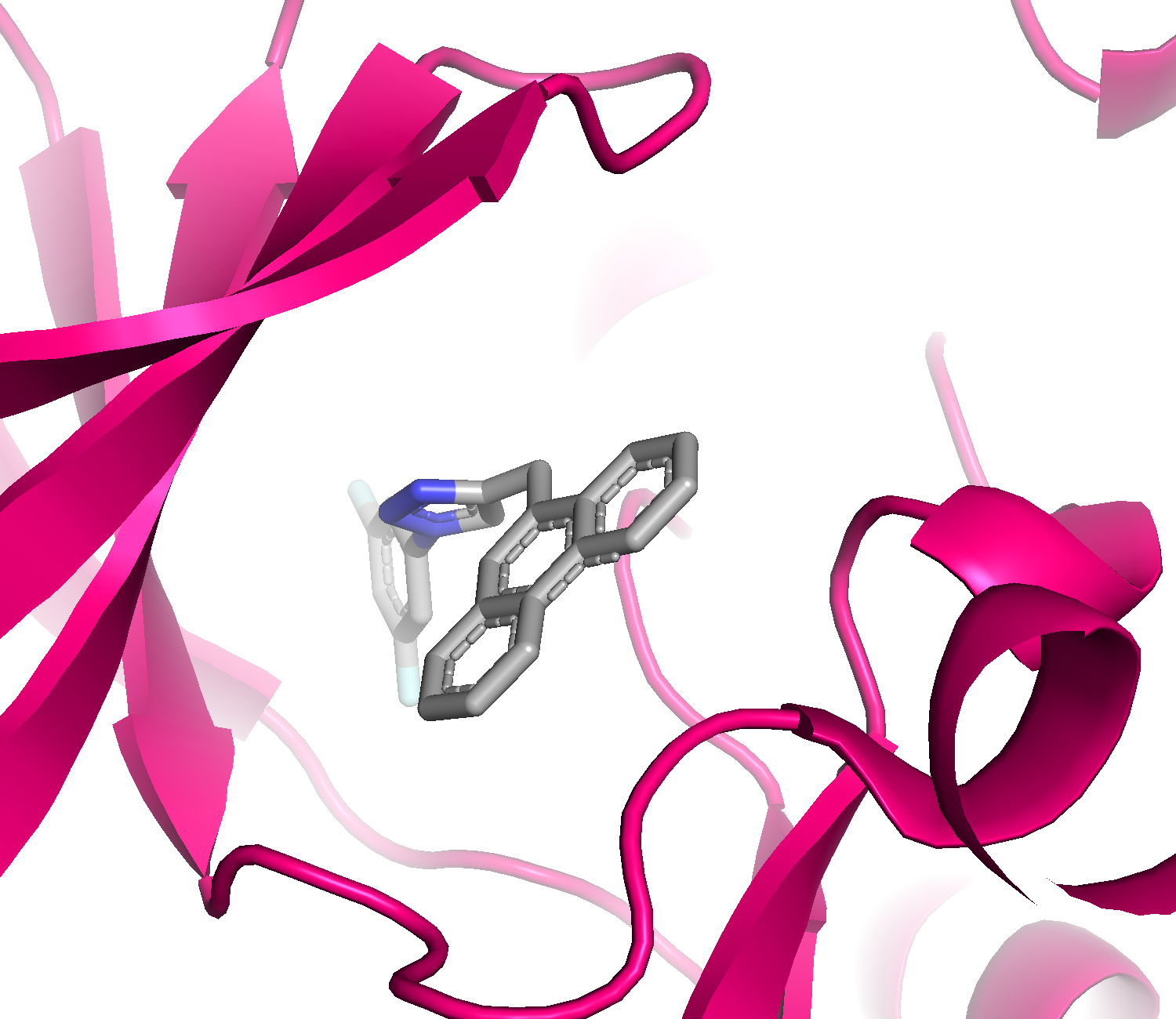}}
\subfigure[-12.7 kcal/mol]{
\includegraphics[width=0.31\linewidth]{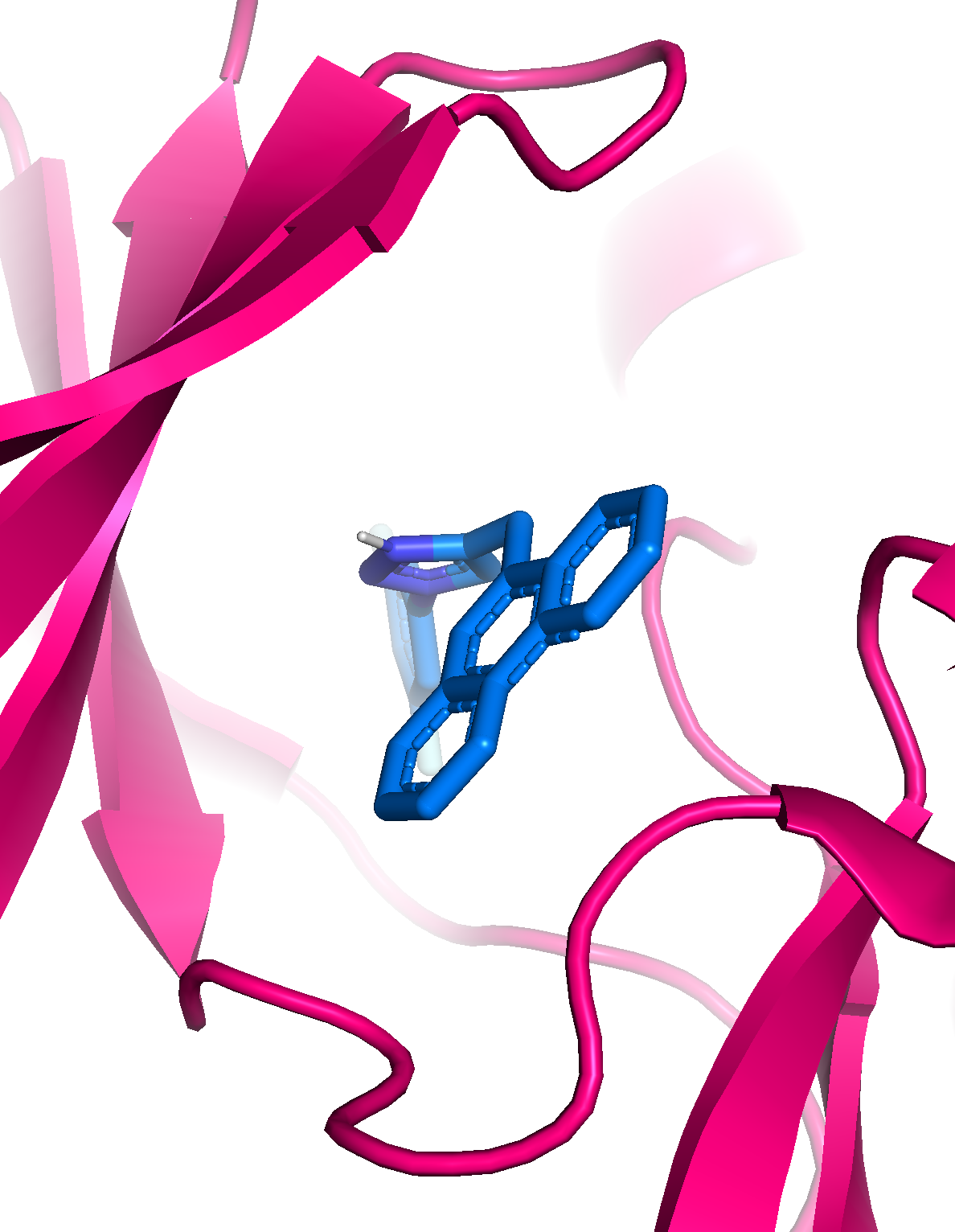}}
\subfigure[-12.6 kcal/mol]{
\includegraphics[width=0.31\linewidth]{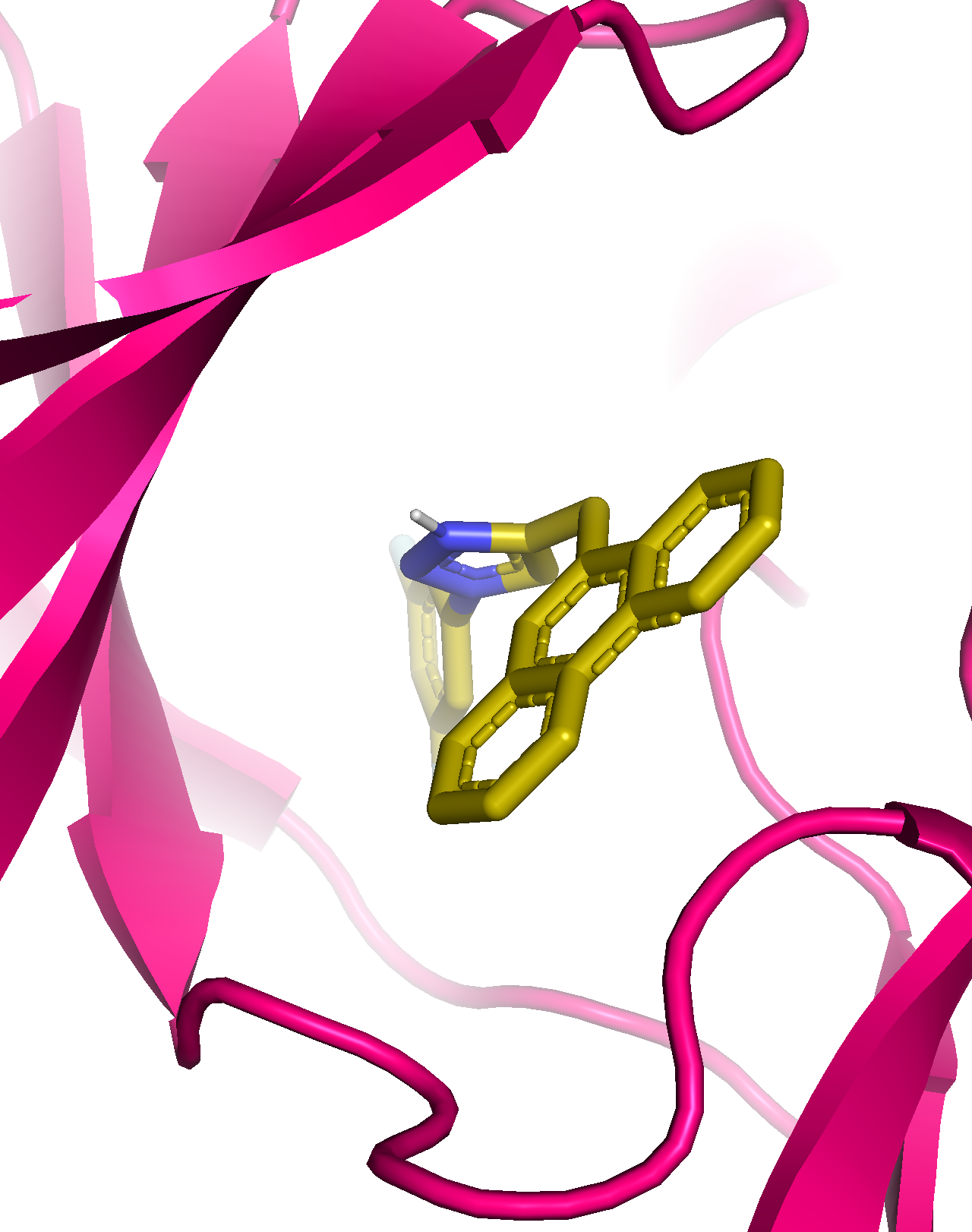}}
\subfigure[-12.6 kcal/mol]{
\includegraphics[width=0.31\linewidth]{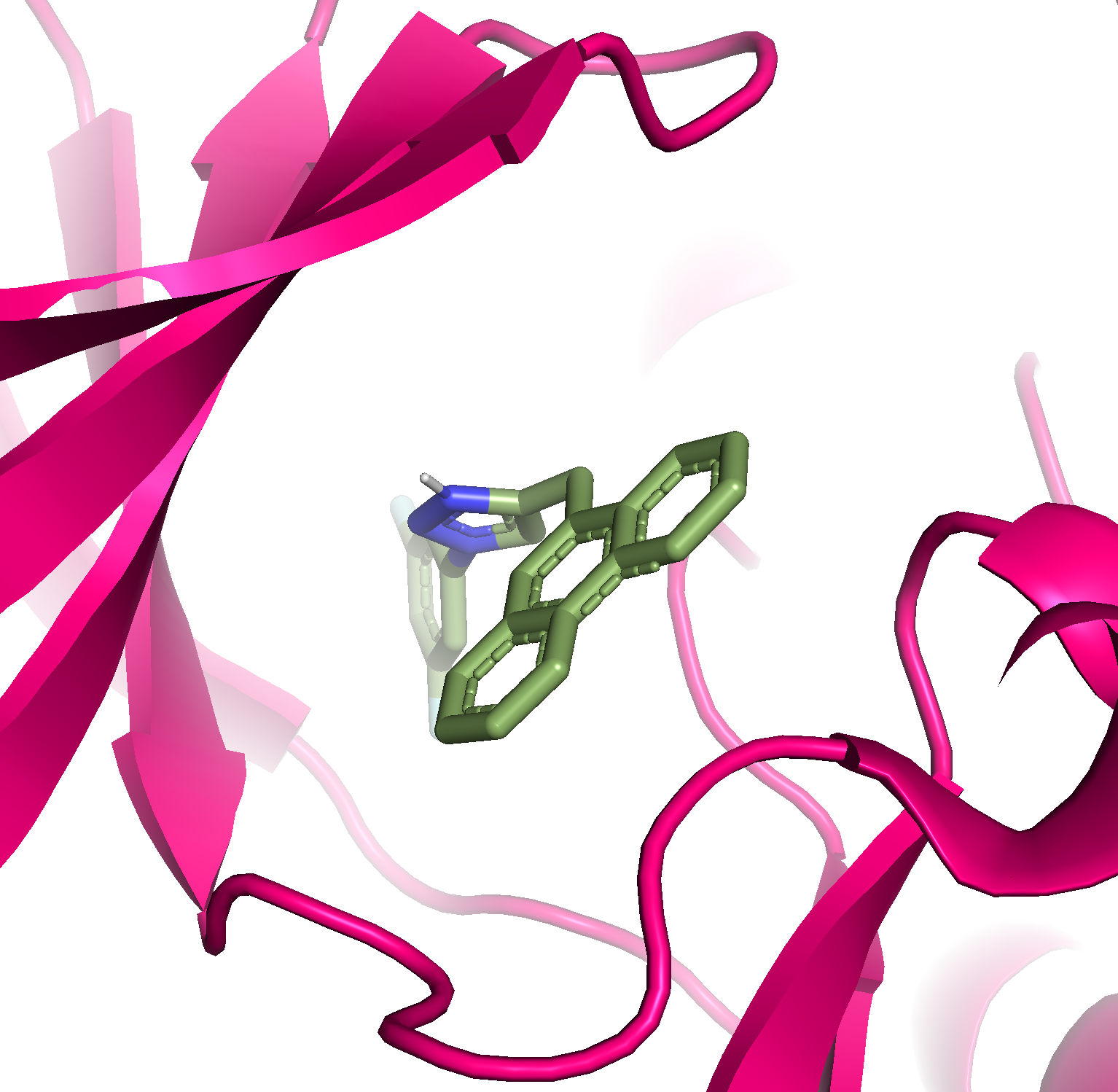}}
\subfigure[-12.6 kcal/mol]{
\includegraphics[width=0.31\linewidth]{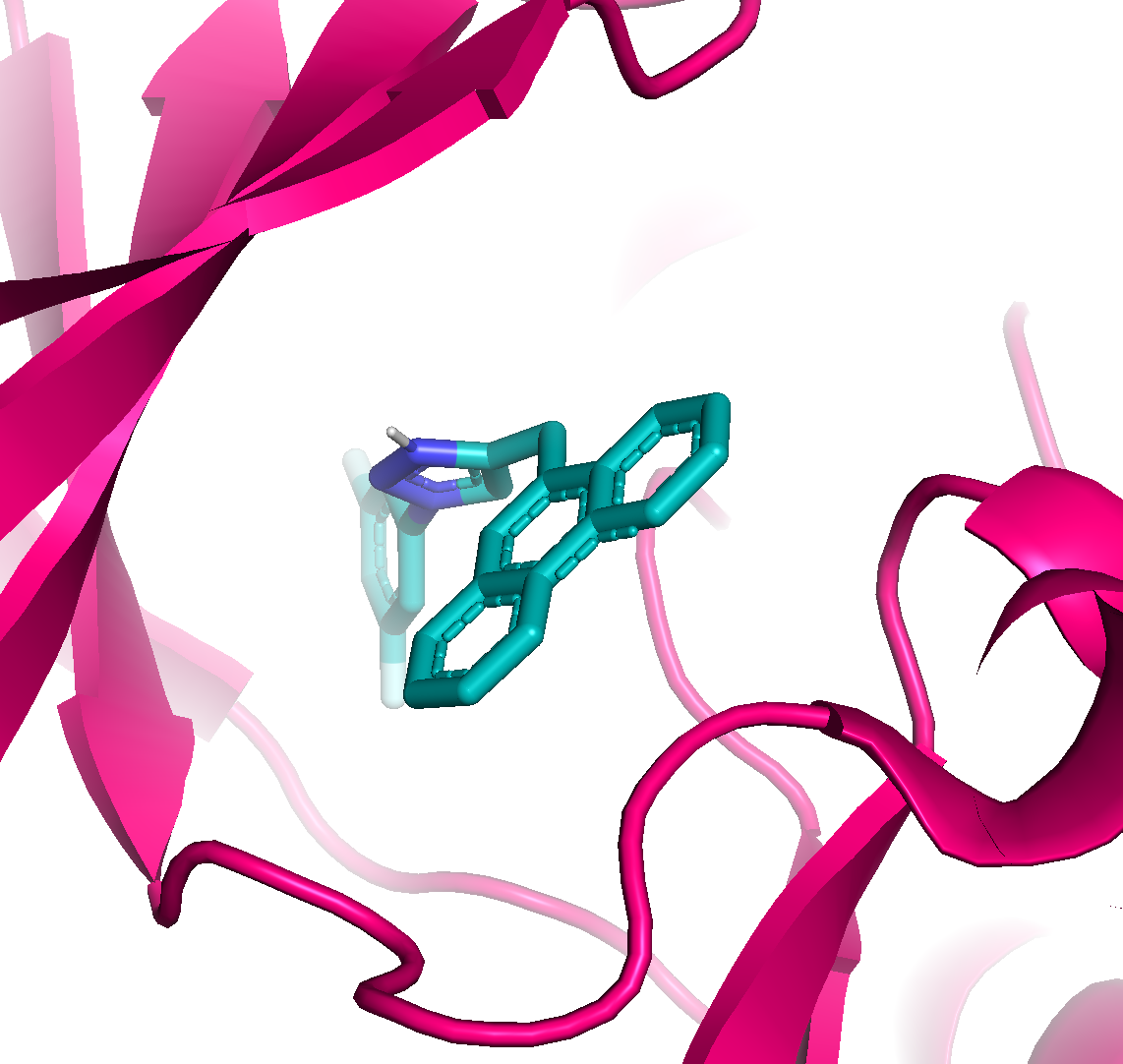}}
\subfigure[-12.6 kcal/mol]{
\includegraphics[width=0.31\linewidth]{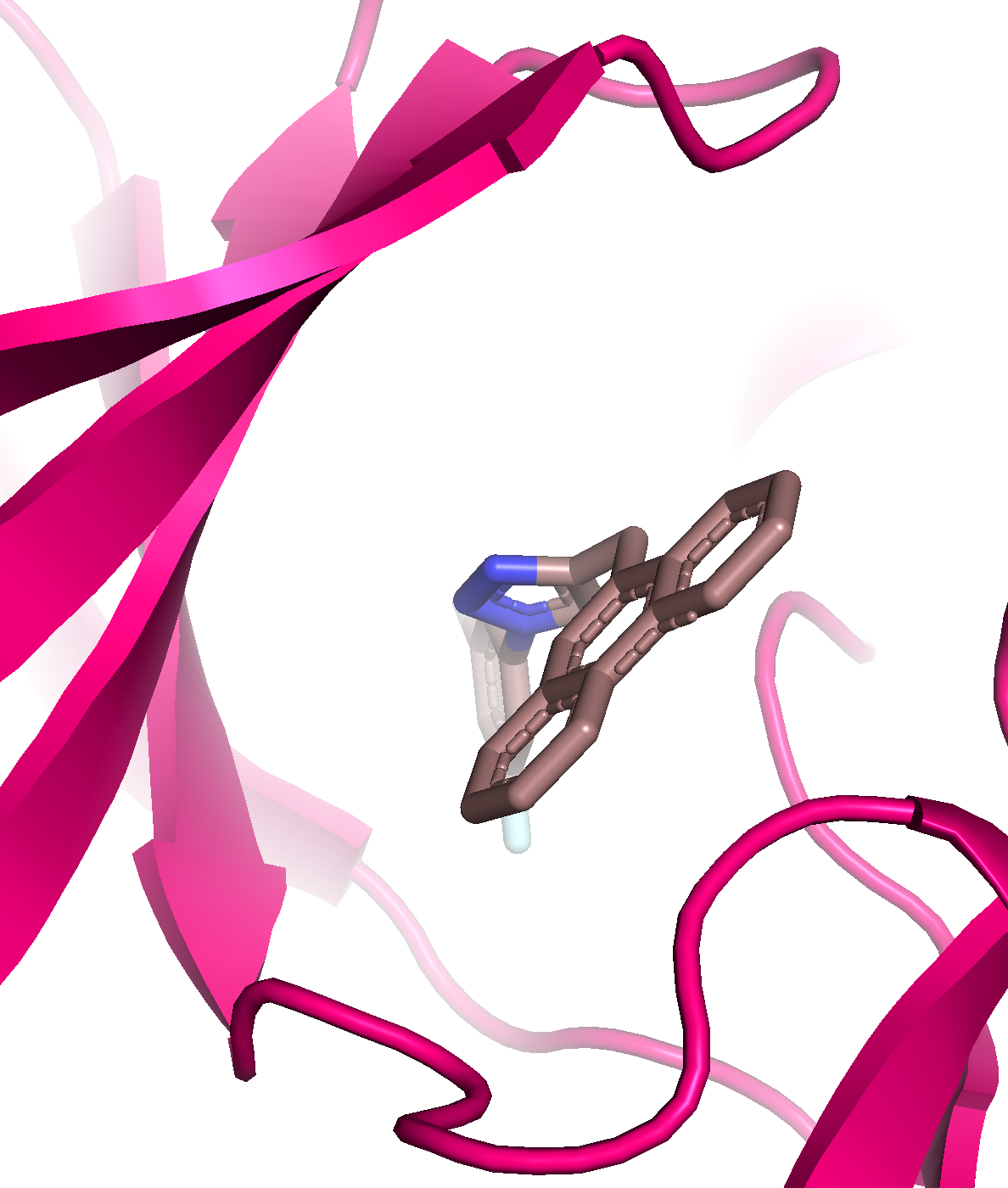}}
\caption{
\marked{
Example of ligand poses (generated by \mname) and binding sites of target structures ``2rgp''. 
}}
\label{fig:example_2rgp}
\end{figure}

\begin{figure}[t!]
\centering
\subfigure[-13.4 kcal/mol]{
\includegraphics[width=0.31\linewidth]{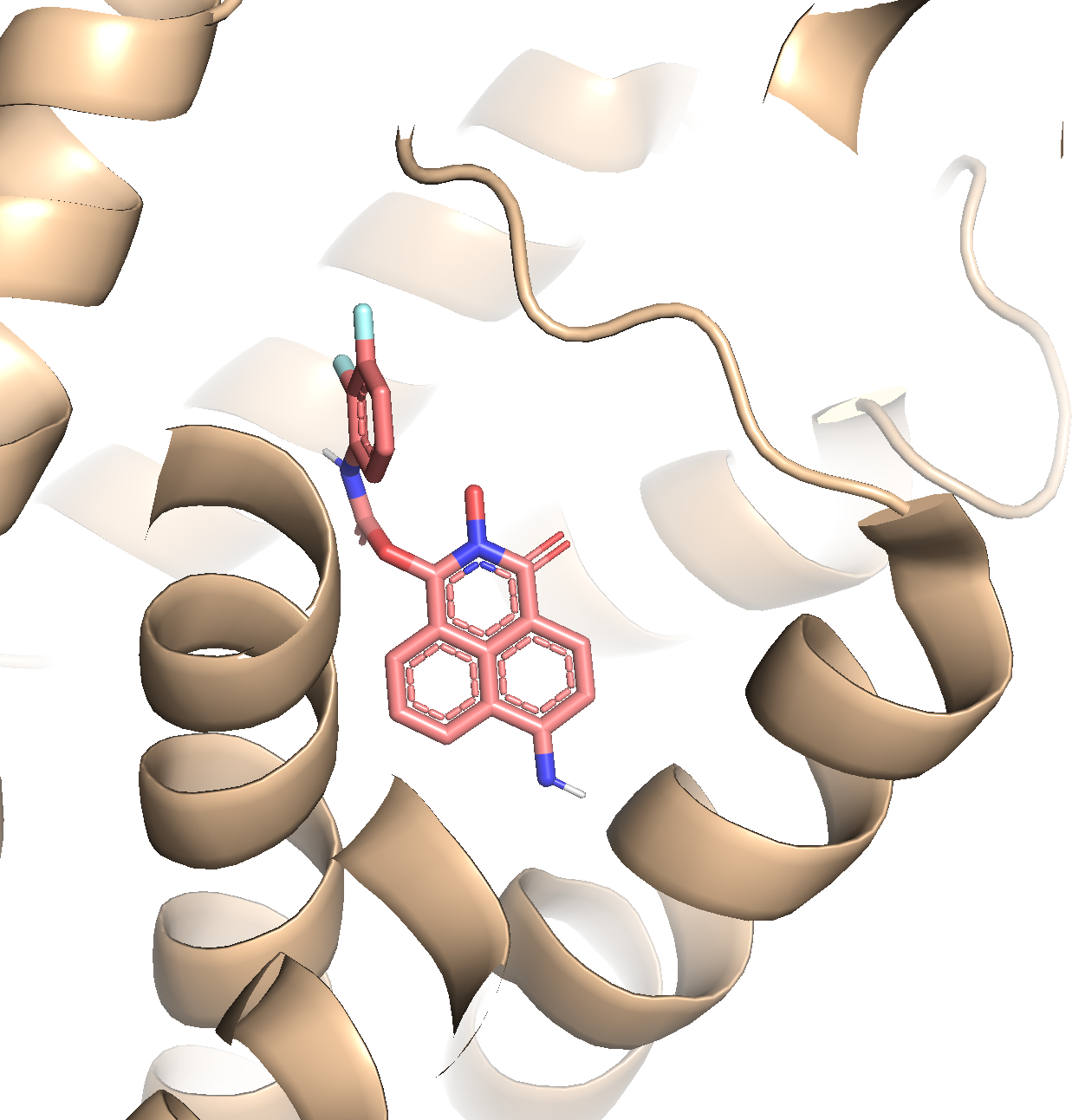}}
\subfigure[-13.4 kcal/mol]{
\includegraphics[width=0.31\linewidth]{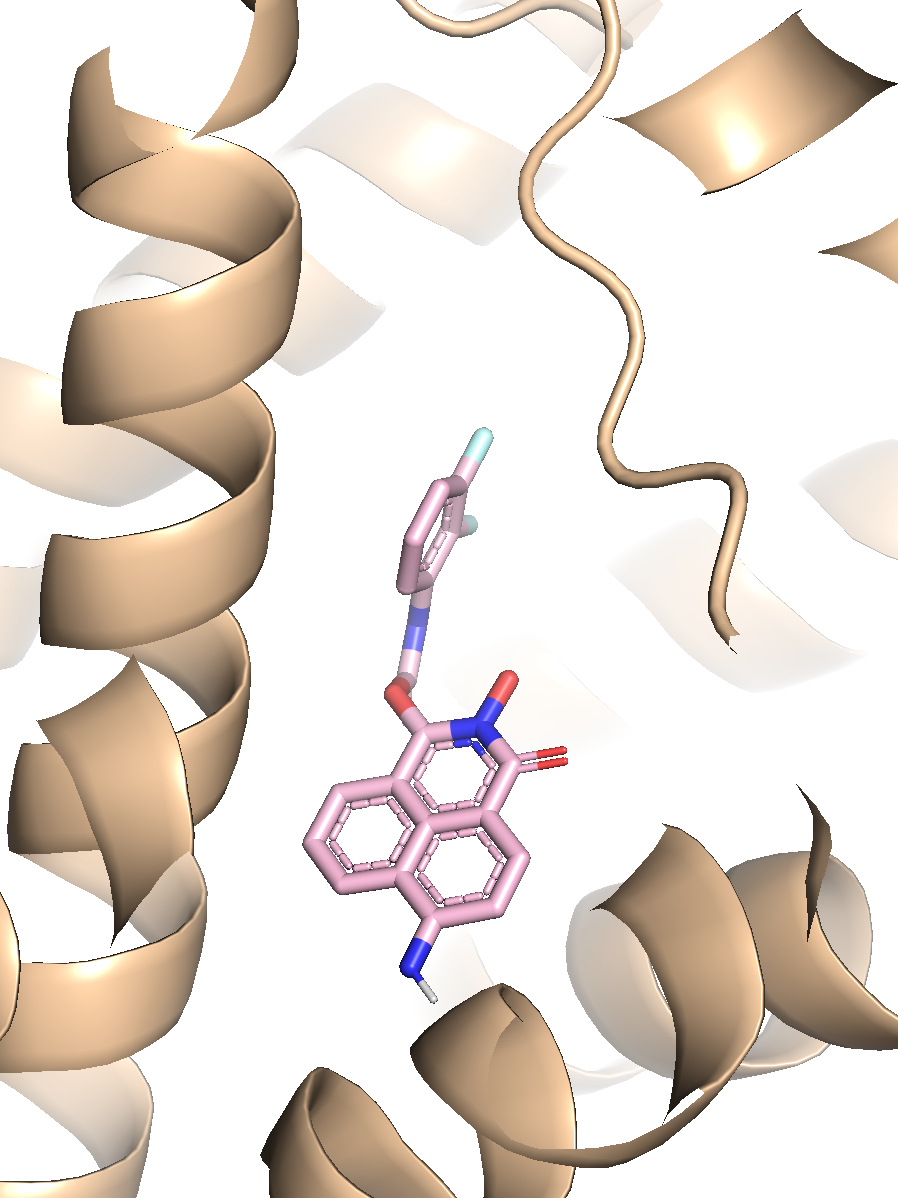}}
\subfigure[-13.4 kcal/mol]{
\includegraphics[width=0.31\linewidth]{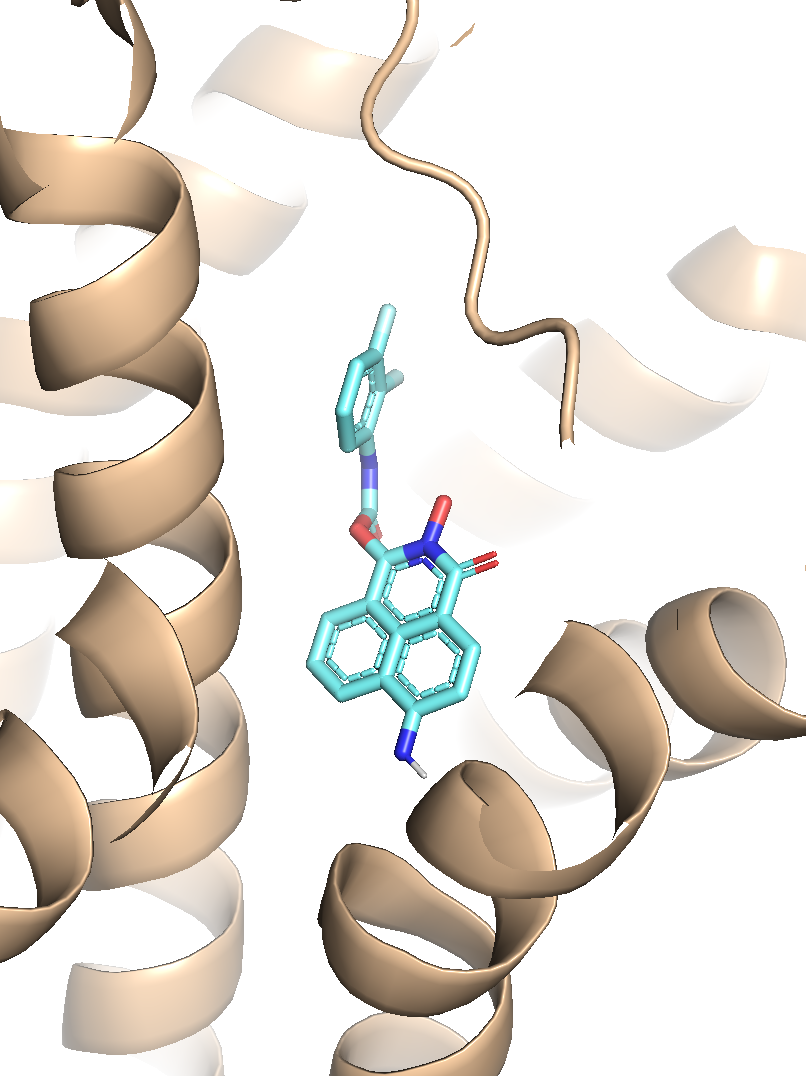}}
\subfigure[-13.4 kcal/mol]{
\includegraphics[width=0.31\linewidth]{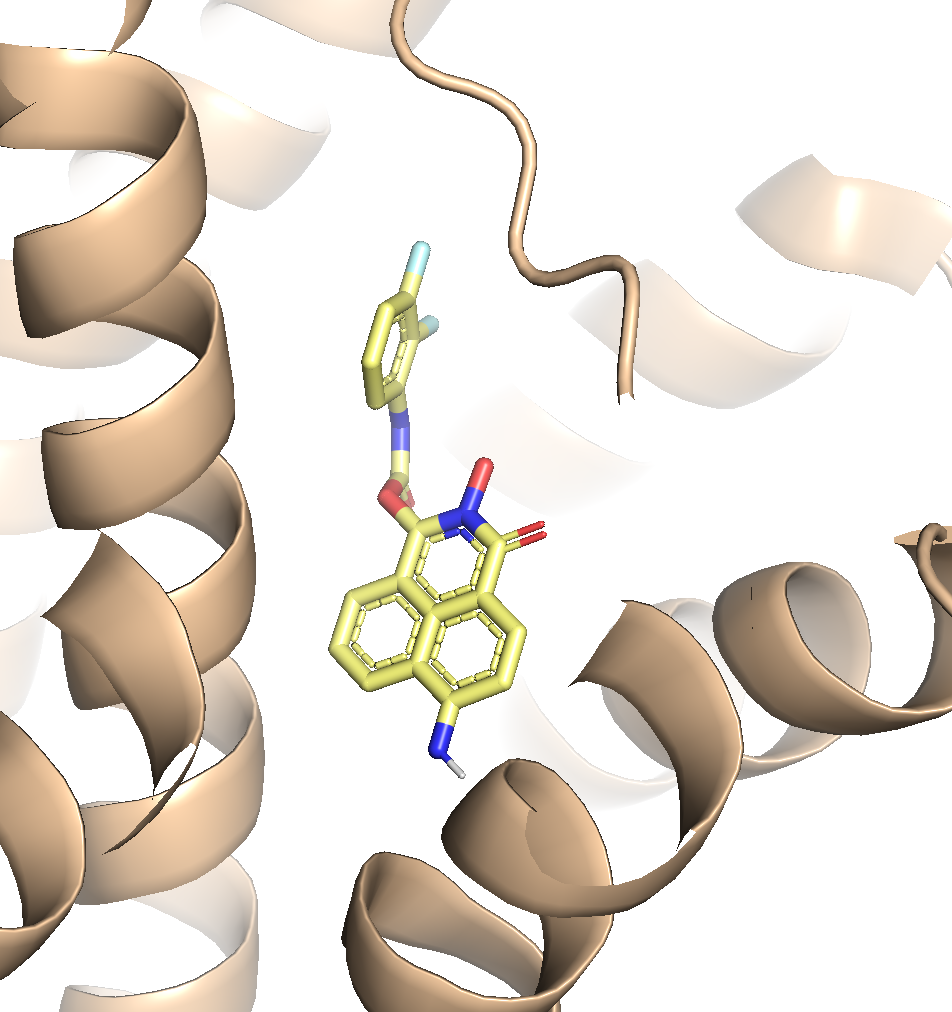}}
\subfigure[-13.3 kcal/mol]{
\includegraphics[width=0.31\linewidth]{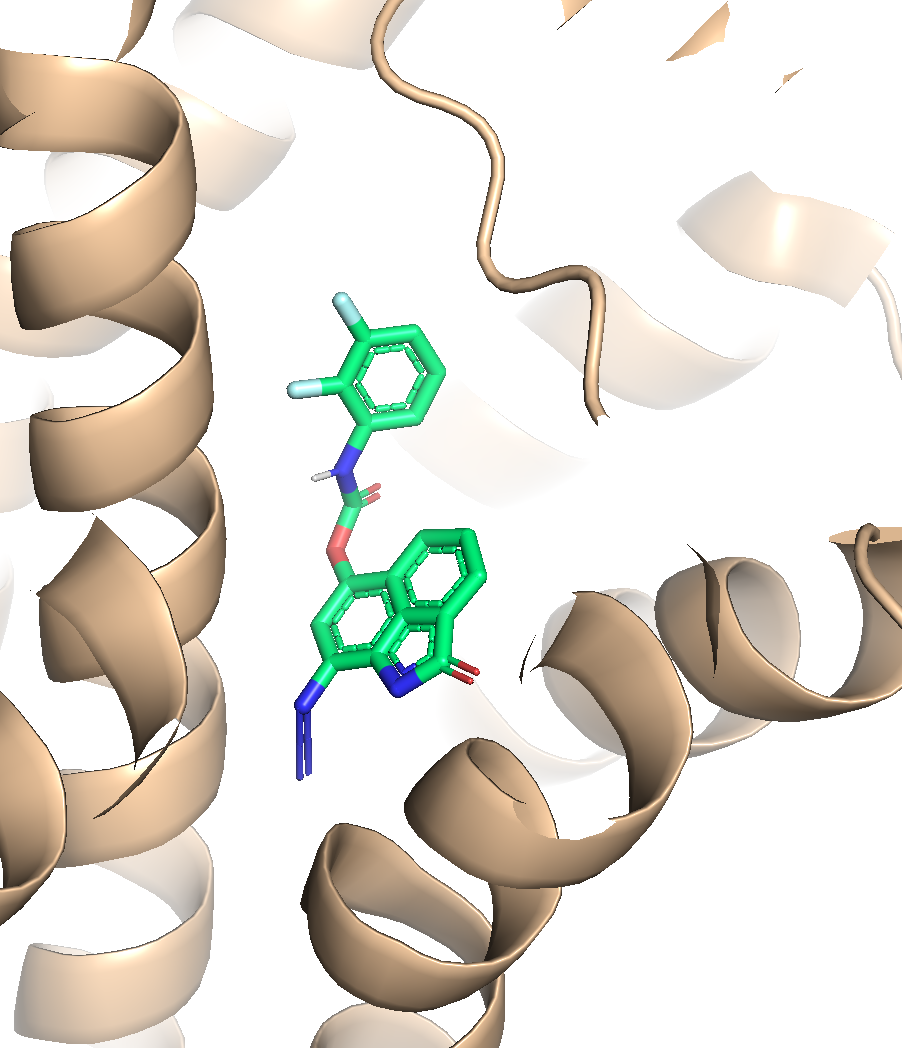}}
\subfigure[-13.3 kcal/mol]{
\includegraphics[width=0.31\linewidth]{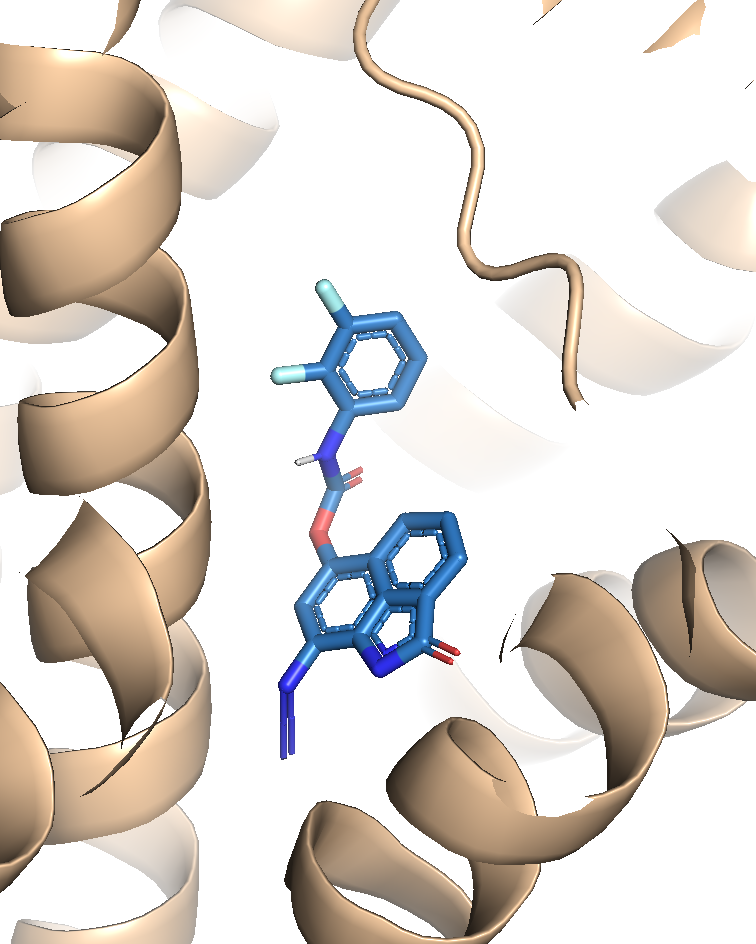}}
\subfigure[-13.3 kcal/mol]{
\includegraphics[width=0.31\linewidth]{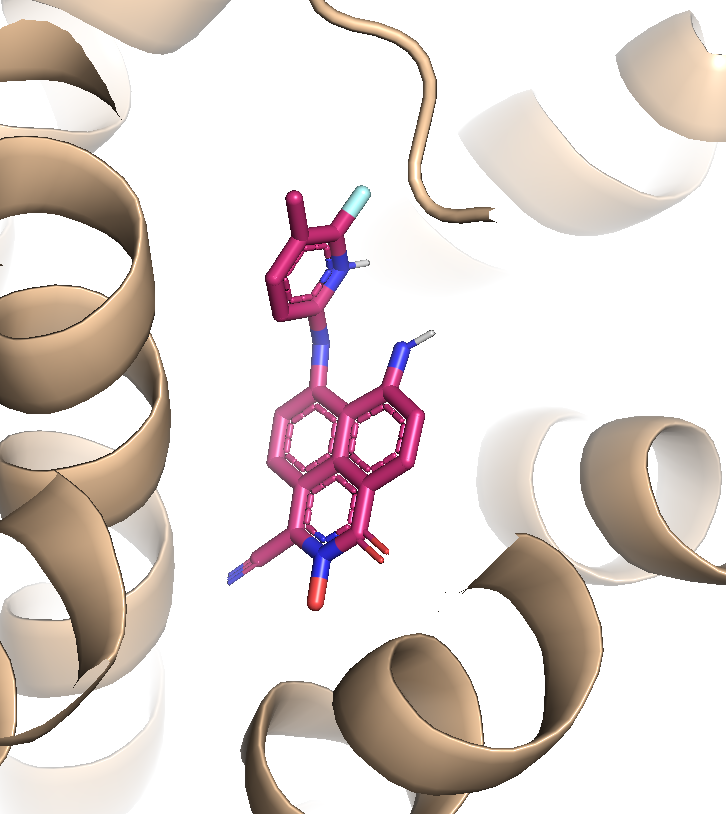}}
\subfigure[-13.3 kcal/mol]{
\includegraphics[width=0.31\linewidth]{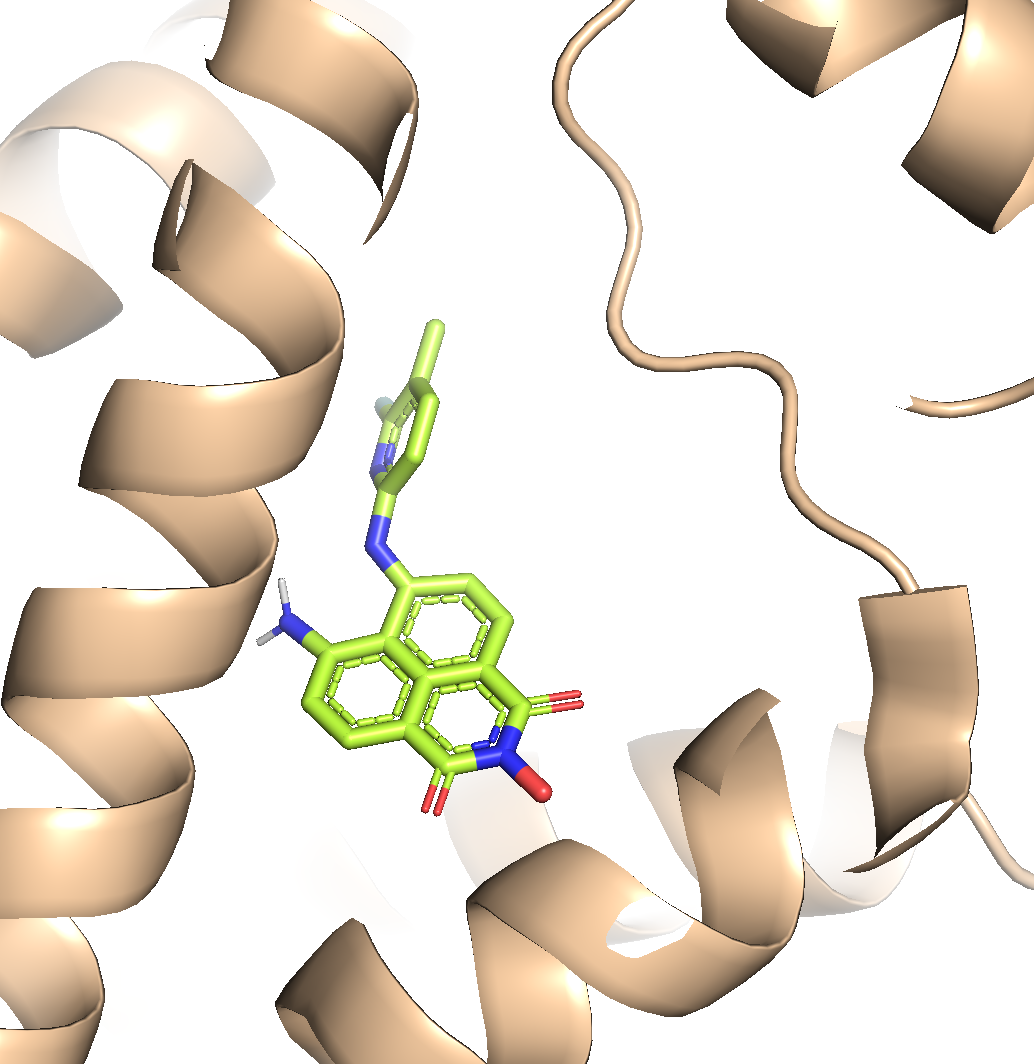}}
\subfigure[-13.3 kcal/mol]{
\includegraphics[width=0.31\linewidth]{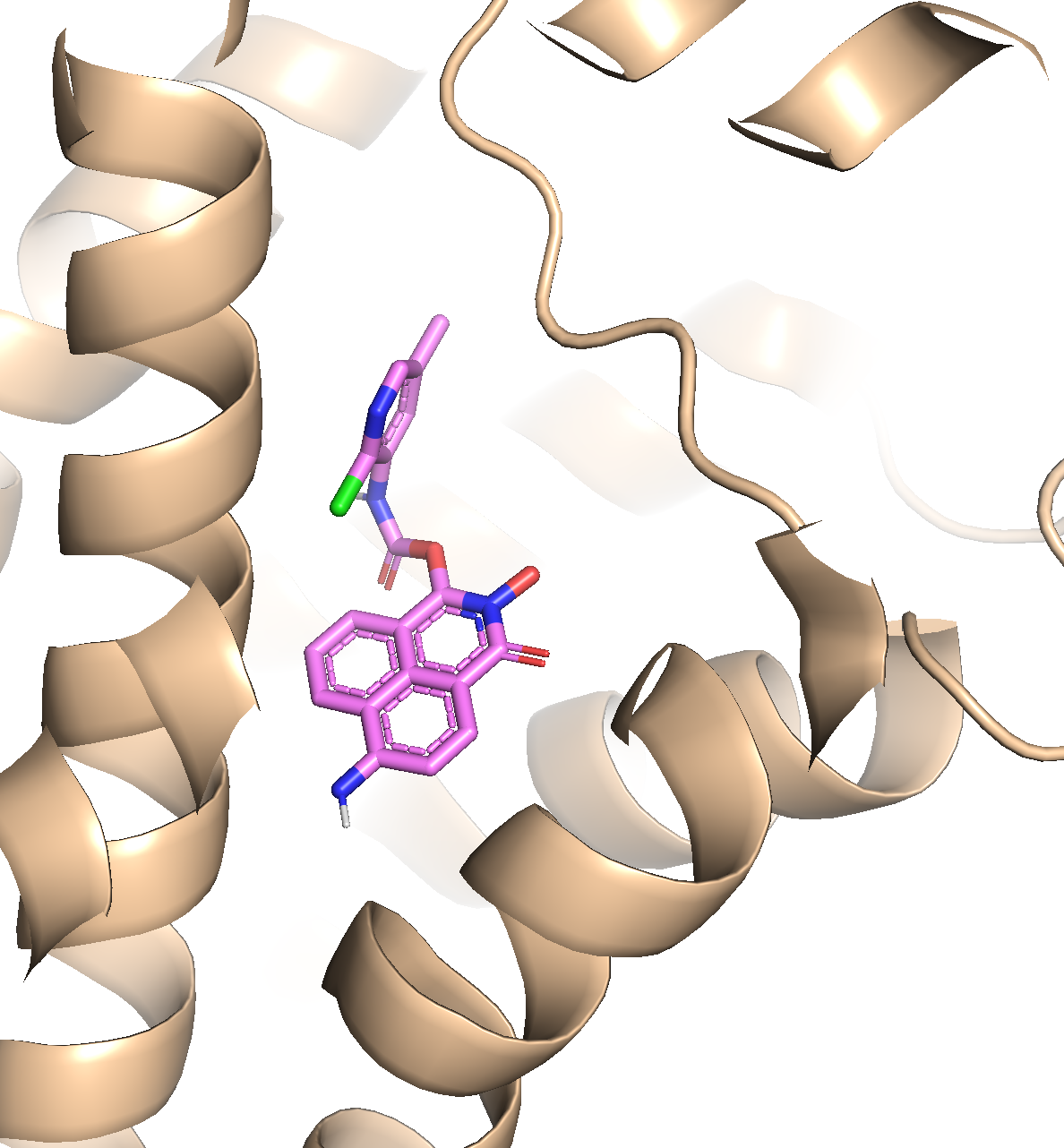}}
\caption{
\marked{
Example of ligand poses (generated by \mname) and binding sites of target structures ``3ny8''. 
}}
\label{fig:example_3ny8}
\end{figure}

\begin{figure}[t!]
\centering
\subfigure[-13.1 kcal/mol]{
\includegraphics[width=0.31\linewidth]{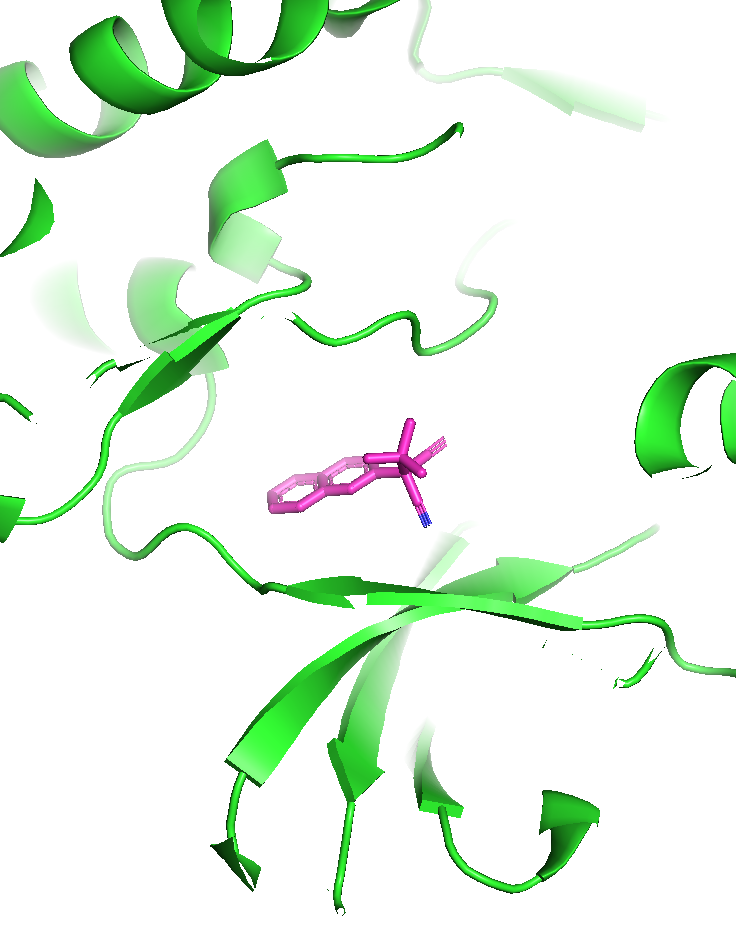}}
\subfigure[-13.1 kcal/mol]{
\includegraphics[width=0.31\linewidth]{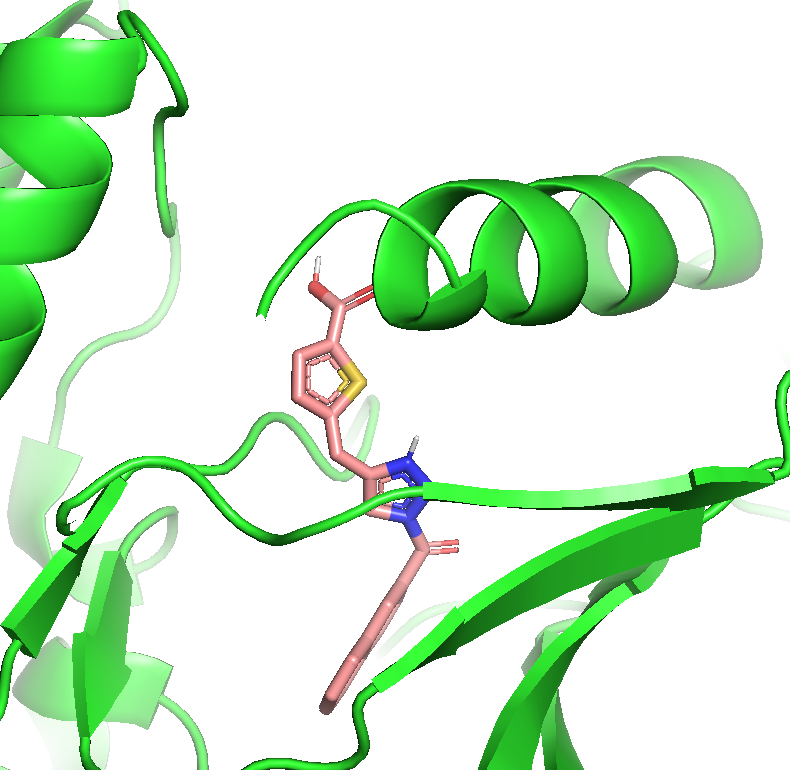}}
\subfigure[-13.1 kcal/mol]{
\includegraphics[width=0.31\linewidth]{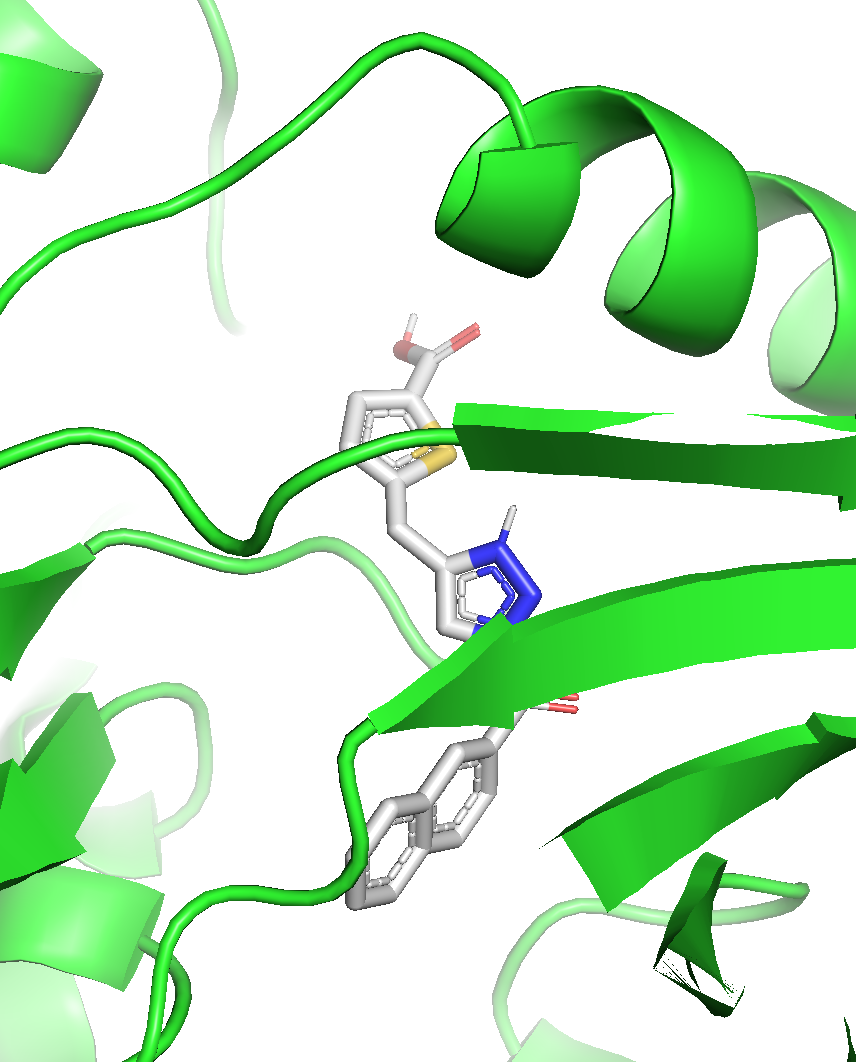}}
\subfigure[-13.0 kcal/mol]{
\includegraphics[width=0.31\linewidth]{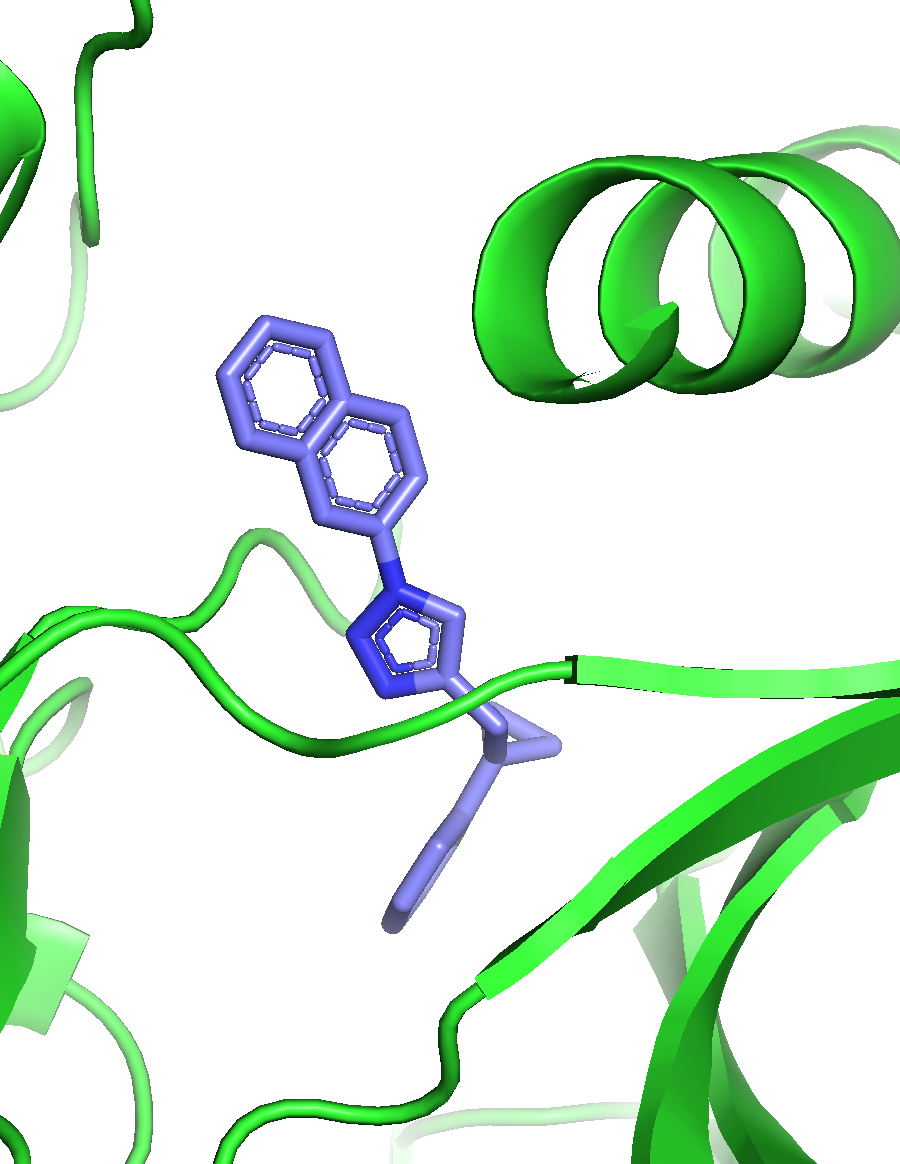}}
\subfigure[-12.9 kcal/mol]{
\includegraphics[width=0.31\linewidth]{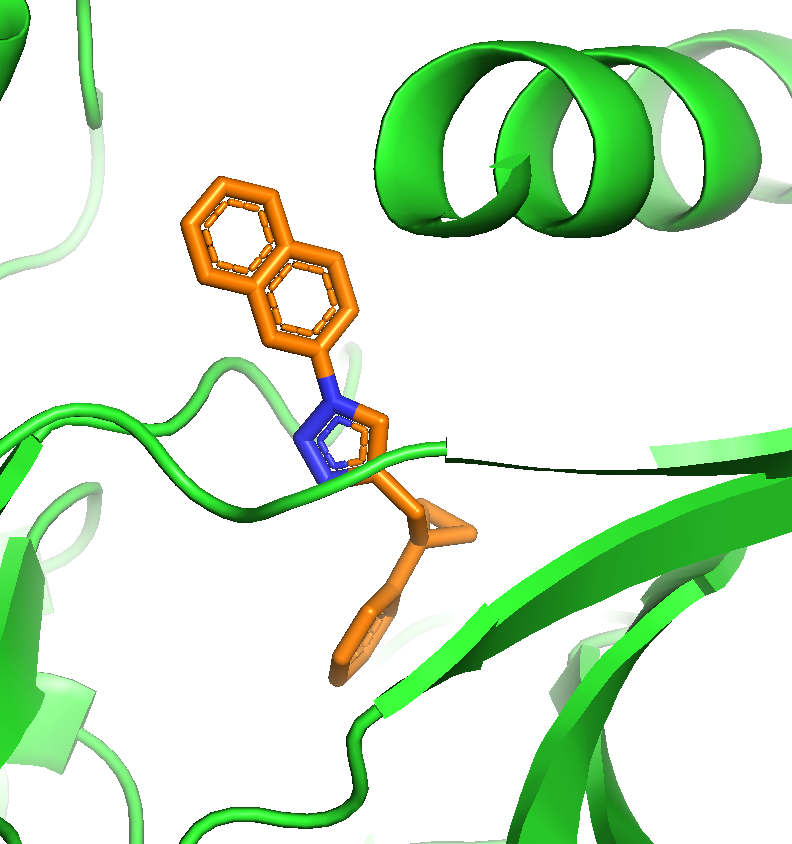}}
% \subfigure[-12.9 kcal/mol]{
% \includegraphics[width=0.31\linewidth]{figure/1iep_7.png}}
\subfigure[-12.8 kcal/mol]{
\includegraphics[width=0.31\linewidth]{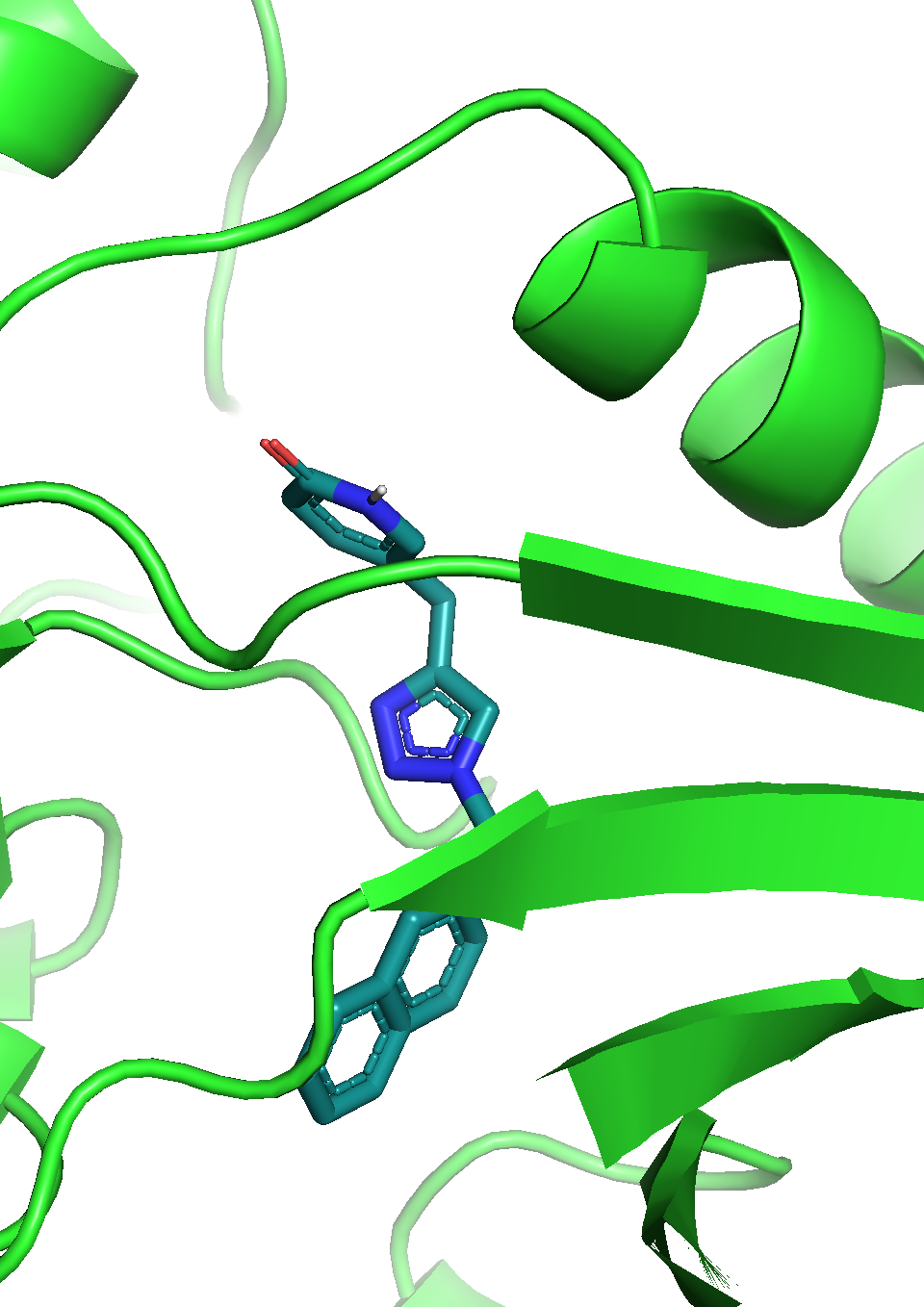}}
\caption{
\marked{
Example of ligand poses (generated by \mname) and binding sites of target structures ``1iep''. 
}}
\label{fig:example_1iep}
\end{figure}

\begin{figure}[t!]
\centering
\subfigure[-12.4 kcal/mol]{
\includegraphics[width=0.31\linewidth]{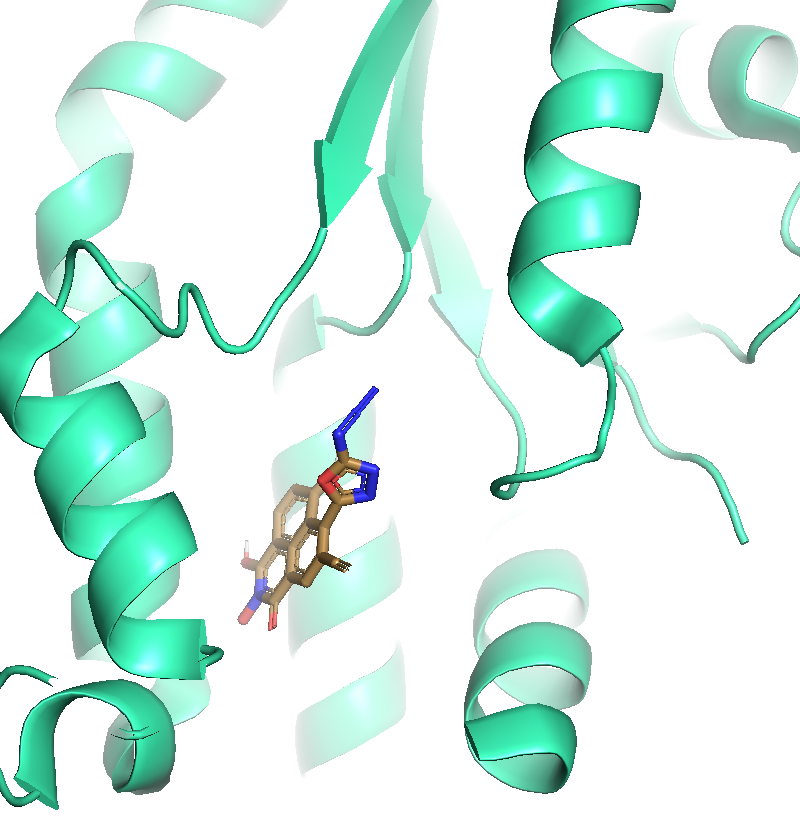}}
\subfigure[-12.4 kcal/mol]{
\includegraphics[width=0.31\linewidth]{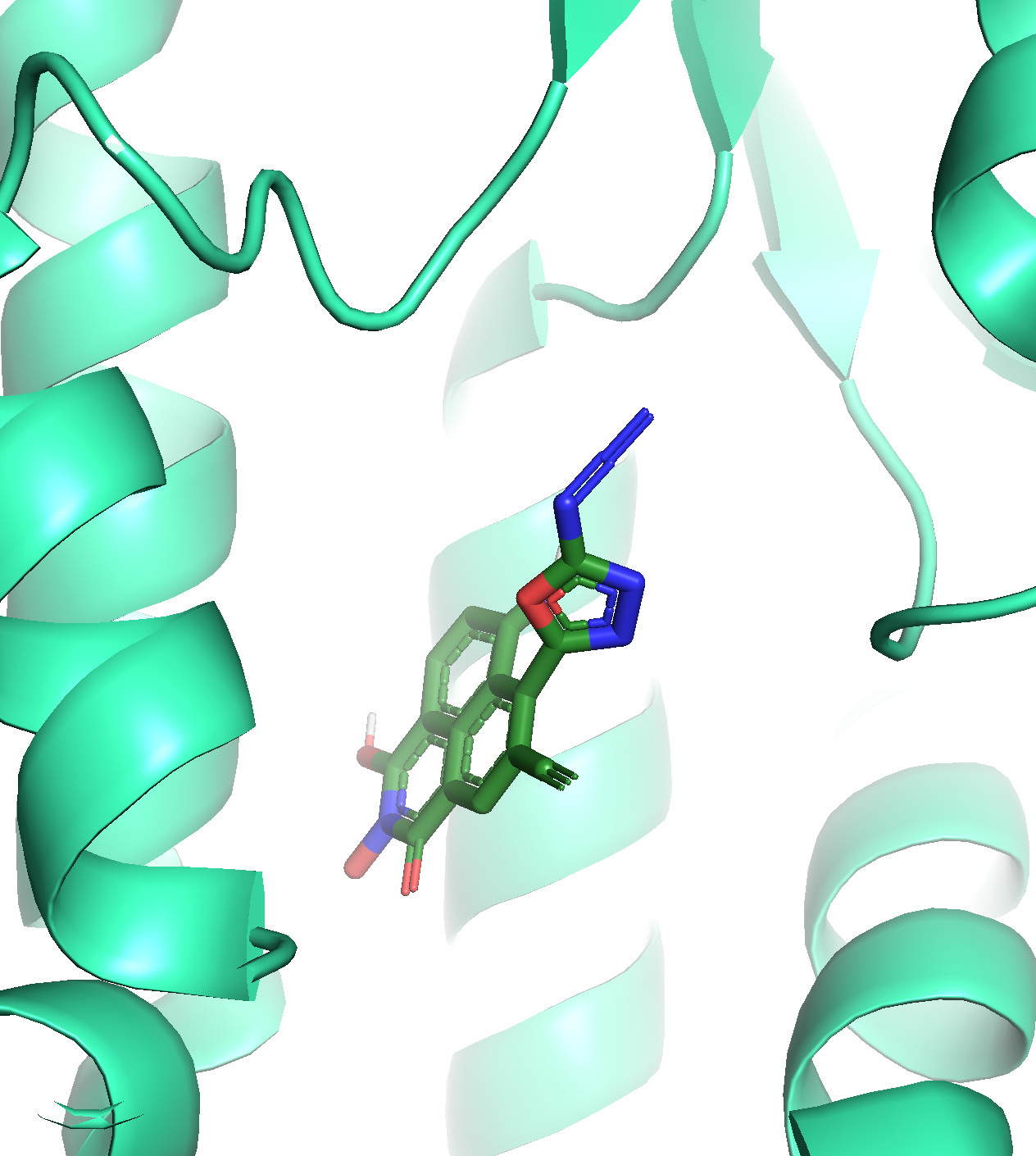}}
\subfigure[-12.3 kcal/mol]{
\includegraphics[width=0.31\linewidth]{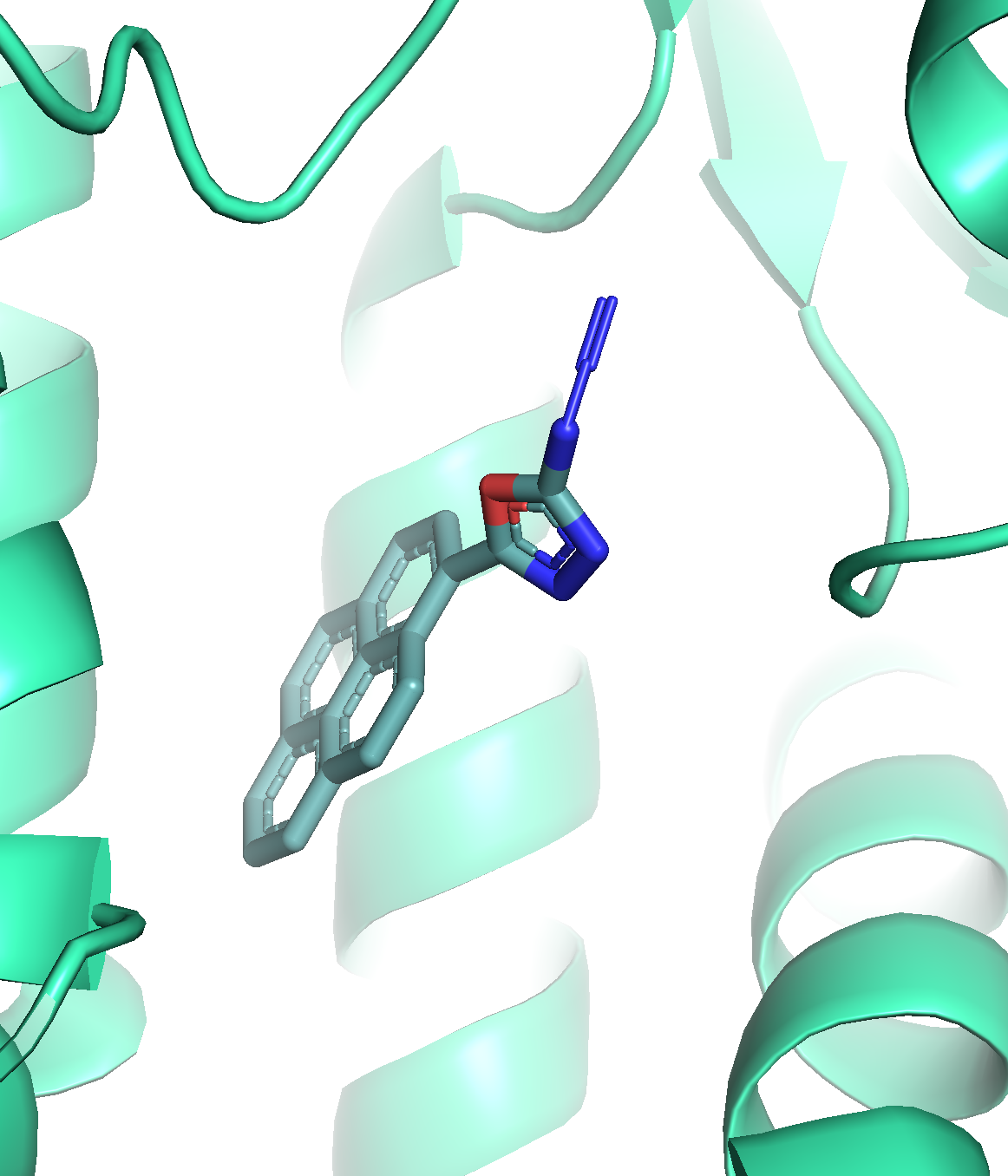}}
\subfigure[-12.3 kcal/mol]{
\includegraphics[width=0.31\linewidth]{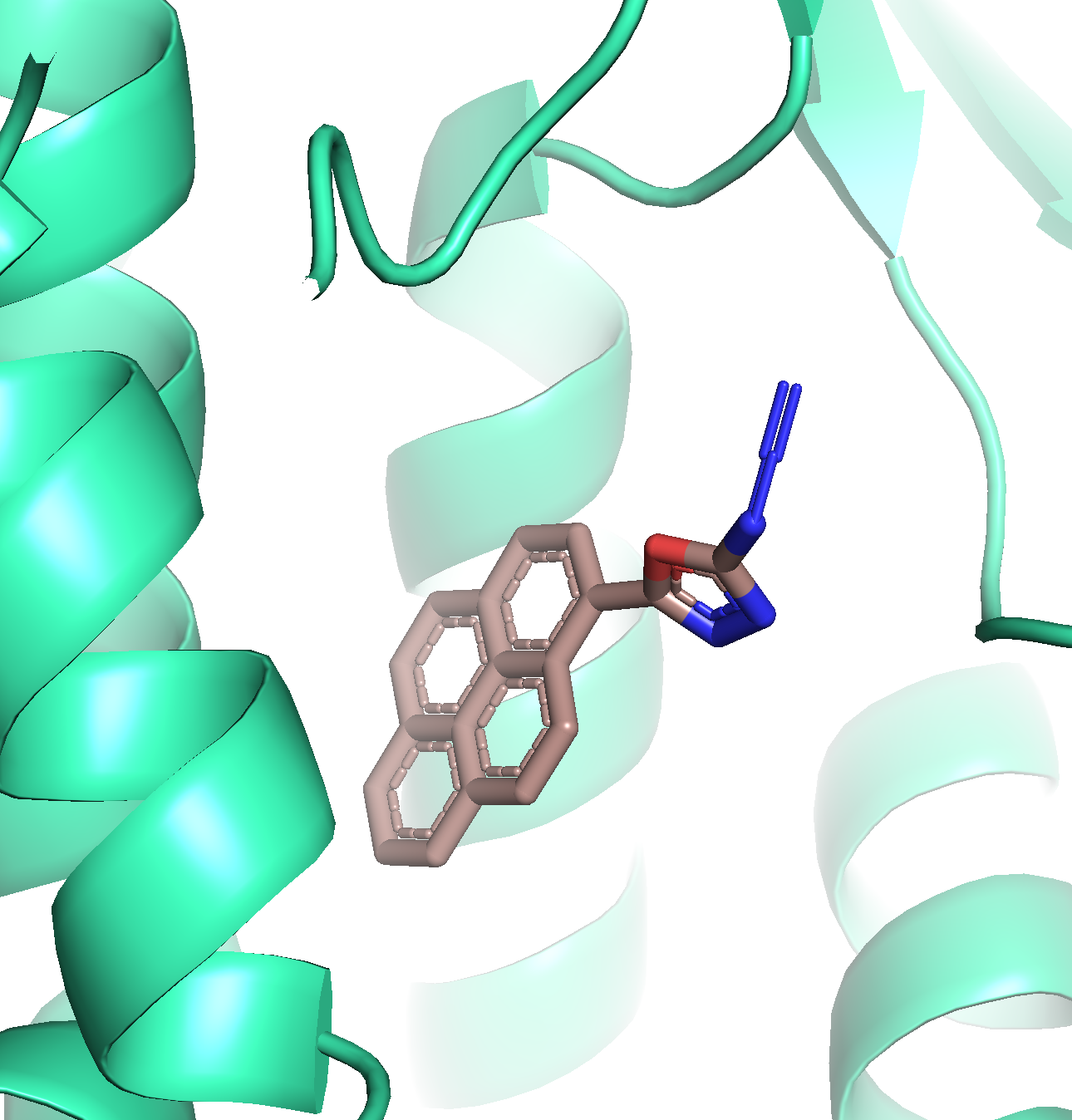}}
\subfigure[-12.3 kcal/mol]{
\includegraphics[width=0.31\linewidth]{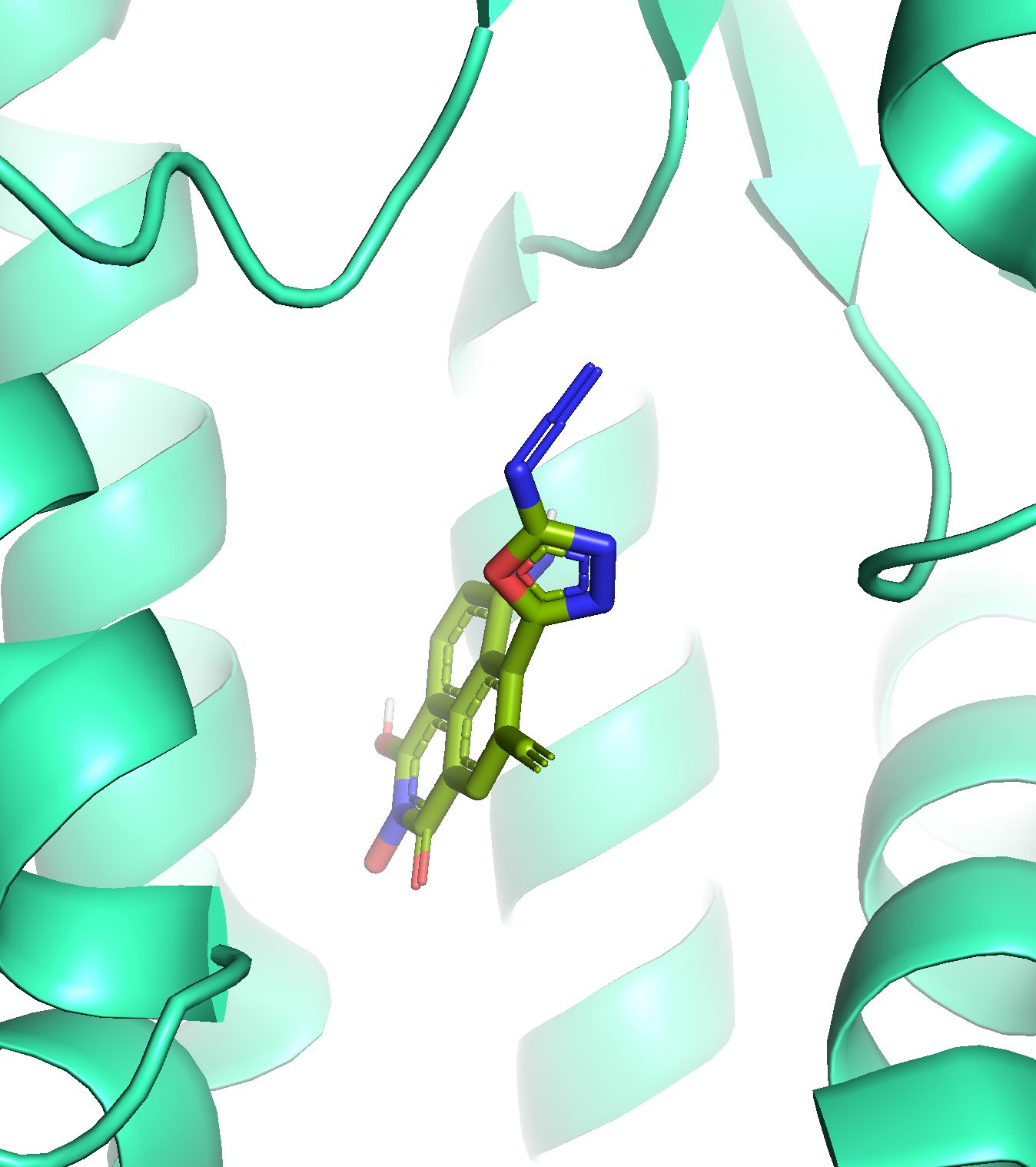}}
\subfigure[-12.2 kcal/mol]{
\includegraphics[width=0.31\linewidth]{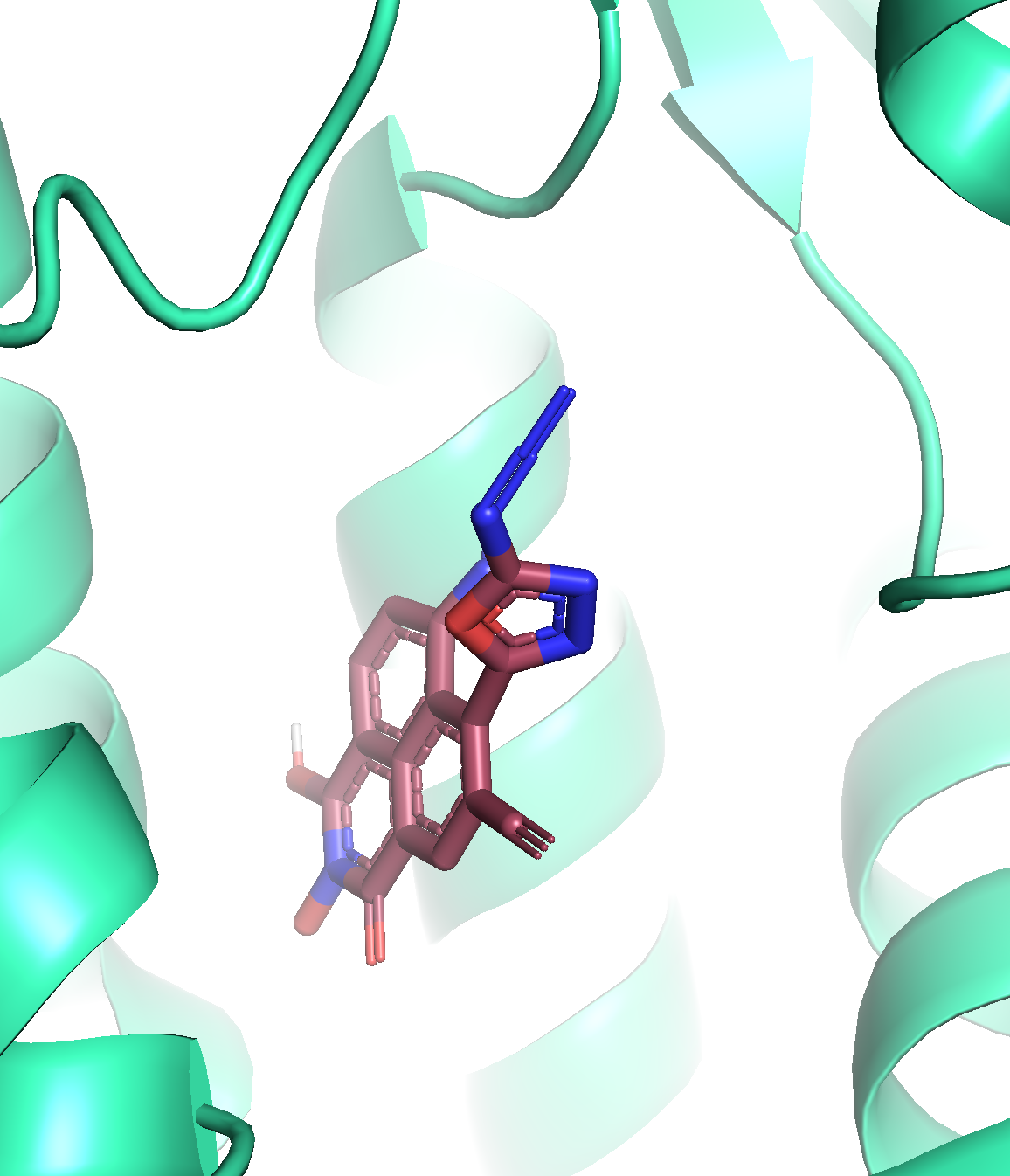}}
\subfigure[-12.2 kcal/mol]{
\includegraphics[width=0.31\linewidth]{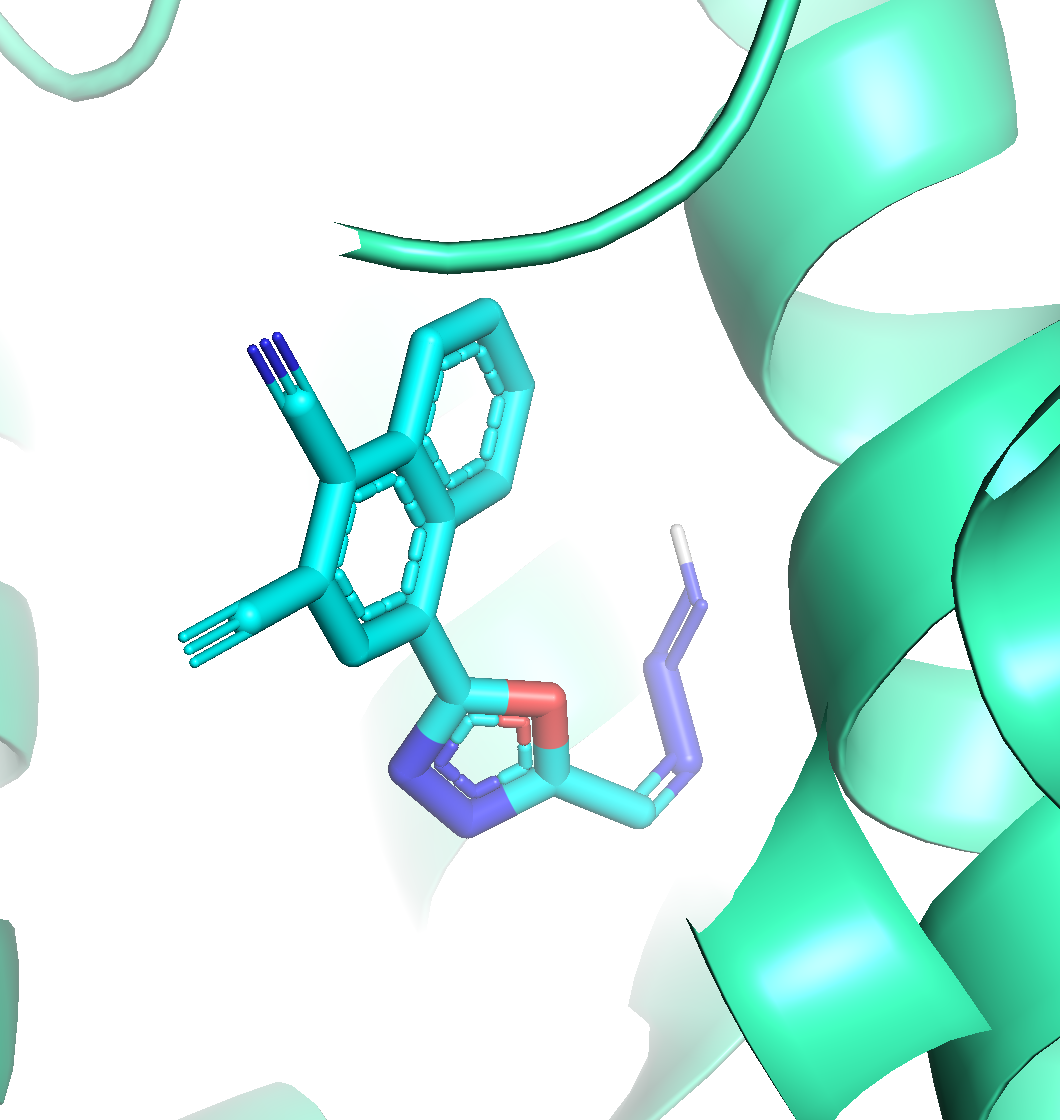}}
\subfigure[-12.1 kcal/mol]{
\includegraphics[width=0.31\linewidth]{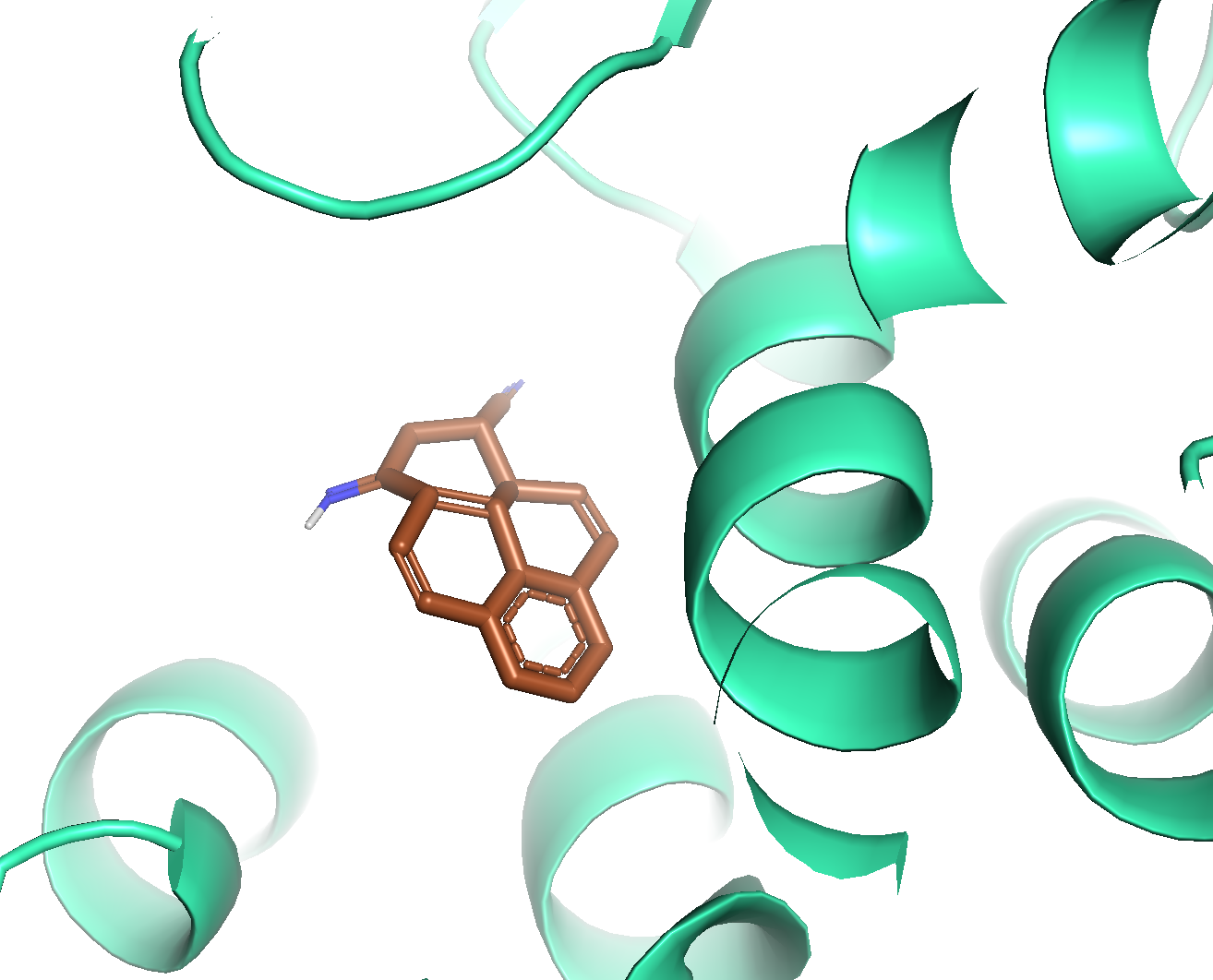}}
\subfigure[-12.1 kcal/mol]{
\includegraphics[width=0.31\linewidth]{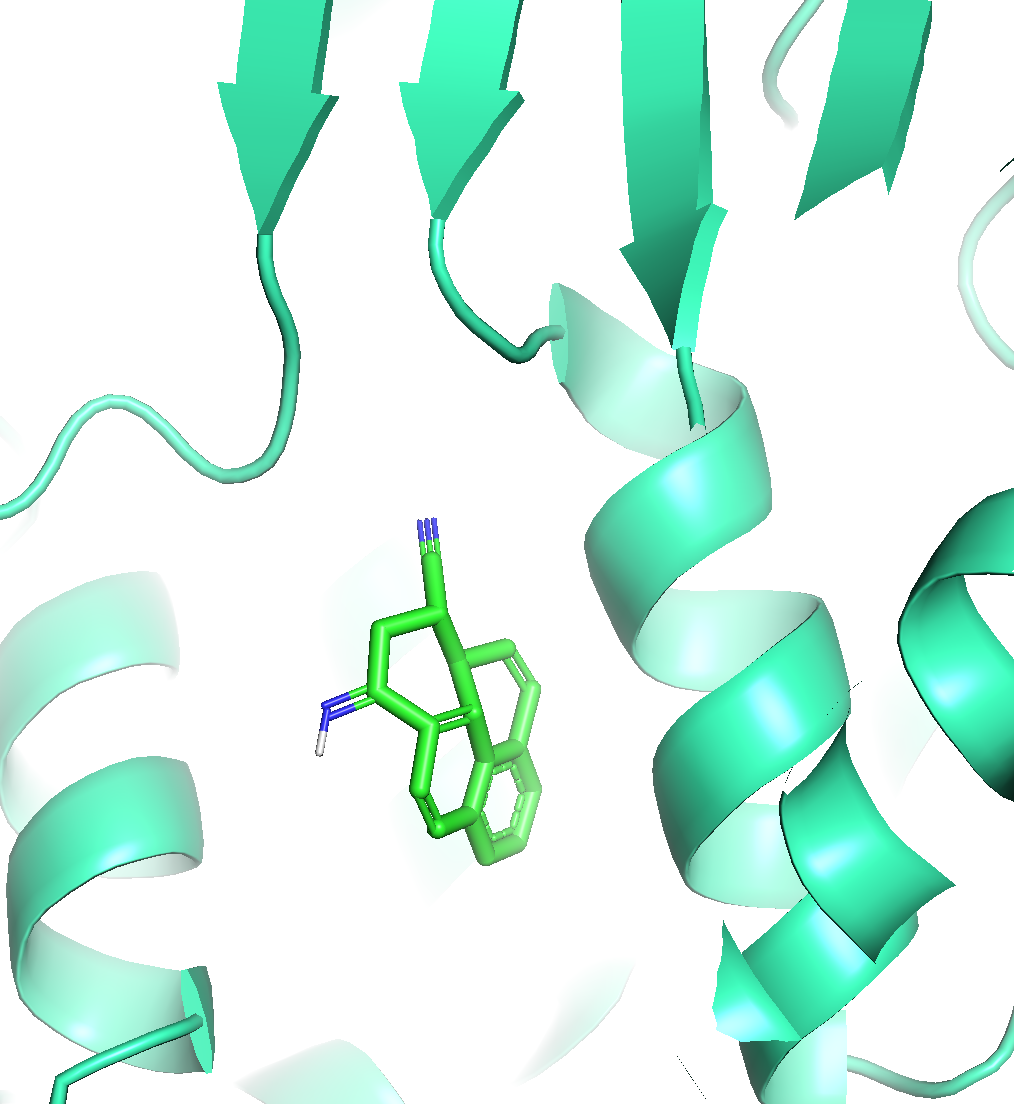}}
\caption{
\marked{
Example of ligand poses (generated by \mname) and binding sites of target structures ``4unn''. 
}}
\label{fig:example_4unn}
\end{figure}

\subsection{Example of the generated ligand}
\label{sec:example_ligand}

\marked{
This section shows some examples of the generated ligands with desirable binding affinity for various target proteins in Figure~\ref{fig:example_2rgp},~\ref{fig:example_3ny8}, ~\ref{fig:example_1iep} and~\ref{fig:example_4unn}, respectively. We observe that the generated ligands bind tightly with the target proteins.  }

\section{Additional Discussion on Related Work}

\noindent\textbf{Methodology}. 
The molecule generations methods can be divided into two categories. The first one is deep generative models (DGMs), which leverage the continuous representation to estimate the data distribution using various kinds of deep neural networks, including variational autoencoder (VAE)~\cite{gomez2018automatic,jin2018junction}, generative adversarial network (GAN)~\cite{guimaraes2017objective,cao2018molgan}, normalizing flow model~\cite{shi2019graphaf,zang2020moflow,luo2021graphdf}, energy based model~\cite{liu2021graphebm} and diffusion model~\cite{xu2021geodiff}, etc. The second one is combinatorial optimization methods, which directly search the discrete chemical space, including genetic algorithm (GA)~\cite{jensen2019graph,nigam2019augmenting,gao2021amortized}, reinforcement learning approaches (RL)~\cite{Olivecrona,You2018-xh,zhou2019optimization,jin2020multi}, Bayesian Optimization (BO)~\cite{korovina2020chembo,moss2020boss}, Monte Carlo Tree Search (MCTS)~\cite{yang2017chemts,yang2020practical,jin2020multi} and Markov Chain Monte Carlo (MCMC)~\cite{fu2021mimosa,xie2021mars,bengio2021gflownet}. 
Deep generative models (DGMs) usually require a large amount of data to fit, which impedes their usage in low data regime. 
In contrast, combinatorial optimization methods  require less training data, while the trade-off is the need to call the optimization oracles during the exploration in the chemical space~\cite{zhou2019optimization,fu2021differentiable,gao2020synthesizability,gao2021amortized,brown2019guacamol}.

Among all the machine learning methods, Genetic algorithm (GA) exhibits superior performance in some standard benchmarks~\cite{brown2019guacamol,huang2021therapeutics,gao2022sample}. The key reason is GA's global assembling strategy. Specifically, in each generation (iteration), GA maintains a population of molecule candidates (a.k.a. parents), and conducts the crossover operation between two (random-selected) parent candidates, which enables relatively large exchanges on molecular sub-graph between molecular graphs. 
However, GA is leveraging random-walk based mutation and crossover operations~\cite{jensen2019graph,spiegel2020autogrow4} and is essentially based on brute-force trial and error.

On the other hand, Reinforcement learning (RL) methods are good at navigating the discrete space via prioritizing the promising searching branches and circumventing brute-force search. 
The current RL-based  methods~\cite{Olivecrona,You2018-xh,zhou2019optimization,jin2020multi} slightly left behind other state-of-the-art combinatorial optimization methods~\cite{fu2021differentiable,huang2021therapeutics}. The main reason is that current RL based molecule optimization approaches are based on auto-regressive assembling strategy, i.e., growing molecules iteratively via adding a basic building block one time,  
where the building block can be either a token in SMILES representation~\cite{Olivecrona} or a substructure in molecular graph representation~\cite{You2018-xh,zhou2019optimization,jin2020multi}. 
Such assembling strategy are essentially local search methods, which hinders the algorithm's ability to overcome the rough optimization landscape (or energy barrier) and is easy to be stuck in the local optimum~\cite{conti2018improving,liu2021decoupling}.

\noindent\textbf{Discussion}. 
Among all the machine learning methods, molecular graph level genetic algorithm (GA) exhibits state-of-the-art performance in some standard molecule optimization benchmarks~\cite{brown2019guacamol,huang2021therapeutics,gao2022sample}. The key reason is GA's assembling manner. Specifically, in each generation (iteration), GA maintains a population of possible candidates (a.k.a. parents), and conducts the crossover operation between two candidates to generate new offspring, which enables thorough exploration to the chemical space. 
However, there is still improvement space for GA. GA are leveraging random-walk based mutation and crossover operations~\cite{jensen2019graph} and suffers from brute-force trial and error strategy. 

On the other hand, reinforcement learning approaches are good at navigating the discrete space via prioritizing the promising decisions that are worth investigating, for example, AlphaGo successfully applied RL to defeat a professional human Go player~\cite{silver2017mastering}. However, the current RL based drug design methods~\cite{Olivecrona,You2018-xh,zhou2019optimization} slightly left behind other state-of-the-art combinatorial optimization methods. The main reason lies at the inferior assembling strategy, which grows molecule in an auto-regressive fashion. It is hard for this kind of local search strategy to overcome the barrier of the objective, so it is easy to be trapped into the local optimum. 

\marked{
Deep learning methods can also enhance genetic algorithm. 
Due to the random selection used in genetic algorithm, it is challenging to apply deep learning methods in the generation of new candidates (molecules in this paper). Deep learning can be used to compose fitness evaluation to select the offspring. For example, GA+D~\cite{nigam2019augmenting} leverages deep neural network as a discriminator to measure the drug’s proximity to the training data, which is incorporated as a scorer in fitness evaluation. \cite{kwon2021evolutionary} train a deep neural network-based property predictor and leverage it to enhance the evolutionary algorithm.
}

In this paper, we attempt to enhance genetic algorithm using reinforcement learning technique. Specifically, we propose \fullname (\mname), which inherits the assembling manner from genetic algorithm and use reinforcement learning to guide the search over the chemical space. 
\cite{ahn2020guiding} also combine RL and GA, which uses LSTM (guided by RL agent) to imitate GA process, however, it is unable to inherit the GA's flexible assembling manner due to the auto-regressive essence of LSTM.

\end{document}